\documentclass[10pt,preprint2]{aastex} 
\usepackage{epsfig}
\usepackage{amsmath}
\usepackage[figuresleft]{rotating}
\usepackage{lscape}
\usepackage{epsfig}
\usepackage{url}
\newcommand{\kms}{\hbox{km~s$^{-1}$}}

\newcommand{\msun}{\hbox{$M_{\odot}$}}

\newcommand{\cmtwo}{\hbox{cm$^{-2}$}}
\newcommand{\HI}{\hbox{{\rm H}\kern 0.1em{\sc i}}}
\newcommand{\HII}{\hbox{{\rm H}\kern 0.1em{\sc ii}}}
\newcommand{\CIV}{\hbox{{\rm C}\kern 0.1em{\sc iv}}}

\newcommand{\nhi}{\hbox{N$_{\mbox{\scriptsize{\HI}}}$}}
\newcommand{\snhi}{\hbox{$\epsilon_{N_{\mbox{\tiny{\HI}}}}$}}
\newcommand{\mhi}{\hbox{$\Sigma_{M_{\mbox{{\tiny{\HI}}}}}$}}
\newcommand{\smhi}{\hbox{$\epsilon_{\Sigma_{{\mbox {\tiny {\rm MHI}}}}}$}}

\newcommand{\shear}{\hbox{$dv_{{\rm los}}/dr_{\perp}$}}
\newcommand{\vuse}{\hbox{$v_{{\rm los}}$}}
\newcommand{\vdisp}{\hbox{$\sigma_{{\rm los}}$}}

\newcommand{\ke}{\hbox{$\Sigma_{{\rm KE}}$}}

\newcommand{\MVcomp}{\hbox{$M_{{\rm V,50\%}}$}}

\newcommand{\kmskpc}{\hbox{km s$^{-1}$ kpc$^{-1}$}}

\newcommand{\boldHI}{\hbox{\textbf{H\footnotesize{I}}}}

\newcommand{\noprint}[1]{}

\shorttitle{Star Clusters and \HI\ in Tidal Tails}
\shortauthors{B.\ Mullan et al.\,}

\begin{document}
\sloppy

\newcounter{figcount}
\setcounter{figcount}{1}


\title{Under Pressure: Star Clusters and the Neutral Hydrogen Medium of Tidal Tails} 

\author{B.\ Mullan\altaffilmark{1}, A.\ A.\ Kepley\altaffilmark{2,}\altaffilmark{3}, A.\ Maybhate\altaffilmark{4}, J.\ English\altaffilmark{5}, K.\ Knierman\altaffilmark{6}, J.\ E.\ Hibbard\altaffilmark{3}, N.\ Bastian\altaffilmark{7}, J.\ C.\ Charlton\altaffilmark{1}, P.\ R.\ Durrell\altaffilmark{8}, C.\ Gronwall\altaffilmark{1,9}, D.\ Elmegreen\altaffilmark{10}, I.\ S.\ Konstantopoulos\altaffilmark{11}}

\altaffiltext{1}{Pennsylvania State University, Department of Astronomy \& Astrophysics, 525 Davey Lab University Park PA 16803; mullan@astro.psu.edu}
\altaffiltext{2}{Department of Astronomy, University of Virginia, P.O. Box 400325, Charlottesville, VA 22904-4325}
\altaffiltext{3}{National Radio Astronomy Observatory, 520 Edgemont Road, Charlottesville, VA 22903-2475}
\altaffiltext{4}{Space Telescope Science Institute, 3700 San Martin Drive, Baltimore, MD 21218}
\altaffiltext{5}{University of Manitoba, Department of Physics and Astronomy, Winnipeg, Manitoba R3T 2N2, Canada}
\altaffiltext{6}{Arizona State University, School of Earth and Space Exploration, Bateman Physical Sciences Center F-wing Room 686,Tempe, AZ 85287-1404}
\altaffiltext{7}{Excellence Cluster Universe, Technische Universit\"{a}t M\"{u}nchen, Boltzmannstra{\ss}e 2, Garching, Gernmany}
\altaffiltext{8}{Youngstown State University, Department of Physics and Astronomy, Youngstown, OH 44555}
\altaffiltext{9}{Institute for Gravitation and the Cosmos, The Pennsylvania State University, University Park, PA 16802}
\altaffiltext{10}{Vassar College, Department of Physics \& Astronomy, Box 745, Poughkeepsie, NY 12604}
\altaffiltext{11}{Australian Astronomical Observatory, PO Box 915, North Ryde NSW 1670, Australia}

\slugcomment{Accepted for publication in the Astrophysical Journal}




\begin{abstract}

Using archival data from ATCA, WSRT, and the VLA, we have analyzed the \HI\ emission of 22 tidal tail regions of the Mullan et al.\ sample of pairwise interacting galaxies. We have measured the column densities, line-of-sight velocity dispersions, and kinetic energy densities on $\sim$kpc scales. We also constructed a tracer of the line-of-sight velocity gradient over $\sim$10 kpc scales. We compared the distributions of these properties between regions that do and do not contain massive star cluster candidates ($M_{{\rm V}}$ $<$ -8.5; $\sim$ 10$^4$--10$^6$ \msun\ as observed in \textit{HST} WFPC2 \textit{VI} data). In agreement with Maybhate et al., we find that a local, $\sim$kpc-scale column density of log \nhi\ $\gtrsim$ 20.6 \cmtwo\ is frequently required for detecting clustered star formation. This \HI\ gas also tends to be turbulent, with line-of-sight velocity dispersions \vdisp\ $\approx$ 10--75 \kms, implying high kinetic energy densities (log \ke\ $>$ 46 erg pc$^{-2}$). Thus high \HI\ densities and pressures, partly determined by the tail dynamical age and other interaction characteristics, are connected to large-scale cluster formation in tidal tails overall. Lastly, we find that the high mechanical energy densities of the gas are likely not generally due to feedback from star formation. Rather, these properties are more likely to be a cause of star formation than a result.

\end{abstract}




\section{Introduction}
\label{sec:intro}
\renewcommand{\thefootnote}{\arabic{footnote}}


Tidal tails are common products of disk galaxy interactions (e.g., \citealp{toomre77}; \citealp{schweizer78}). These extended, pronounced structures begin their lives shortly after the initial encounter of their progenitor galaxies, drawn from a combination of ISM and stars of these systems' outer regions. While relatively short-lived ($\lesssim$1 Gyr), tidal debris exhibits an extraordinary variety of morphological and kinematic phenomena. These range from long, thin tails, to diffuse plumes, to kinematically distinct tidal dwarf galaxies (TDGs), all characteristic of the numerous interactions that populate the Local Volume \citep{duc04}.


Tidal tails are composed of $\sim$10$^9$ \msun\ of gas---and some small underlying population of stars---pulled from the outer regions of their host galaxies by gravitational torques and the tidal field induced by an encounter with another system (see \citealp{bournaud11} for a review). Although there are wide variations in metallicity by galaxy, studies of non-interacting disk galaxies indicate that gas from these regions prior to the interaction are generally \HI-rich and metal/H$_2$-poor \citep{bigiel10}. The neutral hydrogen medium is thus the \textit{de facto} tracer of the kpc-scale ISM in tidal tails, and has a long history of observation (e.g., \citealp{vdh79}; \citealp{simkin87}; \citealp{NGC2444_pub}; \citealp{yun94}; \citealp{NGC6872_pub}; \citealp{sengupta11}). Decades of study have revealed that the \HI\ in tidal debris is a generally inhomogeneous, kinematically disturbed medium, subject to a variable tidal potential throughout the course of the interaction that can be both locally ``extensive" (pulling material apart) and ``compressive" (pushing it together; \citealp{renaud09}).


The importance of \HI\ column density, \nhi, in star formation has observational precedent in many extragalactic environments. In a self-gravitating gas disk, stability analysis shows there is likely a threshold \nhi\ that permits it \citep{kennicutt89}. This may be true across a variety of cosmic venues. On $\sim$500 pc$^2$ scales within dwarf irregular galaxies, this threshold may be $\approx$10$^{21}$ cm$^{-2}$ (\citealp{skillman86}; \citealp{skillman87}). Moreover, UV emission---a key diagnostic for recent star formation in the past $\sim$100 Myr---is observed where the \HI\ column density reaches or exceeds $\approx$2.5 $\times$ 10$^{20}$ cm$^{-2}$ (\citealp{neff05}; \citealp{thilker05}; \citealp{gdp05}). This is consistent with \citet{schaye}, who determine that a minimum surface density of $\approx$3 \msun\ pc$^{-2}$ is necessary to first form a cold neutral phase, with some dispersion from metallicity and the interstellar radiation field. The production of the cold molecular hydrogen necessary for gravitational collapse and star formation would then occur, the amount of which is controlled largely by the \HI\ density \citep{krumholz09}. The ratio of H$_2$ to \HI\ surface densities increases almost linearly with interstellar pressure and the \HI\ densities that promote it, and saturates at a mass density of $\sim$10 \msun\ pc$^{-2}$ (\citealp{wong02}; \citealp{blitz04}; \citealp{blitz06}; \citealp{things}; \citealp{leroy08}). Star formation may then scale with respect to the density and amount of molecular gas available for collapse (e.g., \citealp{KSlaw}; \citealp{liu11} and references therein).


In tidal tails, large distances ($\lesssim$100 Mpc) and low surface brightnesses can make observing this star formation difficult. Compact star clusters and their populations are therefore convenient surrogates for this task; they can be considered high-signal, luminous tracers of the local star formation history of their environments (\citealp{iraklis09}; \citealp{RdG09}). The most massive ($\sim$10$^5$--10$^6$ \msun) of these star clusters or cluster candidates have been observed in some tidal tails (\citealp{RdG}; \citealp{tran}; \citealp{bastian05}), but not all (\citealp{M11}; \citealp{K03}). This implies that the conditions required for extensive star cluster formation (and the detection of the most massive clusters by the size-of-sample effect; e.g., \citealp{gieles08})  are not always present in these environments. Consequently, tails may occupy the edge of the physical parameter space that allows these structures to form and/or survive.


The content and character of the observable \HI\ medium may have a role to play in star cluster formation. \citet{aparna07} examined a sample of tidal debris from late-stage ($\gtrsim$400 Myr old) interactions and determined that a local, $\sim$kpc-scale \HI\ column density of $\approx$10$^{20.6}$ cm$^{-2}$ or higher is required for finding significant populations of luminous star clusters. This is similar to the results of \citet{walter06}, who find that \HII\ regions are only observed where log \nhi\ $\geq$ 21.0 cm$^{-2}$ on 200-pc$^{2}$ scales for the tidal arms of NGC~3077.


However, \citet{aparna07} stipulate that \nhi\ is not a sufficient condition for finding bright, massive star clusters. They cite the debris of NGC~3921 as an environment where a non-negligible fraction of neutral hydrogen exists above their fiducial threshold value, but massive star clusters are unlikely to exist. Other properties related to the kinematics of the \HI\ medium may also be important. For instance, disruptive shear might influence a star-forming environment to ``prefer" forming associations over bound clusters \citep{weidner10}. This may be observed in the central regions of the Antennae, where \citet{whitmore99} do not find massive clusters where the line-of-sight velocity gradient is high.


Morphological asymmetries in the optical and \HI\ debris components also indicate that other, large-scale physics may influence the \HI-star cluster connection.  Dust obscuration, ram-pressure stripping by the galaxy interaction, and photoionization from a central galactic superwind or locally by a previous generation of efficient star formation \citep{hibbard2000} may additionally confuse the issue. Some of these phenomena, along with the bulk properties of the tidal tail, will be dependent on age, initial galactic orientations and speeds, and other properties of the interactions themselves \citep{duc04}.   


The inconsistent production of star cluster populations in tidal tails may also be partially due to the kinematic character of the $\sim$kpc-scale \HI\ medium. Under ``normal" circumstances in many parts of isolated galaxies, the \HI\ medium generally consists of a mixture of cold (T $\sim$ 100 K) \HI\ clouds embedded in a warm \HI\ medium (T $\sim$ 8000 K) with characteristic thermal linewidths $\approx$6--8 km s$^{-1}$ (\citealp{mckee77}; \citealp{wolfire95}; \citealp{wolfire03}). Past the canonical optical boundary $R_{25}$, the warm ISM may be heated by UV radiation from the interior galaxy or the extragalactic background \citep{schaye}. In tidal tails, however, observed linewidths often exceed thermal values. These motions are typically attributed to turbulence in the gas, which may be generated by feedback from star formation \citep{tamburro} or magnetic effects (e.g., \citealp{sellwood99}). Perhaps most importantly, turbulence can also be produced by the large-scale shocks and gravitational instabilities triggered by the interaction \citep{bournaud10}.  


The \textit{pressures} exerted by turbulent warm \HI\ and the reservoir of high-density gas they produce may therefore engender the conditions necessary for pockets of the cold neutral medium \citep{wolfire03} to exist on unobservable $\sim$pc scales. If provided ample dust, these regions become sites for H$_2$ production and subsequent star formation. Observationally, interacting systems do indeed appear to generally contain more molecular gas than their quiescent counterparts \citep{casasola04}. If gas becomes too turbulent, however, it may interfere with its own cooling processes \citep{wolfire03} and prevent high-density molecular gas from either forming, or cooling efficiently if already present. For example, \citet{guillard12} detect highly turbulent CO (with velocity dispersions $\gtrsim$100 \kms) in collisionally shocked regions of Stephan's Quintet. These regions also have low PAH-to-CO surface luminosities, indicating low star formation efficiency from an evidently high gas heating rate. Moreover, sufficient \HI\ density must be required to build enough H$_2$ to self-shield from photodisassociation by the extragalactic background or relatively nearby stellar sources. Thus, it is likely that star formation is contingent on \HI\ density as well as kinematics, dictated by large ($\sim$kpc)-scale phenomena.


Interstellar pressure not only enhances star formation in general, but it may also help dictate the types of stellar ``packaging" that are produced. In tidal tails, the presence of dense, turbulent gas may conceivably encourage the formation of tidal dwarf galaxies (\citealp{elmegreen93}; \citealp{elmegreen97}) or other bound structures like compact star clusters. High pressures help prevent a cluster from disruption in its early gas expulsion phase (e.g., \citealp{whitmore07}; \citealp{bastian08}; and references therein). If star formation depends on gas density, then high pressures promote clustered star formation at a rate that increases as the local star formation rate (SFR) increases \citep{larsen00}. In addition, \citet{elmegreen08} finds that bound clusters can form at lower average densities at high Mach numbers than at low Mach numbers; thus the threshold density needed to form clusters is effectively reduced in highly turbulent media and clusters can become a favored ``mode" of star formation.

Tidal tails appear to offer a unique environment where the individual \HI-related variables that may influence star cluster formation can be systematically tested. In this paper, we explore the complex relationship between galaxy interactions, their effects on the \HI\ medium in their tidal tails, and the star clusters that are selectively found within. This study is the first detailed investigation of the kpc-scale column densities and kinematic character of the neutral hydrogen medium and its relationship to star cluster populations in this type of extragalactic environment. We extend the sample of merging galaxies of \citet{aparna07} and \citet{K03} with archival Very Large Array (VLA), Australia Telescope Compact Array (ATCA), and Westerbork Synthesis Radio Telescope (WSRT) 21-cm observations of the tidal tail sample studied by \citet{M11}. The 22 tails total span a wide range of dynamical ages, merging mass ratios, and previously established star cluster population sizes and characteristics.  An overview of the \HI\ observations and reduction is provided in Section \ref{sec:obs}, along with selection of star cluster candidate positions and photometry from optical data. In Section \ref{sec:analysis}, we present our methods of measuring various \HI\ quantities, e.g., column density and velocity dispersion. We examine the local \HI\ characteristics of the tails with respect to the cluster populations in Section \ref{sec:results}. We present our results in Section \ref{sec:tailcomp}, and discuss trends and differences in the \HI\ content among established types of interactions. In this section we also examine turbulence and its role as a cause or result of star formation. Lastly, we offer conclusions and final remarks in Section \ref{sec:conclusions}.



\section{Observations}
\label{sec:obs}

\subsection{The Tidal Tail Sample and Their Star Cluster Candidates}
\label{sec:hstdata}

We obtained the \textit{HST} WFPC2 data for the tidal tail regions studied by \citet{M11}; hereafter M11, \citet{aparna07}, and \citet{K03}. These include contributions from cycles 7 (GO-7466) and 16 (GO-11134), with imaging performed in the F555W/F814W and F606W/F814W bands, respectively. The systems that produced these tails were originally selected to represent a large parameter space of observable and dynamical characteristics of galaxy interactions present in the local universe, from optical/\HI\ properties, to progenitor mass ratios, to interaction ages (M11). In this study, we use the same naming convention for debris regions as these publications; e.g., NGC~1487E and NGC~1487W refers to the eastern and western tails of NGC~1487, respectively, while the three WFPC2 pointings of NGC~4038/9 are denoted by NGC~4038A, NGC~4038B, and NGC~4038C. 

We also acquired the source catalogs of J2000 positions and photometry of compact objects detected in \citet{aparna07} and M11; the latter contains sources detected by \citet{K03}. Source photometry had been transformed from F555W/F606W and F814W to standard $V$ and $I$ bandpasses with the \citet{holtzman95} prescription\footnote{M11 employed updated methods for recent data and instrumental characteristics documented in \url{http://purcell.as.arizona.edu/wfpc2\_calib/2008\_07\_19.html}.} in these studies to facilitate comparison with the earlier photometry published in standard magnitudes. Photometry had also been corrected for Galactic extinction using \citet{schlegel} and \citet{girardi}; see M11 and \citet{K03} for details. Objects detected and measured with this methodology were defined as star cluster candidates (SCCs) if they met the following color-magnitude criteria:

\begin{itemize}
	\item $M_V$ $<$ -8.5. Many studies use a similar magnitude cutoff ($M_V$ = -8 to -9: \citealp{K03}; \citealp{K07}; \citealp{schweizer96}; \citealp{whitmore99}) because of detection limits and the threat of contamination from non-cluster sources at fainter magnitudes. Source detection within the tail regions of the M11 sample was previously found to be more than 50\% complete at $M_V <$ -8.5, so we implement this criterion here. M11 also reaffirmed the findings of \citet{efremov87}; i.e.\ that this magnitude limit sufficiently protects against contamination from single main sequence and post main sequence stars.
 	\item $V-I$ $<$ 1.0. This allows for a wide range of possible metallicities, extinctions, and ages in our cluster population (e.g., \citealp{Peterson}; \citealp{Ajhar}, \citealp{whitmore95}; \citealp{Kundu}). This $V-I$ limit was also selected to reflect uncertainties or additional reddening from dynamical \citep{pzwart10}, stochastic \citep{maiz09}, and disruptive \citep{anders} effects.
\end{itemize}
	
For similar reasons, M11 and \citet{aparna07} used $V-I$ $<$ 2.0 as a color limit in their work. In practice, comparing the completeness curves of in-tail vs. out-of-tail areas reveal relatively limited tail extinctions (A$_V \sim$ 0.5; M11), and stochastic effects should be restricted to $V-I$ changes at the $\lesssim$5\% level \citep{anders} for the $M_V <$ -8.5, $\sim$10$^5$--10$^6$ \msun\ clusters we study. Furthermore, an examination of the M11 sources and their color-magnitude diagrams reveals that a more stringent cutoff helps reduce the number of out-of-tail and presumably background sources detected, which tend to be redder ($V-I$ $\gtrsim$ 1). These results are substantiated by the multiwavelength observations and selection criteria of \citet{trancho12}.

\subsection{\boldHI\ Calibration and Reduction}
\label{sec:hicalib}

\mbox{Table 1} lists the tails from M11, \citet{K03}, and \citet{aparna07} for which archival \HI\ data exist (22 tails total). We include the dynamical ages and mass ratios of these interacting systems for later reference. Our mass ratio values and tail ages are taken directly from M11 and sources therein. The mass ratio is defined as $M_1$/$M_2$, where $M_1$ and $M_2$ are the respective masses of the perturbing and main galaxies. Upon further review, we inverted the reported mass ratios of NGC~2782E and NGC~4747 (to 4 and 20) to reflect the understanding in the literature that the tail regions examined here are likely pulled from the smaller-massed galaxies of their interactions.

Neutral hydrogen data for NGC~1614, NGC~2782, NGC~2992/3, NGC~2444, and NGC~2535 were taken from the VLA archives\footnote{The National Radio Astronomy Observatory is a facility of the National Science Foundation operated under cooperative agreement by Associated Universities, Inc}. The data were reduced in AIPS using the standard procedures detailed in the AIPS Cookbook\footnote{\url{http://www.aips.nrao.edu/cook.html}}. Here we provide a brief outline of this methodology. In addition to the program source, a bright primary (gain) calibrator and a secondary (phase) calibrator were observed. The primary calibrator was usually observed at least once in the observing program and the secondary calibrator observed every approximately 40 minutes to an hour of time on source. Using the channel-0 data, the flux density of the primary calibrator was found using {\em setjy}. Then the amplitudes and phases of both the primary calibrator and the secondary calibrator were determined using {\em calib}. The flux density of the secondary calibrator was determined from the primary calibrator using the task {\em getjy}. The calibration was applied to the source using {\em clcal}. The channel-zero calibration solutions were copied to the spectral line data. Finally, the spectral line data was bandpass calibrated in {\em bpass} using the primary calibrator.

The calibrated data were test imaged in {\em imagr} and the line-free channels determined. Using this information, the data were continuum subtracted using {\em uvlsf}. The final, continuum subtracted images were produced in {\em imagr}; the data were imaged with varying values of the ``robust" weighting parameter \citep{robust}, with values from 0--5. A larger robust parameter enhances detection of diffuse \HI\ emission at the expense of angular resolution. In most cases, the diffuse emission was resolved out of the high resolution image. In the case of two overlapping intermediate frequencies (IFs), the images in the overlapping channels were averaged and a combined cube formed.

We acquired a reduced data cube for NGC~4747 from the Westerbork Synthesis Radio Telescope (WSRT) web archive\footnote{The Westerbork Synthesis Radio Telescope is operated by the Netherlands Institute for Radio Astronomy ASTRON, with support of NWO.}. These data were originally taken, calibrated, and reduced as a part of the WHISP program \citep{NGC4747_pub}. Here, we use the cube generated to $\approx$60$\arcsec$ resolution, which we found was an appropriate balance between spatial resolution and detecting low-signal \HI\ in the tidal regions of NGC~4747.

The neutral hydrogen data for NGC~6872 and NGC~1487 were obtained from the ATCA archives. The NGC~1487 data were processed using Miriad \citep{miriad} for all steps, including the calibration (with the tasks {\em uvflag}, {\em blflag}, and {\em mfcal}) and removal of continuum (with the task {\em uvlin}).  The primary calibrators were 1934-638 and 1151-348, while the secondary was 0332-403.  We produced images by combining the array configurations listed in \mbox{Table 1}---using standard tasks {\em invert}, {\em clean}, and {\em restor}. The data for NGC~6872 were calibrated with a combination of AIPS and Miriad. AIPS was used to flag ({\em spflag}) and image the data ({\em imagr}), while data was calibrated (using {\em mfcal} because the nearby primary calibrator 1934-638 was used as the secondary calibrator) in Miriad. The ATCA does not doppler track, so the correct velocity definition was determined in Miriad and applied to the data in AIPS using the task {\em cvel}.

We also obtained the reduced cubes used by \citet{aparna07} in their \HI\ analysis. Their reduction strategies are similar, optimizing diffuse \HI\ emission in the VLA-observed tails of NGC~4038, NGC~7252, and NGC~3921, and the ATCA-observed tails of NGC~3256. Data for NGC~520 were obtained from \citet{NGC520_pub}. A brief description of all radio data is presented in \mbox{Table 1}. \mbox{Table 1} includes source distances, beam dimensions ($b_1$ and $b_2$), and beam position angles (P.A.). We also list angular and physical pixel areas ($A_{{\rm pix}}$), channel widths ($dv$), 1$\sigma$ detection limits in single channel integrated flux densities $\epsilon_{S_i}dv$ (10$^{-4}$ Jy km s$^{-1}$ channel$^{-1}$) and column densities \snhi\, array setups, and original publications and/or project codes for the observations. We use luminosity distances for all systems (assuming $H_0$ = 73 km s$^{-1}$ kpc$^{-1}$), except for NGC~4038 where we adopt the distance measurement of \citet{saviane}. This was done for consistency with M11. All cubes were precessed to J2000 for direct comparison with SCC positions.

A cursory glance at \mbox{Table 1} shows that the pixel sizes of the cubes are not a fixed fraction of the beam sizes. The size and dimensions of the synthesized beam depend on the sampling (and weighting/tapering) of the $uv$ plane in observing these sources, which is a function of the array configurations used, observation program design, and source declination. Works cited in the Original Publications column contain additional information on these issues for each observation. The pixel size, meanwhile, is chosen to avoid undersampling the beam (fewer than $\approx$2--3 pixels across the beam) and avoid large computational times ($\gtrsim$10 pixels across the beam, depending on the original map size). Different pixel sizes are selected between these extremes  to maintain sufficient signal within pixels and individual channels for the varying source intensities of the tidal tail sample.


\begin{sidewaystable*}[p]
{\scriptsize
 \centering
\begin{tabular*}{1.0\textwidth}{@{\extracolsep{\fill}}lrrr|rrrcc|rrrr}
\multicolumn{13}{c}{\textbf{Table 1.}}\\
\multicolumn{13}{c}{Tidal Tail Observations}\\
\hline\hline
\multicolumn{1}{c}{Tail} & 
\multicolumn{1}{c}{Dist.$^a$} & 
\multicolumn{1}{c}{Age$^b$} & 
\multicolumn{1}{c|}{Mass} & 
\multicolumn{1}{c}{b$_{1}$$^d$} & 
\multicolumn{1}{c}{b$_{2}$$^d$} & 
\multicolumn{1}{c}{Beam} & 
\multicolumn{1}{c}{A$_{pix}$$^f$} & 
\multicolumn{1}{c|}{A$_{pix}$$^g$} & 
\multicolumn{1}{c}{d$v$$^h$} & 
\multicolumn{1}{c}{$\epsilon_{S_i}$ d$v$$^i$} & 
\multicolumn{1}{c}{log} &  
\multicolumn{1}{r}{Project} \\

\multicolumn{1}{c}{} &
\multicolumn{1}{c}{(Mpc)}  &
\multicolumn{1}{c}{(Myr)} &
\multicolumn{1}{c|}{Ratio$^c$} &
\multicolumn{1}{c}{}  &
\multicolumn{1}{c}{}  &
\multicolumn{1}{c}{P.A.$^e$} &
\multicolumn{1}{c}{(arcsec$^2$)} &
\multicolumn{1}{c|}{(kpc$^2$)}  &
\multicolumn{1}{c}{} &
\multicolumn{1}{c}{}  &
\multicolumn{1}{c}{\snhi$^j$}  &
\multicolumn{1}{r}{Codes$^l$} \\
\hline

\multicolumn{4}{c|}{}&
\multicolumn{5}{l|}{Instruments and Arrays$^k$}&
\multicolumn{4}{l}{Original Publications}\\
\hline\hline
\multicolumn{13}{c}{The \citet{M11} Sample} \\
\hline\hline
NGC 1487E & 10.8 & 500 & 0.25 & 27.0 & 24.1 & 16.1 & 9 & 0.02 & 10 & 0.93 & 19.11  & C531 \\
\multicolumn{4}{r|}{} & \multicolumn{5}{l|}{ATCA 375 + 750A + 1.5A + 1.5B + 1.5D + 6A} & \multicolumn{4}{l}{\citet{getts01}; \citet{NGC1487_pub}} \\
\hline
NGC 1487W & 10.8 & 500 & 0.25 & 27.0 & 24.1 & 16.1 & 9 & 0.02 & 10 & 0.93 & 19.11 & C531 \\
\multicolumn{4}{r|}{} & \multicolumn{5}{l|}{ATCA 375 + 750A + 1.5A + 1.5B + 1.5D + 6A} & \multicolumn{4}{l}{\citet{getts01}; \citet{NGC1487_pub}} \\
\hline
NGC 4747 & 20.2 & 320 & 20 & 61.9 & 48.5 & 0.00 & 400 & 3.84 & 16.4 & 39.47 & 19.09  & - \\
\multicolumn{4}{r|}{} & \multicolumn{5}{l|}{WSRT} & \multicolumn{4}{l}{\citet{NGC4747_pub}}  \\
\hline
NGC 520 & 27.2 & 300 & 0.05 & 24.5 & 23.8 & 38.3 & 25 & 0.43 & 10.5 & 1.13 & 18.75  & AH412/ AH417 \\
\multicolumn{4}{r|}{} & \multicolumn{5}{l|}{ VLA C + D} & \multicolumn{4}{l}{ \citet{NGC520_pub}} \\
\hline
NGC 2992 & 36.6 & 100 & 1 & 23.9 & 17.9 & -14.9 & 16 & 0.50 & 20.5 & 2.78 & 19.34  & AD402 \\
\multicolumn{4}{r|}{} & \multicolumn{5}{l|}{VLA C} & \multicolumn{4}{l}{\citet{NGC2992_pub}} \\
\hline
NGC 2993 & 36.6 & 100 & 1 & 23.9 & 17.9 & -14.9 & 16 & 0.50 & 20.5 & 2.78 & 19.34 & AD402 \\
\multicolumn{4}{r|}{} & \multicolumn{5}{l|}{VLA C} & \multicolumn{4}{l}{\citet{NGC2992_pub}} \\
\hline
NGC 2782E & 38.1 & 200 & 4 & 19.8 & 17.9 & 86.4 & 4 & 0.14 & 10.3 & 0.53 & 19.22  & AS389/ AS453 \\
\multicolumn{4}{r|}{} & \multicolumn{5}{l|}{VLA B + C + D} & \multicolumn{4}{l}{\citet{Smith97}; \citet{Smith94}; \citet{Smith91}} \\
\hline
NGC 2782W & 38.1 & 200 & 0.25 & 19.8 & 17.9 & 86.4 & 4 & 0.14 & 10.3 & 0.53 & 19.22 & AS389/ AS453 \\
\multicolumn{4}{r|}{} & \multicolumn{5}{l|}{VLA B + C + D } & \multicolumn{4}{l}{\citet{Smith97}; \citet{Smith94}; \citet{Smith91}} \\
\hline
NGC 2444 & 58.2 & 100 & 0.50 & 27.9 & 27.1 & -34.2 & 16 & 1.28 & 20.5 & 1.70 & 19.12  & AA63 \\
\multicolumn{4}{r|}{} & \multicolumn{5}{l|}{VLA C + D} & \multicolumn{4}{l}{\citet{NGC2444_pub}} \\
\hline
NGC 2535 & 59.8 & 100 & 0.30 & 20.4 & 18.4 & 16.6 & 9 & 0.76 & 5.1 & 0.87 & 19.08 & AK327 \\
\multicolumn{4}{r|}{} & \multicolumn{5}{l|}{VLA C + D} & \multicolumn{4}{l}{\citet{NGC2535_pub}} \\
\hline
NGC 6872E & 62.6 & 150 & 0.20 & 53.5 & 51.6 & 25.1 & 16 & 1.46 & 19.8 & 1.64 & 19.11  & C979 \\
\multicolumn{4}{r|}{} & \multicolumn{5}{l|}{ATCA 750D + EW352 + 1.5G + 750F} & \multicolumn{4}{l}{\citet{NGC6872_pub}} \\
\hline
NGC 6872W & 62.6 & 150 & 0.20 & 53.5 & 51.6 & 25.1 & 16 & 1.46 & 19.8 & 1.64 & 19.11 & C979 \\
\multicolumn{4}{r|}{} & \multicolumn{5}{l|}{ATCA 750D + EW352 + 1.5G + 750F} & \multicolumn{4}{l}{\citet{NGC6872_pub}} \\
\hline
\multicolumn{13}{l}{Continued on next page\ldots}\\
\end{tabular*} 
}
\end{sidewaystable*}

\begin{sidewaystable*}[p]
{\scriptsize
 \centering
\begin{tabular*}{1.0\textwidth}{@{\extracolsep{\fill}}lrrr|rrrcc|rrrr}
\multicolumn{13}{c}{\textbf{Table 1.} Continued}\\
\hline\hline
\multicolumn{1}{c}{Tail} & 
\multicolumn{1}{c}{Dist.$^a$} & 
\multicolumn{1}{c}{Age$^b$} & 
\multicolumn{1}{c|}{Mass} & 
\multicolumn{1}{c}{b$_{1}$$^d$} & 
\multicolumn{1}{c}{b$_{2}$$^d$} & 
\multicolumn{1}{c}{Beam} & 
\multicolumn{1}{c}{A$_{pix}$$^f$} & 
\multicolumn{1}{c|}{A$_{pix}$$^g$} & 
\multicolumn{1}{c}{d$v$$^h$} & 
\multicolumn{1}{c}{$\epsilon_{S_i}$ d$v$$^i$} & 
\multicolumn{1}{c}{log} &  
\multicolumn{1}{r}{Project} \\

\multicolumn{1}{c}{} &
\multicolumn{1}{c}{(Mpc)}  &
\multicolumn{1}{c}{(Myr)} &
\multicolumn{1}{c|}{Ratio$^c$} &
\multicolumn{1}{c}{}  &
\multicolumn{1}{c}{}  &
\multicolumn{1}{c}{P.A.$^e$} &
\multicolumn{1}{c}{(arcsec$^2$)} &
\multicolumn{1}{c|}{(kpc$^2$)}  &
\multicolumn{1}{c}{} &
\multicolumn{1}{c}{}  &
\multicolumn{1}{c}{\snhi$^j$}  &
\multicolumn{1}{r}{Codes$^l$} \\
\hline

\multicolumn{4}{c|}{}&
\multicolumn{5}{l|}{Instruments and Arrays$^k$}&
\multicolumn{4}{l}{Original Publications}\\
\hline\hline
\multicolumn{13}{c}{The \citet{M11} Sample} \\
\hline\hline
NGC 1614N & 65.6 & 750 & 1 & 15.0 & 10.0 & -83.0 & 9 & 0.91 & 10.6 & 3.09 & 19.63 & AH0527 \\
\multicolumn{4}{r|}{} & \multicolumn{5}{l|}{VLA C + D} & \multicolumn{4}{l}{\citet{NGC1614_pub} } \\
\hline
NGC 1614S & 65.6 & 750 & 1 & 15.0 & 10.0 & -83.0 & 9 & 0.91 & 10.6 & 3.09 & 19.63  & AH0527 \\
\multicolumn{4}{r|}{} & \multicolumn{5}{l|}{VLA C + D} & \multicolumn{4}{l}{\citet{NGC1614_pub}} \\
\hline\hline
\multicolumn{13}{c}{The \citet{K03}/\citet{aparna07} Sample} \\
\hline\hline
NGC 4038A & 13.8 & 420 & 1 & 20.7 & 15.4 & 24.3 & 25 & 0.11 & 5.2 & 3.17 & 19.20  & AG0516 \\
\multicolumn{4}{r|}{} & \multicolumn{5}{l|}{VLA C + D} & \multicolumn{4}{l}{\citet{NGC4038_pub}} \\
\hline
NGC 4038B & 13.8 & 420 & 1 & 20.7 & 15.4 & 24.3 & 25 & 0.11 & 5.2 & 3.17 & 19.20 & AG0516 \\
\multicolumn{4}{r|}{} & \multicolumn{5}{l|}{VLA C + D} & \multicolumn{4}{l}{\citet{NGC4038_pub}} \\
\hline
NGC 4038C & 13.8 & 420 & 1 & 20.7 & 15.4 & 24.3 & 25 & 0.11 & 5.2 & 3.17 & 19.20 & AG0516 \\
\multicolumn{4}{r|}{} & \multicolumn{5}{l|}{VLA C + D} & \multicolumn{4}{l}{\citet{NGC4038_pub}} \\
\hline
NGC 3256E & 42.8 & 400 & 1 & 25.7 & 19.3 & 6.94 & 16 & 0.69 & 33.5 & 13.06 & 20.01  & C061 \\
\multicolumn{4}{r|}{} & \multicolumn{5}{l|}{ATCA 1.5B + 1.5C + 1.5D} & \multicolumn{4}{l}{\citet{NGC3256_pub}} \\
\hline
NGC 3256W & 42.8 & 400 & 1 & 25.7 & 19.3 & 6.94 & 16 & 0.69 & 33.5 & 13.06 & 20.01  & C061 \\
\multicolumn{4}{r|}{} & \multicolumn{5}{l|}{ATCA 1.5B + 1.5C + 1.5D} & \multicolumn{4}{l}{\citet{NGC3256_pub}} \\
\hline
NGC 7252E & 62.2 & 730 & 1 & 26.9 & 16.1 & 18.6 & 25 & 2.28 & 42.5 & 4.53 & 19.35 & AH0372/ AH0412 \\
\multicolumn{4}{r|}{} & \multicolumn{5}{l|}{VLA C + D} & \multicolumn{4}{l}{\citet{NGC7252_pub}} \\
\hline
NGC 7252W & 62.2 & 730 & 1 & 26.9 & 16.1 & 18.6 & 25 & 2.28 & 42.5 & 4.53 & 19.35 & AH0372/ AH0412 \\
\multicolumn{4}{r|}{} & \multicolumn{5}{l|}{VLA C + D} & \multicolumn{4}{l}{\citet{NGC7252_pub}} \\
\hline
NGC 3921S & 84.5 & 460 & 1 & 19.3 & 18.1 & -85.1 & 25 & 4.20 & 10.7 & 1.61 & 18.91 & AH0417 \\
\multicolumn{4}{r|}{} & \multicolumn{5}{l|}{VLA C + D} & \multicolumn{4}{l}{\citet{NGC520_pub}} \\

\hline\hline
\end{tabular*} 
}
{\footnotesize
\textbf{Notes.}
\\
$^a$Luminosity distance (Mpc)
\\
$^b$Dynamical age from M11 and sources therein (Myr)
\\
$^c$Defined as $M_1$/$M_2$, with $M_1$ and $M_2$ the mass of the progenitor and perturbing galaxy, respectively
\\
$^d$Beam dimensions (arcsec)
\\
$^e$Beam position angle (deg)
\\
$^f$Single pixel area (arcsec$^2$)
\\
$^g$Single pixel area (kpc$^2$)
\\
$^h$Channel width (\kms)
\\
$^i$1$\sigma$ detection limit in the single channel integrated flux density (10$^{-4}$ Jy \kms\ channel$^{-1}$)
\\
$^j$1$\sigma$ detection limit in column density (cm$^{-2}$ channel$^{-1}$)
\\
$^k$Instruments (ATCA, WSRT, or VLA) and array configurations used in original data acquisition (see original publications for details)
\\
$^l$ATCA and VLA archive project and proposal codes of acquired data
\\
}
\end{sidewaystable*}





\section{Analysis}
\label{sec:analysis}

Our reduced and calibrated cubes contain three dimensions: the typical x/y astrometric components, and frequency. Each element of the cube registers the intensity (Jy beam$^{-1}$) at a certain position and over a particular channel of observed frequency, which tracks the \HI\ emission over velocity space. These elements and their errors can be transformed into measurements of flux density (Jy) by multiplying the intensity by the ratio of the pixel and beam angular sizes, $A_{{\rm pix}}~A_{{\rm beam}}^{-1}$, whose values are recorded in Table~1. 

Our immediate objective is to probe the \HI\ properties on the smallest physical scales afforded by our data cubes, specifically where there are established cluster candidates. \citet{aparna07} produced maps of the \HI\ column density for their sample of tidal tails and determined the underlying \nhi\ values for each of their SCC positions. Given the differences in resolution scales between their optical and radio data, this was in essence a pc-to-kpc comparison. Here, we instead identify areas in our \HI\ data that ``contain" SCCs and compare those to the areas that do not. This is a favorable kpc-to-kpc spatial comparison.

We are therefore concerned with the spectrum of \HI\ emission of different xy positions within the smallest regions possible. Individually the resolution of each cube is set by the synthesized beam ($\sim$ several kpc in this sample), which for the purposes of interferometric imaging is sampled by the selected pixel size ($\sim$0.1--2 kpc). In the sections below we map out \HI\ characteristics from the cubes on a pixel-by-pixel basis. Because this project is a synthesis of optical and radio data and terminology, we will continue to use the term ``pixel" to describe the xy components of our data cubes. Each pixel thus contains a one-dimensional cross section of the cube corresponding to all frequency (velocity) channels of an individual spatial position. Extracting and characterizing these spectra for every pixel with sufficient \HI\ signal in an established tail region is our critical task, requiring careful pixel and channel selection. Not all channels of a given pixel will contain \HI\ signal, which necessitates careful masking of ``noise" channels and selection of ``signal channels" to properly ascertain \HI\ properties for that pixel. We describe our technique, implemented with the Interactive Data Language (IDL), below.

\subsection{Channel Selection}
\label{sec:masking}

We first apply channel selection criteria to all pixels, to identify the channels in which \HI\ emission is confidently identified. There are a variety of methods to identify signal in individual channels of an \HI\ pixel and to mask channels without. For instance, THINGS (The \HI\ Nearby Galaxy Survey; \citealp{things}) required that a signal be registered by at least three contiguous channels (each of a width $\lesssim$5 km s$^{-1}$). Their instrument setup was designed to Nyquist sample the warm neutral medium, and they report that this method sufficiently highlights real 21-cm emission and eliminates noise.

Our data exhibit a much wider range of channel widths, from $\approx$5.13 to 42.5 km s$^{-1}$, so we must avoid such a constraint for cubes where multiple channels would span an unrealistically broad velocity space. Moreover, our data are inhomogeneous in sensitivity, so the appearance of an \HI\ line profile may vary depending on these instrumental constraints and orientation/projection effects. Thus, we customized the channel selection criteria for each tail after examining the spectra of many pixels across the tidal debris. \mbox{Table 2} lists the channel selection requirement as a signal-to-noise ratio for single channel signal and equivalent integrated flux density for each tail, along with the minimum number of adjacent channels $N_{{\rm chan}}$ needed at that flux density for a pixel to count as signal. 

The values in \mbox{Table~2} were selected after an extensive trial-and-error process for each tail that produced reasonable spectra and maps of the properties described in Section \ref{sec:measprops} below. We varied both $N_{{\rm chan}}$ and the necessary signal-to-noise ratio until no single, isolated pixels in maps of the line-of-sight velocity dispersion (\vdisp) registered physically unrealistic values within the tail regions (i.e.\ $\gtrsim$150 \kms, depending on the tail). Pixels within a beam size are statistically correlated, so finding occasional pixels with discrepant values in velocity dispersion maps indicated that noisy channels outside the velocity space of actual emission were unwittingly selected and thus tended to increase \vdisp\ relative to the surrounding pixels. The spectra of these locations were also visually studied to optimize our masking parameters. 

We then masked out, i.e.\ ignored, the noise channels in the cube, with the exception of noise channels alongside channels containing signal. These were left unmasked, as were all adjacent channels until a flux density level consistent with 0 Jy ($\pm$ 1$\sigma$) was found along either side of those original signal channels. This helped to prevent clipping of broad spectral features present across multiple channels. If any of these additional channels registered flux densities below 0 Jy, they were then masked to discourage biasing any measurements toward lower values. When integrated, the unmasked pixels provide a total signal typically $>$3$\sigma$ over the channel-integrated r.m.s.\ noise.



\begin{table}[tbp]
{\normalsize
\centering
 
\begin{tabular}{p{0.13\textwidth}|p{0.085\textwidth}p{0.085\textwidth}p{0.05\textwidth}}
\multicolumn{4}{c}{\textbf{Table 2.}}\\ 
\multicolumn{4}{c}{Single Channel Signal Detection Criteria} \\
\hline\hline
Tail & S/N$^a$ & $S_{{\rm min}}$$dv$$^b$& $N_{{\rm chan}}$$^c$ \\
\hline
 
NGC 1487E & 2.75 & 2.56 & 2 \\
NGC 1487W & 2.75 & 2.56 & 2 \\
NGC 4747 & 3.00 & 118.42 & 1 \\
NGC 520 & 2.25 & 2.55 & 3 \\
NGC 2992 & 2.50 & 6.95 & 2 \\
NGC 2993 & 2.50 & 6.95 & 2 \\
NGC 2782E & 2.25 & 1.20 & 2 \\
NGC 2782W & 2.25 & 1.20 & 2 \\
NGC 2444 & 2.00 & 3.40 & 2 \\
NGC 2535 & 2.25 & 1.96 & 3 \\
NGC 6872E & 2.25 & 3.70 & 2 \\
NGC 6872W & 2.25 & 3.70 & 2 \\
NGC 1614N & 2.25 & 6.95 & 2 \\
NGC 1614S & 2.25 & 6.95 & 2 \\
NGC 4038A & 2.25 & 7.14 & 3 \\
NGC 4038B & 2.75 & 8.73 & 2 \\
NGC 4038C & 2.75 & 8.73 & 2 \\
NGC 3256E & 3.50 & 45.70 & 1 \\
NGC 3256W & 3.50 & 45.70 & 1 \\
NGC 7252E & 3.50 & 15.86 & 1 \\
NGC 7252W & 3.50 & 15.86 & 1 \\
NGC 3921S & 2.00 & 3.23 & 3 \\
\hline\hline
\end{tabular} 
}
{\footnotesize
\textbf{Notes.}
\\
$^a$Signal-to-noise ratio required for a single channel
\\
$^b$Minimum integrated flux density required for a single channel (10$^{-4}$ Jy \kms)
\\
$^c$The number of adjacent channels needed at the indicated flux to register \HI\ emission in a pixel
\\
}
\end{table}



\begin{figure*}[tb]
\centering
\includegraphics[width=0.9\textwidth]{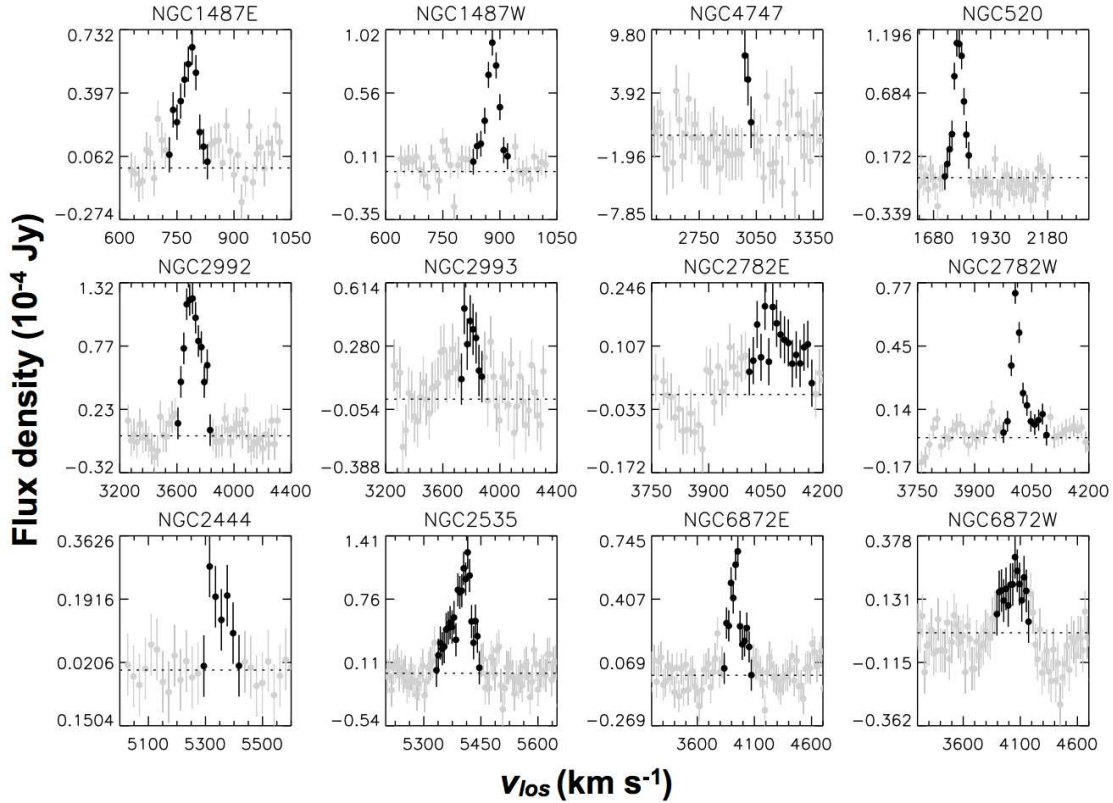}
\caption{Randomly selected example spectra of pixels in the \HI\ debris of NGC~1487E--NGC~6872W. Gray points indicate data for all channels; black points indicate channels that have been selected as \HI\ emission with the criteria and procedure of \S\ref{sec:masking}. Error bars (r.m.s.\ noise) are also shown. \label{fig:tailspex1}}
\addtocounter{figcount}{1}
\end{figure*}

\begin{figure*}[tb]
\centering
\includegraphics[width=0.9\textwidth]{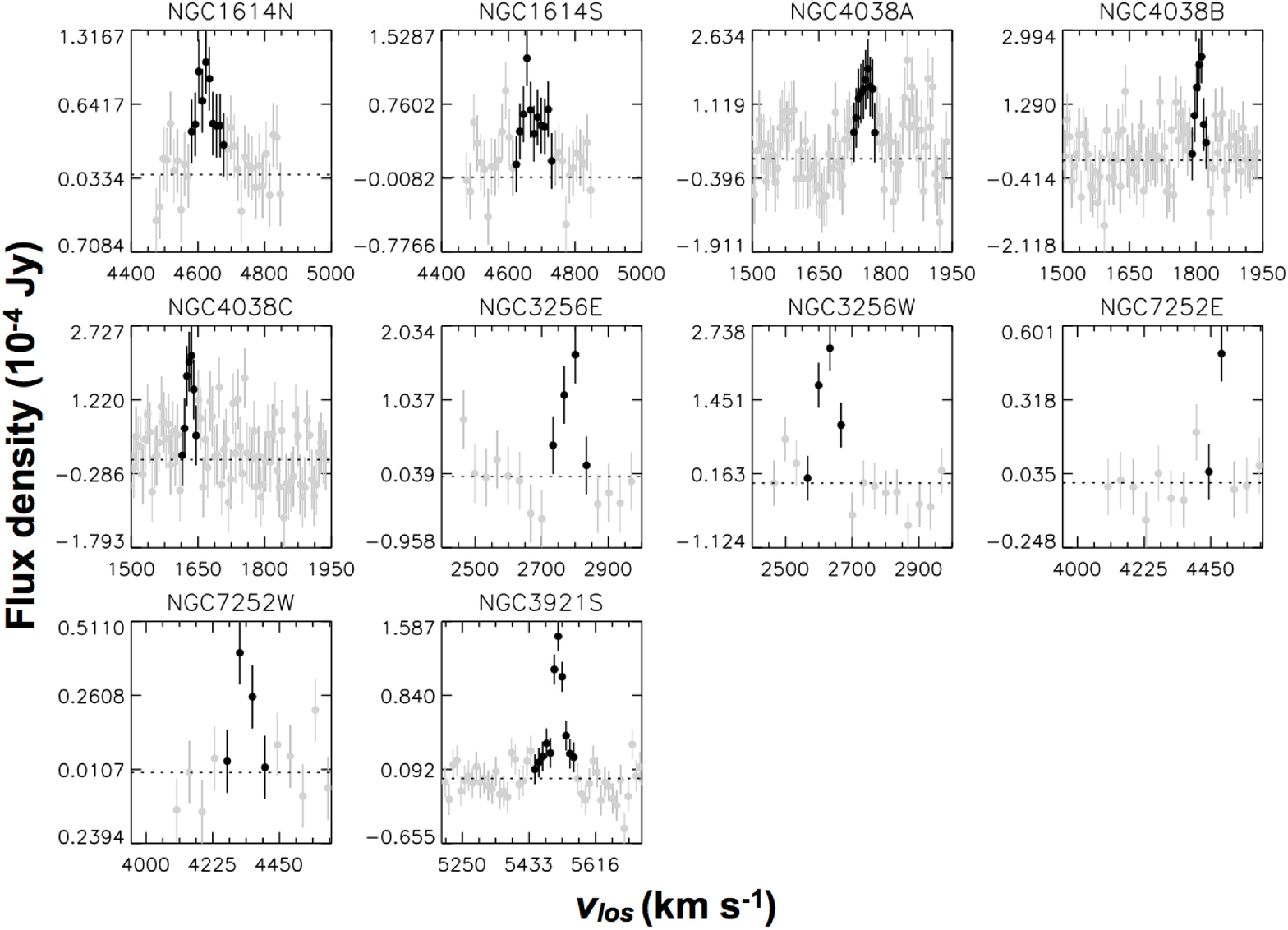}
\caption{Randomly selected example spectra of pixels in the \HI\ debris of NGC~1614N--NGC~3921S. Gray points indicate data for all channels; black points indicate channels that have been selected as \HI\ emission with the criteria and procedure of \S\ref{sec:masking}. Error bars (r.m.s.\ noise) are also shown. \label{fig:tailspex2}}
\addtocounter{figcount}{1}
\end{figure*}

We display several examples of results from our signal selection procedure in \mbox{Figure \ref{fig:tailspex1}} and \mbox{Figure \ref{fig:tailspex2}}. A spectrum from one randomly selected pixel from each tail is shown in these figures, plotted in gray; the overplotted error bars indicate the r.m.s.\ noise. Signal channels are indicated as black points. The heterogenous channel widths and noisy quality of some of the randomly chosen \HI\ profiles are clearly evident in these figures, highlighting the need for a varied signal selection criteria across the sample.  

%
%

%


\subsection{Pixel Selection}
\label{sec:pixselect}

We also limit our analysis to regions in our \HI\ data encompassed by previous optical observations, i.e., selecting \HI\ pixels that are ``in-tail" and within the WFPC2 FOV of the previous studies. Within the WFPC2 images of their datasets, M11 and \citet{K03} defined areas that were at least one count above the sky level in F606W or F555W as ``in-tail." These groups used the WF2--4 chips in their analysis, choosing to ignore the planetary camera with its prohibitively high readnoise. \citet{aparna07} applied this definition and the optical boundaries it demarcates to the \HI\ maps of their tidal tail sample to study the local column densities within. In M11, the expanded tidal tail sample afforded a glimpse at tails with different \HI\ and optical morphologies, whose optically bright regions did not always coincide with accumulations of \HI\ (c.f. \citealp{hibbard2000}). Consequently, we redefine ``in-tail" to refer to an area of the \HI\ data that is confined to the WFPC2 footprint of its corresponding optical observations, and is either within the optical boundaries set by M11 or met the \HI\ criteria of Section \ref{sec:masking}.

For each tail, we determine the physical extent of the \HI\ emission by convolving a map of the integrated flux density with an artificial beam twice the size of the synthesized beam and masking out pixels that do not satisfy the minimum criterion for \HI\ signal. We found that this technique successfully suppresses artifacts in the cube that are beyond the true large-scale emission in the tidal debris, as opposed to using no smoothing kernel or kernels of sizes $\gtrsim$3 times that of the beam. Regions determined to reside at the WFPC2-imaged edges of the host galaxies by M11 were rejected.

\subsection{Measured Properties}
\label{sec:measprops}

\subsubsection{Primary Properties}
\label{sec:priprops}

 The three-dimensional nature of the \HI\ cubes provides an opportunity to examine not only the distribution of local \HI\ surface densities, but also the velocities and kinematics of the gas. Thus, for each in-tail pixel, we compute the total integrated intensity $I_{{\rm tot}}$ (Jy beam$^{-1}$ \kms), the mean line-of-sight velocity \vuse\ (km s$^{-1}$), and the line-of-sight velocity dispersion \vdisp\ (km s$^{-1}$). In each pixel, we calculate $I_{{\rm tot}}$ as the zeroth moment of the velocity-intensity distribution: 
 
\begin{equation}
I_{{\rm tot}} = \sum_{i=1}^N I_i~dv,
\end{equation} 
\noindent where $I_i$ is the intensity in each signal channel and $dv$ is the channel width in km s$^{-1}$. Again, The corresponding integrated flux density $S_{{\rm tot}}$ can be obtained by multiplying $I_{{\rm tot}}$ by the ratio of pixel and beam angular sizes $A_{{\rm pix}}~A_{{\rm beam}}^{-1}$ (both in arcsec$^2$). The line-of-sight velocity and (squared) velocity dispersion is given by the first and second moments:
\begin{equation}
\vuse = \frac{1}{\displaystyle \sum_{i=1}^N I_i} \sum_{i=1}^N  v_i I_i~,
\end{equation}
\noindent and:
\begin{equation}
\vdisp^2 = \frac{1}{\displaystyle \sum_{i=1}^N I_i} \sum_{i=1}^N  (v_i - \vuse)^2 I_i~,
\end{equation}
\noindent respectively. In Equation (2) and (3), $v_i$ is the velocity value for the $i$-th channel. Summations were performed only on channels left unmasked as described in Section \ref{sec:masking}--\ref{sec:pixselect} for all equations, and errors in $I_i$ ($\epsilon_{I_i}$) were propagated through to all quantities. We adopt a minimum value of 50\% of the given channel width for \vdisp\ (an instrumental limit), as well as for errors in both \vuse\ and \vdisp. For pixels defined as in-tail by virtue of their optical properties alone, these moments were not computed and the \HI\ properties were classified as ``unfitted." In occasional cases of marginally detectable \HI\ emission (a moment-zero measurement), higher moments were unable to be reliably determined, and such pixels were therefore unfitted in \vuse\ and \vdisp.

\begin{figure*}[tb]
\centering
\includegraphics[width=0.7\textwidth]{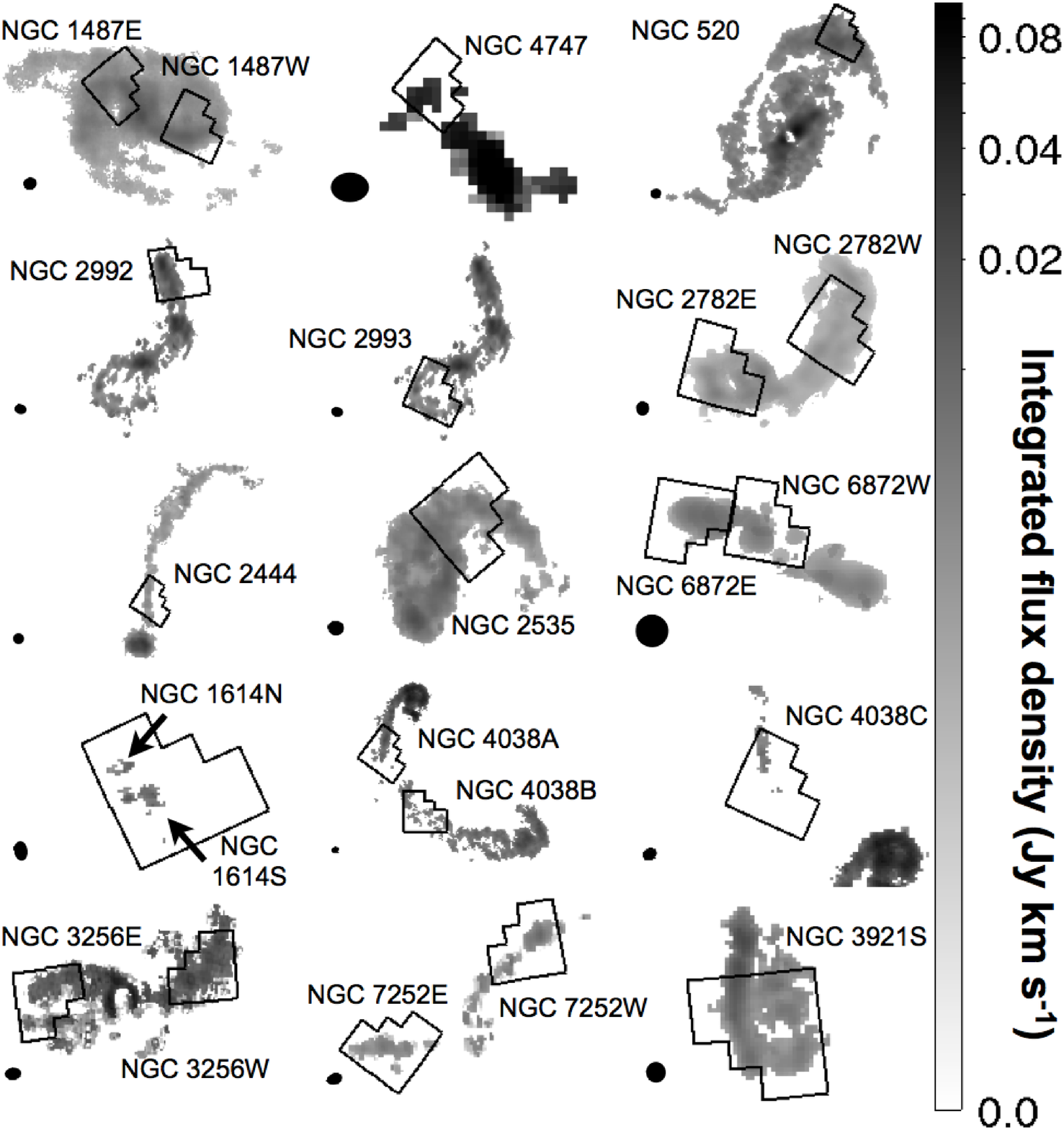}
\caption{Channel-integrated flux density maps of the tidal tails of our sample (see Figure 1 of M11 and K03 for optical counterparts). WFPC2 FOVs are overlaid as black footprints; these are all $\approx$ 2$\farcm$47 on a side. The sizes and orientations of the synthesized radio beams are represented in the bottom left of each frame. A full resolution version of this figure is available in {\it The Astrophysical Journal.} \label{fig:alltails}}
\addtocounter{figcount}{1}
\end{figure*}

For visual reference, we display moment-zero integrated flux density maps of the full extent of each data cube in \mbox{Figure \ref{fig:alltails}}, with WFPC2 fields of view overlaid. In all images, north is up, east is to the left of the frame, and every WFPC2 footprint is $\approx$2$\farcm$47 on a side. The sizes and orientations of the synthesized radio beams are represented in the bottom left of each frame in this figure.


%

%

We also experimented with finding the equivalent quantities by fitting the \HI\ spectra offered by each pixel with a series of gaussian functions. In practice, we found that our values of $S_{{\rm tot}}$ for both methods agreed very well. However, spectral fitting often yielded multiple gaussian components, so in many cases it could not be easily determined which component was ``best" associated with the tidal debris and which \vuse\ and \vdisp\ should be compared to the moment equivalents (c.f.\ the complex line structures evinced in \mbox{Figures \ref{fig:tailspex1}--\ref{fig:tailspex2}}). The intensity-weighted mean of the individual components facilitated a comparison of the \vuse\ measurements, but there was no clear prescription for comparing the number of \vdisp\ contributions from the gaussian fits to the moment-based measurement. Many pixels throughout the tail sample can be well fit by multiple gaussian components that overlap in varying amounts, so summing the gaussian-fit velocity dispersions would be an inaccurate universal tracer of kinematic complexity.

In all, because the results for integrated flux density were very similar, and for simplicity of analysis we wish to use single quantities to describe the velocity structure at positions along a tidal tail, we choose to employ the moment measurements for these primary properties. Similar work for galaxy disks can be found in \citet{tamburro}. While this method loses sight of interesting kinematics in some locations along tidal tails, e.g.\ multiple overlapping gaussian fits and extended or asymmetric profiles, complex line structures are still reflected in large second moments of the intensity distributions. Our measurement of \vdisp\ is thus still a tracer of such tail \HI\ kinematics.

\subsubsection{Secondary Properties}
\label{sec:secprops}

Further \HI\ characteristics can be calculated from the moment-based primary properties, e.g., the \HI\ column density, mass surface density, and large-scale velocity gradient.The total integrated intensity $I_{{\rm tot}}$ is related to the \HI\ column density \nhi\ (cm$^{-2}$) by the simplified relation (e.g., \citealp{HIeqn}):
\begin{equation}
\nhi\ [{\rm cm^{-2}}] = 1.247 \times 10^{24}~\frac{I_{\rm tot}}{A_{{\rm beam}}}~.
\end{equation}
\noindent Here, $A_{{\rm beam}}$ is the area of the beam in square arcseconds. The error in \nhi\ is given by:
\begin{equation}
\snhi [{\rm cm^{-2}}] = 1.247 \times 10^{24}~\frac{\sqrt{N_{\rm u}}}{A_{{\rm beam}}}~\epsilon_{I_i}~dv~.
\end{equation}
\noindent $N_{{\rm u}}$ denotes the number of unmasked channels for the specified pixel, $dv$ is the channel width (\kms), and $\epsilon_{I_i}$ is the single channel r.m.s.\ error in $I_i$. The corresponding error in $S_i$ ($\epsilon_{S_i}$) is then simply ($A_{{\rm pix}}~A_{{\rm beam}}^{-1}$)~$\epsilon_{I_i}$.  

We moreover measure the mass surface density, kinetic energy density, and velocity gradient of each pixel. The mass surface density \mhi\ and its uncertainty \smhi\ are calculated with the following equations (from \citealp{HIeqn}, with some alterations to change masses to mass surface densities and include unit conversions):
\begin{equation}
\mhi\ [\msun~{\rm pc^{-2}}] = 1.00 \times 10^4~\frac{I_{\rm tot}}{A_{{\rm beam}}},
\end{equation}
\noindent and:
\begin{equation}
\smhi\ [\msun~{\rm pc^{-2}}]  = 1.00 \times 10^4~\frac{\sqrt{N_{\rm u}}}{A_{{\rm beam}}}~\epsilon_{I_i}~dv~,
\end{equation}
\noindent where $\epsilon_{I_i}$ is the single channel r.m.s.\ error in intensity as above.

To characterize the dynamical state of the \HI\ gas, we further define the kinetic energy density of the gas in a single pixel as $\ke~=~\alpha \mhi\vdisp^2$, where the factor $\alpha$=3/2 assumes the velocity distribution is isotropic. \citet{tamburro} similarly calculate a pixel-by-pixel kinetic energy for their sample of non-interacting galaxies. For streaming \HI\ features at a variety of inclinations and projection effects, this constant may differ in reality by a factor of a few. Thus we consider this only an order-of-magnitude tracer of the actual, local turbulent kinetic energy per unit area. 

Lastly, we produce a diagnostic of the locally centered line-of-sight velocity gradient across the plane of the sky \shear\ (km s$^{-1}$ kpc$^{-1}$) from the root mean square of the velocity gradients between the pixel of interest and others in its immediate neighborhood. We write this as:  
\begin{equation}
\shear = \frac{1}{\sqrt{N_{{\rm pix}}}}~\left(\sum_{i=1}^{N_{{\rm pix}}} \frac{(v_i - v)^2}{dr_{\perp,i}^2}\right)^{\frac{1}{2}}~,
\label{eqn:shear}
\end{equation}
\noindent where $v$ is the \HI\ velocity for the central pixel of interest and  $v_i$ is the equivalent velocity for the $i$-th nearby pixel with which we construct a velocity gradient over $dr_{\perp,i}$, the distance in kpc between the pixels. $N_{{\rm pix}}$ is the number of pixels that surround the central pixel of interest and are used in the calculation. Here, we use all pixels with measurable \HI\ emission that surround the central pixel in an annulus whose inner and outer radii are set by 0.5 and 1.5 times the maximum dimension of the beam. This allows a comparison between the central pixel of interest and the closest statistically uncorrelated pixels 1 beam away. Thus, while previous quantities pertain to pixel scales ($\sim$kpc), this velocity gradient is measured over several beams, and probes gas motions and kinematics on larger ($\sim$10 kpc) scales. If \vuse\ and \vdisp\ were unable to be determined in a given pixel, then that pixel was also deemed ``unfitted" by \shear.

\begin{figure*}[p]
\centering
\includegraphics[width=1.0\textwidth]{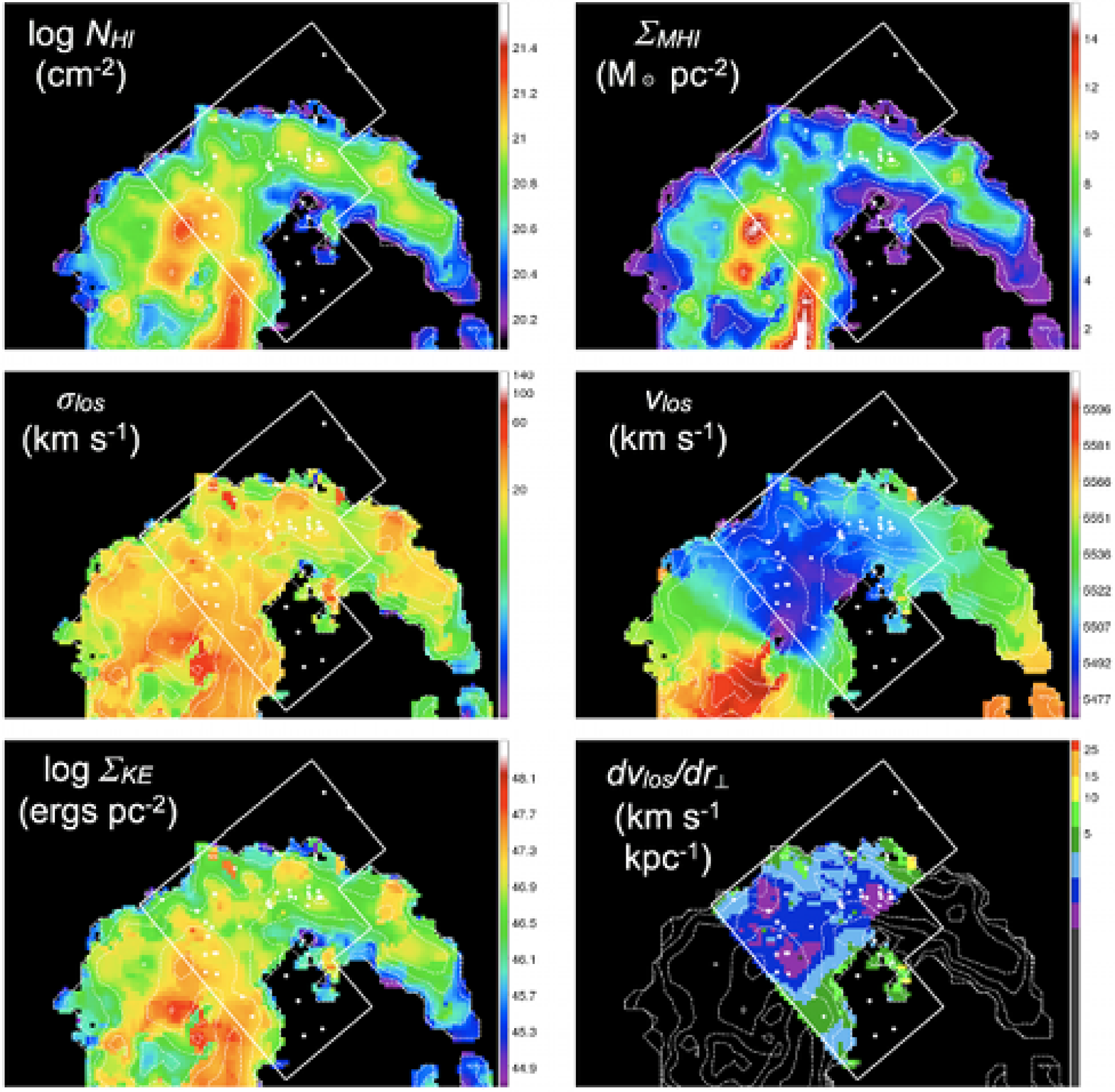}
\caption{{Maps of several derived quantities - \nhi, \mhi, \vdisp, \vuse, \ke, and \shear\ fit for the tail NGC 2535. Contours of log \nhi\ = 20.0--21.4 (cm$^{-2}$) in steps of 0.2 dex are overlaid on all maps. Although we only consider pixels within the WF2--WF4 chips of the overlaid WFPC2 FOV, the rest of the system has been mapped for clarity in some cases. Full resolution Figures 4.1--4.22 for all tails are available in the electronic edition of {\it The Astrophysical Journal.}}\label{fig:tailmap}}
\addtocounter{figcount}{1}
\end{figure*}

As an example of these calculations, we present a montage of \nhi, \mhi, \vdisp, and \vuse\ maps in \mbox{Figure \ref{fig:tailmap}} for the tail NGC~2535.  Corresponding maps of the remaining tail sample can be found in full resolution in the Online Figure Set \ref{fig:tailmap}. As before, we include the WFPC2 FOV, but also indicate positions of SCCs with circles. We furthermore introduce maps of \ke\ and \shear\ in this figure. Contours of log \nhi\ from 20.0 to 21.4 (cm$^{-2}$) in steps of 0.2 dex are overlaid in all maps. 

%

%



\section{Relating \HI\ and SCC Distributions}
\label{sec:results}

Our motivation in this section is to determine whether there are observable environmental differences between regions of tidal tails that foster star cluster formation and those that do not, specifically to the best resolution each data set affords. For each tail we therefore separated \HI\ pixels by their proclivity for harboring SCCs and compare their distributions of properties measured in Section \ref{sec:analysis}. By contrasting the distributions of properties of tail pixels that contain no SCCs with tail pixels that contain at least one SCC, we assess possible environmental differences on a $\sim$kpc scale within a given tail.

However, we realize that not all SCCs are star clusters, and some care must be taken in interpreting differences or similarities between \HI\ characteristics for these two categories. Fortunately, if there are quantifiable differences between pixels with SCCs and those without despite contamination from non-cluster carrying pixels, these are still likely substantial indicators of environmental dissimilarity. Null results, given this caveat, are not as easily understood. 

We reiterate that the pixel size across the sample is not a constant fraction of the beam sizes. Because we are not yet concerned with rigorously comparing the tails (that is deferred to Section~\ref{sec:tailcomp}), this is not an immediate concern. Rather, because the resolution remains set by the beam size, variations in pixel size for a given tail only affect the coarseness of any considered distribution of an \HI\ property and not the distribution itself. Our plots of tail properties we discuss later in this section and the Appendix use bin sizes that are larger than the pixel-to-pixel variations between correlated pixels, so these are not affected.

\subsection{Individual SCC-Harboring Probabilities}
\label{sec:probs}

In this section, we concentrate on the pixels of a given tail that contain at least one SCC. The sources detected in M11 are not verified clusters, but cluster candidates, so each of these pixels has a probability that it contains at least one actual, luminous cluster. We therefore first examine the source statistics in a manner consistent with M11 to determine how likely a source is a cluster rather than a background contaminant. From this we compute the probability that the pixel containing that SCC has at least one cluster.   

Similar to M11, we determine the relative areas within and outside the tail regions surveyed within the WFPC2 footprint, $A_{{\rm in}}$ and $A_{{\rm out}}$, respectively. The main difference in this study, aside from pixel sizes, is the optical/\HI\ tail definition used here. Within these areas, we find the number of SCCs $N_{{\rm in}}$ and $N_{{\rm out}}$. We divide $N_{{\rm in}}$ by $A_{{\rm in}}$ to find the overall in-tail source density $\Sigma_{{\rm in}}$; these results are listed in \mbox{Table 3} for each tail, along with the in-tail completeness limits found in M11. 

M11 also calculated the out-of-tail density $\Sigma_{{\rm out}} = N_{{\rm out}}/A_{{\rm out}}$ and subtracted it from $\Sigma_{{\rm in}}$ to find the background-subtracted, average in-tail density of SCCs in these tidal tail regions. However, \mbox{Table 3} shows that many, particularly close systems have extensive \HI\ and optical coverage and small out-of-tail areas $A_{{\rm out}}$. The number of detected out-of-tail objects within these regions is statistically uncertain, or not at all known ($N_{{\rm out}} = 0$) in some cases. 

We therefore determine $\Sigma_{{\rm out}}$ with results from star counts generated by the Galactic population synthesis model of \citet{starcounts}\footnote{\url{http://www.model.obs-besancon.fr}}. In a 100 arcsec$^2$ solid angle centered at coordinates in each tail, we generate the number of foreground stars visible within the photometric criteria of Section \ref{sec:obs}. This includes a $V-I$ $<$ 1 cutoff and an apparent $V$-band magnitude range matching the sources detected in, and completeness limits of, M11 (19 $\lesssim V \lesssim$ 26). We then determine the resulting number of contaminants per out-of-tail area $\Sigma_{{\rm out}}$ using the distance to each tail. We find that variations of solid angle (1--100 arcsec$^2$) introduce variations of only a few percent.

In this calculation, we also roughly model the photometric errors of the returned catalogs on the errors in photometry in M11. Here, errors in magnitude $m$, $\sigma_{{\rm mag}}$ are calculated by a simple linear model $\sigma_{{\rm mag}}$ = $A + B m$, where $A$ and $B$ are 0.01 and 0.0035, respectively. Moreover, we employ a mean $V$-band diffuse absorption of 0.7 mag kpc$^{-1}$, which is recommended for intermediate-to-high Galactic latitudes \citep{starcounts}. It should be noted that this may introduce some uncertainty for NGC~3256E/W, which is a particularly low-latitude ($b \approx\ 11^{\circ}$) object. But ultimately, it is a more reliable method that is more free of statistical uncertainties with low counts of detected objects.  In addition, these sources were screened for extended sources in M11, so contamination by background galaxies---already a minor issue from the stringent $V-I$ cutoff---is minimal. In short, the primary contamination should be foreground stars, and this is better assessed with the current method.

We want to concentrate on individual in-tail SCCs and their local environments, so we then calculate the probability $p_{{\rm real}}$ that a SCC-containing pixel contains at least one cluster based on the probable degree of in-tail contamination by out-of-tail sources. To do so we first find the area-adjusted number of probable contaminants in the in-tail source list, $\Sigma_{{\rm out}} \times A_{{\rm in}}$.  This quantity divided by $N_{{\rm in}}$ is simply the fraction of sources within the tail that are likely contaminants, or $\Sigma_{{\rm out}}/\Sigma_{{\rm in}}$. The corresponding fraction of sources that are likely star clusters (SCs) is then $f_{{\rm SC}} = 1 -( \Sigma_{{\rm out}}/\Sigma_{{\rm in}})$. 

We record various steps of this process in \mbox{Table 3}, along with the calculated $V$-band 50\% completeness limits for each tail (M11; \citealp{K03}; \citealp{aparna07}) for reference. For pixels containing one SCC, $p_{{\rm real}}$ is just the value of $f_{{\rm SC}}$ for that tail. More generally, and especially for pixels with multiple SCCs, $p_{{\rm real}}$ is the result of applying the binomial distribution, using $n$ SCCs in a pixel, each with a probability $p_{{\rm real}} = f_{{\rm SC}}$. Thus, the odds that a pixel with multiple SCCs has at least one actual cluster are higher than those listed for single SCC-harboring pixels listed in the penultimate column of \mbox{Table 3}. We indicate in the last column whether M11 found strong statistical evidence of tail-wide cluster population in each of the tail regions. With the exception of the low-latitude (and highly foreground source-contaminated) NGC 3256W, $f_{{\rm SC}}$ values are all $\gtrsim$0.7. In contrast, tails without global cluster populations exhibit a wider range of values.



\renewcommand{\thefootnote}{\alph{footnote}}
\begin{table*}[bt]
{\scriptsize
 
\begin{tabular*}{1.0\textwidth}{@{\extracolsep{\fill}}l|rrrrrr|rrr|r}
\multicolumn{11}{c}{\textbf{Table 3.}}\\ 
\multicolumn{11}{c}{Source Statistics} \\
\hline\hline
\multicolumn{1}{l}{ } & \multicolumn{6}{l}{Source Detections} & \multicolumn{3}{l}{Star Counts} & \multicolumn{1}{l}{M11 Results} \\
\hline
 
Tail & \MVcomp$^a$ & $N_{{\rm in}}$$^b$ & $N_{{\rm out}}$$^c$ & $A_{{\rm in}}$$^d$ & $A_{{\rm out}}$$^e$ & $\Sigma_{{\rm in}}$$^f$ & $\Sigma_{{\rm out}}$$^g$ & $\Sigma_{{\rm out}} \times A_{{\rm in}}$$^h$ & $f_{{\rm SC}}$$^i$ & Tail-wide \\
 
 & & & & & & & & & & SCCs?$^j$ \\
 
\hline
 
NGC 1487E & -4.6 & 1 & 0 & 40.60 & 5.89 & 0.025 & 0.039 & 1.56 & 0 & \\
NGC 1487W & -4.5 & 0 & 0 & 41.07 & 5.52 & 0 & 0.039 & 1.58 & 1 & \\
NGC 4747 & -6.0 & 6 & 1 & 84.52 & 84.52 & 0.071 & 0.014 & 1.22 & 0.80 & \\
NGC 520 & -6.4 & 2 & 0 & 266.21 & 28.66 & 0.008 & 0.007 & 1.87 & 0.06 & \\
NGC 2992 & -7.2 & 10 & 3 & 305.49 & 229.87 & 0.033 & 0.009 & 2.77 & 0.72 & $\checkmark$ \\
NGC 2993 & -7.3 & 11 & 1 & 367.49 & 169.38 & 0.030 & 0.009 & 3.34 & 0.70 & $\checkmark$ \\
NGC 2782E & -7.5 & 56 & 5 & 413.93 & 165.16 & 0.135 & 0.006 & 2.50 & 0.96 & $\checkmark$ \\
NGC 2782W & -7.3 & 6 & 2 & 472.48 & 105.25 & 0.013 & 0.006 & 2.85 & 0.52 & \\
NGC 2444 & -8.2 & 5 & 7 & 504.38 & 850.42 & 0.010 & 0.004 & 1.83 & 0.63 & \\
NGC 2535 & -8.4 & 36 & 8 & 871.95 & 554.81 & 0.041 & 0.003 & 2.36 & 0.93 & $\checkmark$ \\
NGC 6872E & -8.3 & 111 & 28 & 730.59 & 824.29 & 0.152 & 0.009 & 6.68 & 0.94 & $\checkmark$ \\
NGC 6872W & -8.5 & 101 & 14 & 729.12 & 827.22 & 0.139 & 0.009 & 6.66 & 0.93 & $\checkmark$ \\
NGC 1614N & -8.6 & 19 & 8 & 304.89 & 1408.86 & 0.062 & 0.003 & 1.01 & 0.95 & $\checkmark$ \\
NGC 1614S & -8.6 & 30 & 10 & 325.82 & 1387.93 & 0.092 & 0.003 & 1.07 & 0.96 & $\checkmark$ \\
NGC 4038A & -5.8 & 3 & 0 & 45.79 & 30.30 & 0.066 & 0.050 & 2.30 & 0.24 & \\
NGC 4038B & -5.1 & 1 & 0 & 46.91 & 29.96 & 0.021 & 0.050 & 2.35 & 0 & \\
NGC 4038C & -5.8 & 1 & 0 & 27.50 & 48.71 & 0.036 & 0.050 & 1.38 & 0 & \\
NGC 3256E & -7.8 & 32 & 9 & 460.19 & 271.43 & 0.070 & 0.035 & 16.12 & 0.50 & \\
NGC 3256W & -7.8 & 51 & 4 & 580.74 & 146.74 & 0.088 & 0.035 & 20.34 & 0.60 & $\checkmark$ \\
NGC 7252E & -8.1 & 11 & 16 & 627.03 & 928.00 & 0.018 & 0.003 & 2.18 & 0.80 & \\
NGC 7252W & -8.1 & 17 & 11 & 533.54 & 1016.92 & 0.032 & 0.003 & 1.85 & 0.89 & \\
NGC 3921S & -8.8 & 14 & 5 & 1718.82 & 1147.28 & 0.008 & 0.001 & 1.79 & 0.87 & \\
\hline\hline
\end{tabular*} 
}
{\footnotesize
\textbf{Notes.}
\\
$^a$50\% $V$-band completeness limit (M11)
\\
$^b$Number of SCCs detected within the tail region
\\
$^c$Number of SCC-like objects detected outside the tail region
\\
$^d$Area of in-tail region (kpc$^2$)
\\
$^e$Area of out-of-tail region (kpc$^2$)
\\
$^f$In-tail SCC surface density (kpc$^{-2}$)
\\
$^g$Out-of-tail surface density of SCC-like objects (kpc$^{-2}$)
\\
$^h$Probable number of in-tail SCCs that are contaminants
\\
$^i$Overall fraction of SCCs that are likely star clusters
\\
$^j$This column is checked if M11 found evidence for large-scale populations of star clusters within the tail regions
\\
}
\end{table*}
\renewcommand{\thefootnote}{\arabic{footnote}}

%
%

As a final check, we compared the distribution and properties of star clusters detected in NGC2782W by \citet{TF12} and M11. Because of the instrumental differences between the studies (the former use $GALEX$ imaging and optical spectra, and the latter use $HST$ imaging), there is $\approx$50\% overlap between the source catalogs for that tail. That is, M11 detected about twice as many cluster candidates as \citet{TF12} spectroscopically characterized clusters. This agrees very well with results from the star counts model that predicted 52\% of the M11 detections are contaminants.

\subsection{Comparison of \HI\ Distributions}
\label{sec:histograms}

We now compare the distributions of  \nhi, \shear, \vdisp, and \ke\ for between \HI\ pixels not containing SCCs and those that do. In doing so we must ignore the fraction of pixels that were unfitted by the measuring techniques, $f_{{\rm unfit}}$; these are typically pixels that fall within the $optical$ tail, but have ill-defined \HI\ content. To outline potential contrasts between the pixels that are home to potential star clusters and those that are not, we performed K--S tests between these distributions for each measured property. Here, $p_{{\rm KS}}$ values reveal the probability that the SCC-sporting and SCC-empty pixels come from the same distribution of each measurement. If $p_{{\rm KS}} \lesssim 0.01$, it is likely that these two categories of pixels have two separate distributions of an \HI\ characteristic. The results of these tests are also provided in \mbox{Table 4}; columns indicate the results of this test performed on the indicated property. We indicate where values are below 10$^{-4}$, as specific results below this cutoff are not enlightening.

Overall, it appears that the distributions of our calculated \HI\ column densities, kinematic tracers, and mechanical energy densities for both SCC-carrying and SCC-devoid pixels are indistinct ($p_{{\rm KS}} \gtrsim 0.01$). NGC~2782E and NGC~6872E are remarkable counterexamples, with low computed K--S probabilities and two of the highest tail-wide SCC population densities reported in M11 ($\approx$0.2 SCCs kpc$^{-2}$). A few other tails show evidence of independent distributions of individual properties between these two pixel types (e.g., NGC 2992 with \nhi, NGC~2535 and NGC~6872W with \shear, and NGC~3256E and NGC~7252W with \vdisp).The implications of these results are discussed for individual tails in the Appendix. It is probable that the varying degree of contamination from foreground stars across the tail sample---displayed in Table~3---contributes to a contamination of the \HI\ distributions for the pixels hosting at least one SCC and confuses the K--S results.

It is additionally important to consider that there are minor, inherent shortcomings to K--S tests; namely that they are particularly sensitive to differences in the centers of cumulative distribution functions relative to their edges. Also note that many tails have small samples of SCC-carrying pixels, to which such statistical tests are particularly sensitive. as mentioned at the outset of this section, pixel size variations also introduce differences in the number of pixels considered, and the coarseness of the unbinned \HI\ measurement distributions. This will have some, albeit minor, affect on the results of the K--S tests.  

\begin{figure*}[htb]
\centering
\includegraphics[width=0.9\textwidth]{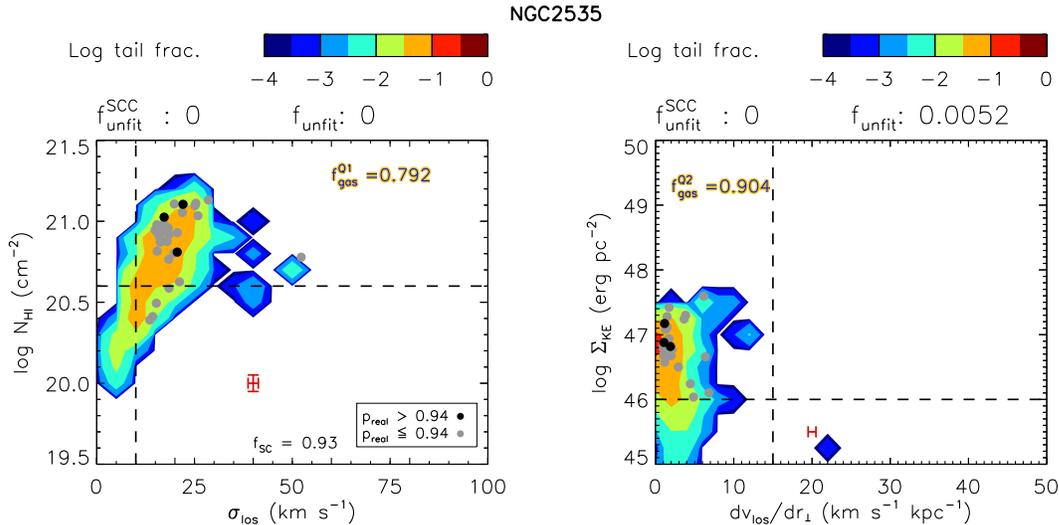}
\caption{Combined, contoured distribution of measured \HI\ quantities for NGC 2535, with levels corresponding to the fraction of the tail within the indicated parameter space. Pixels with SCCs most likely to have star clusters are plotted in black ($p_{{\rm real}} > $ 0.94), and those less likely in gray ($p_{{\rm real}} <$ 0.94). The number of unfitted pixels, and the tail fraction within the specified quadrants Q1 (left) and Q2 (right) are provided. These are defined by the intersection of lines of potential threshold column density \citep{aparna07} with the thermal-turbulent velocity dispersion in the left plots; and their combined energy density with the value of shear in the solar neighborhood in the right plots. See \S4 for details. Full resolution Figures 5.1--5.22 for all tails are available in the electronic edition of {\it The Astrophysical Journal.}\label{fig:contours}}
\addtocounter{figcount}{1}
\end{figure*}

With all uncertainties considered, two distributions with K--S probabilities $\gtrsim$0.01 may still be independent but are difficult to differentiate here. Regardless, it appears that the dependence of star cluster positions (and possibly clues to their formation) on individual \HI\ densities and kinematics cannot be easily ascertained on a tail-by-tail, property-by-property basis like this. We therefore also explore the holistic \HI\ parameter space of the tail pixels, to see if and how all these \HI\ properties collectively relate to SCCs. \mbox{Figure \ref{fig:contours}} shows the combined distributions of \nhi\ and \vdisp\ (left), and \ke\ and \shear\ (right) of NGC~2535. Similar plots for the remaining tail sample are available in the Online Figure Set \ref{fig:contours} . Bin sizes of 0.1 dex in log \nhi\ (cm$^{-2}$), 5.0 km s$^{1}$ in \vuse, 2.0 km s$^{-1}$ kpc$^{-1}$ in \shear, and 0.25 dex in log \ke\ (erg pc$^{-2}$) were employed in contouring. Contour levels are scaled to show the fraction of the total number of tail pixels examined, including pixels that were unfitted by \nhi, \vdisp, \shear, or \ke\ measurements. We record the fraction of unfitted SCC-carrying and SCC-devoid pixels for reference as $f_{{\rm unfit}}^{{\rm SCC}}$ and $f_{{\rm unfit}}$, respectively. Median uncertainties for all unbinned properties are shown as red error bars. 

Dashed lines in the first plot show potentially important values of the plotted properties---the \citet{aparna07} \nhi\ threshold (horizontal line), and the turbulent--thermal transition \vdisp\ $\approx$ 10 km s$^{-1}$ (vertical line; this shows where gas velocity dispersions are driven by turbulent instead of thermal motions). The vertical dashed line on the second plot indicates a constant \shear\ = 15 km s$^{-1}$ kpc$^{-1}$, approximately the value of the Oort constant A, the value of the shear in the solar neighborhood. This line is purely fiducial and meant to guide distinctions between relatively shallow and steep velocity gradients. The horizontal line on that plot shows a \ke\ cutoff of 10$^{46}$ erg kpc$^{-2}$, which considers an energy threshold set by setting a hypothetical critical \HI\ mass surface density for star formation of 3 \msun\ pc$^{-2}$ \citep{schaye} and threshold turbulent velocity dispersion 10 km s$^{-1}$. These demarcations are meant to guide the eye and highlight what fraction of a tail has relatively more dense, turbulent, etc.\ gas.

%
%


We indicate the fraction of the total number of tail pixels that exhibit super-``critical" \HI\ characteristics for the combined distributions with $f_{{\rm gas}}^{{\rm Q1}}$ for the left plot, and $f_{{\rm gas}}^{{\rm Q2}}$ for the right. ``Q1" and ``Q2" refer to particular quadrants delineated by the dashed lines, i.e.\ gas with log \nhi\ $>$ 20.6 cm$^{-2}$ and \vdisp\ $>$ 10 km s$^{-1}$ (left plot), and gas with log \ke\ $>$ 46 erg pc$^{-2}$ and \shear\ $<$ 15 km s$^{-1}$ kpc$^{-1}$ (right plot). Lastly, the SCC-laden pixels are overplotted in both plots as gray points if their calculated $p_{{\rm real}}$ values are $\leq$0.94, and as black points if $p_{{\rm real}} > 0.94$. This value of 0.94 was selected with some foresight and will be addressed further in Section \ref{sec:tailcomp}. Recall that $p_{{\rm real}}$ indicates the likelihood that an \HI\ pixel contains at least one star cluster. The plots in Online Figure Sets 4 and 5 are discussed individually for each tail region in the Appendix. 

A clear correlation is evident between \HI\ column density and line-of-sight velocity dispersion for much of our sample as seen in Online Figure Set \ref{fig:contours}. The example spectra displayed in Figures~\ref{fig:tailspex1} and \ref{fig:tailspex2} show that many of these randomly selected \HI\ profiles are kinematically complex, and many would have wide velocity dispersions as measured by the  second moment of the intensity distribution. The addition of multiple \HI\ components therefore both broadens \vdisp\ and increases \nhi. The degree of overlap between components in a spectrum may depend on the tail kinematics and projection effects, so we would perceive a different relationship---i.e.\ slope of contoured distributions---between these two properties in each tail of Figure Set \ref{fig:contours}. Most importantly, detecting multiple \HI\ components indicates that the actual physical density of \HI\ at the location of a pixel (or SCC) may not be consistently mapped by column density in all tails, so some confusion in interpreting the importance of \nhi\ is expected.

%
%

\section{Results and Discussion}
\label{sec:tailcomp}

In the previous sections, we analyzed the \HI\ properties to the best possible spatial resolution of their individual cubes. We present a thorough discussion of each tail's results and figures in the Appendix. It is the purpose of that Appendix---and the supporting figures presented thus far---to review the SCC distributions and \HI\ characteristics of each tail for the benefit of future work on individual tidal tails and the interplay between their ISM and their star cluster populations. Several intriguing qualitative trends in the $\sim$kpc-scale \HI\ column densities, kinematics, and mechanical energy densities can be seen in this sample that would not be otherwise ascertained from the simple K--S tests of Table~4. These also help direct our approach in Section \ref{sec:smooth}.  

The most obvious trend to note is that several of the systems identified in M11 as having widespread cluster populations (reproduced as the last column of Table~3) also tend to have a large fraction of their $\sim$kpc-scale \HI\ with column densities exceeding the log \nhi\ $>$ 20.6 \cmtwo\ (NGC~2535, NGC~2782E, NGC~1614N/S, and NGC~3256W). This gas is also turbulent (\vdisp\ $\approx$ 10--75 \kms) with velocity dispersions that vary tail-to-tail, though probably in part because of tail orientation, projection effects, and spatial scales. Kinetic energy densities of the tails are therefore also generally high (log \ke\ $>$ 46 erg pc$^{-2}$). The \HI\ pixels with the highest probability of containing at least one cluster (and usually contain more than one SCC) very often lie in these high-\nhi/\ke\ regions; the selection of the $p_{{\rm real}}$=0.94 cutoff reflects this. Mechanical energy density should loosely trace \HI\ {\it pressures} (i.e., integrated along the line of sight), so this would imply that large-scale clustered star formation as detected by M11 requires high pressures (high \ke) and high \nhi\ on the spatial scales probed here. Conversely, a number of tails with low numbers of SCCs and low calculated probabilities of their pixels containing real clusters have little high column density and often also little observably turbulent gas (NGC~1487E/W if de-projected; NGC~4747, NGC~520, and NGC~4038ABC). 

However, many tidal regions with large percentages ($\gtrsim$50\%) of their \HI\ exhibiting high column densities and kinetic energy densities on these $\sim$kpc scales do not have pronounced cluster populations in M11 (e.g., NGC~2993, NGC~3256E, and NGC~2782W). In some cases tail orientations and projection effects may increase the observed \nhi, while the actual physical densities of \HI\ may be low. Kinetic energy densities are also high from similarly inflated turbulent velocity dispersions (\vdisp\ $\approx$ 10--75 \kms). When compared to its eastern counterpart, NGC~2782W presents an interesting case for the hidden role of tail-wide H$_2$ content or dust-to-gas ratios that may also account for the observed discrepancy in cluster formation efficiencies given similar atomic hydrogen content (these two tails can be easily compared because they are present in the same \HI\ cube). The fact that NGC~2782E could be the distorted minor galaxy of the merger \citep{Smith94} and has confirmed CO detections \citep{smith99} would insinuate that the ISM is very different in these two tails. An existing molecular gas reservoir in one tail, and/or different dust-to-gas ratios among the debris regions would allow efficient star formation in one case but not another. \citet{K12} explore this concept further in their assessment of these tails. 

Moreover, several tails have pixels with high probabilities ($p_{{\rm real}}$ $>$ 0.94) of containing at least one cluster in regions of low or undetected  \HI\ column density (NGC~4747, NGC~6872E/W, NGC~1614N/S). Either \nhi\ is not important for these tails (perhaps in concert with differing molecular content), it is important on physical scales smaller than we can currently resolve (note that these tails have some of the largest angular pixel and beam sizes of the sample), or it has been dispersed and/or consumed over $\sim$100 Myr timescales, after the clusters had formed. This is reasonably consistent with overall \HI\ column density behavior with tail age seen in M11. Photoionizing feedback from local or central sources may also be a factor in these tails (e.g., \citealp{hibbard2000}). Given our work with cluster probabilities, we also cannot say with certainty that many SCCs that lie in regions of low \nhi\ or \ke\ are {\em not} clusters (e.g., NGC~7252 and NGC~2444). Feedback, interaction dynamics, and gas dispersal may again have had an effect here, or these tails may be home to smaller-scale star formation events producing few bound, luminous clusters as opposed to the large-scale triggered events our study is better designed to address.

%
%
%
%

\begin{table}[tb]
{\scriptsize
 
\begin{tabular}{lrrrr}
\multicolumn{5}{c}{\textbf{Table 4.}}\\
\multicolumn{5}{c}{K--S Probabilities} \\
\hline\hline
Tail & \nhi\ & \vdisp\ & \shear\ & \ke\ \\
\hline
 
\hline
 
NGC 1487E & - & - & - & - \\
NGC 1487W & - & - & - & - \\
NGC 4747 & - & - & - & - \\
NGC 520 & - & - & - & - \\
NGC 2992 & 1.3e-3 & 0.10 & 0.52 & 0.23 \\
NGC 2993 & 0.17 & 0.039 & 0.45 & 0.086 \\
NGC 2782E & $<$1e-4 & $<$1e-4 & 3.7e-4 & $<$1e-4 \\
NGC 2782W & 0.80 & 0.43 & 0.67 & 0.33 \\
NGC 2444 & 0.53 & 0.37 & 0.090 & 0.61 \\
NGC 2535 & 0.012 & 0.15 & 9.6e-4 & 0.039 \\
NGC 6872E & $<$1e-4 & $<$1e-4 & 0.076 & $<$1e-4 \\
NGC 6872W & 0.24 & 0.26 & 9.8e-3 & 0.33 \\
NGC 1614N & - & - & - & - \\
NGC 1614S & 0.93 & 0.39 & 0.017 & 0.89 \\
NGC 4038A & - & - & - & - \\
NGC 4038B & - & - & - & - \\
NGC 4038C & - & - & - & - \\
NGC 3256E & 0.71 & $<$1e-4 & 0.15 & 0.61 \\
NGC 3256W & 0.16 & 0.28 & 0.011 & 0.77 \\
NGC 7252E & 0.99 & 0.50 & 0.63 & 0.99 \\
NGC 7252W & 0.42 & 7.7e-3 & 0.044 & 0.39 \\
NGC 3921S & 0.41 & 0.27 & 0.87 & 0.25 \\
\hline\hline
\end{tabular} 
}
\end{table}

%
%

A few tails may also highlight the importance of velocity gradients on larger scales. A number of tails show large regions of locally, relatively shallow \shear\ (values vary from the size of the beam and the velocity structure of the \HI). In several cases these are areas of many different beam physical scales where many SCCs are located (NGC~2992, NGC~3256W, NGC~68782E, NGC~2535, NGC~2782E, and NGC~7252W), or there is a prominent, diffuse optical structure that may have lower mass star clusters (NGC~1487E/W). This possibly underscores the importance of large-scale ($\sim$10 kpc) kinematics in star cluster formation, including tail-wide shearing motions (and lack thereof) that our measurement of the line-of-sight velocity gradient might trace in certain tails. 

These results are promising, but we again caution against more rigorously comparing the \HI\ distributions between the debris regions as shown thus far. These tails exhibit a large morphological diversity, both from their range of interaction stages and types, and also intractable projection effects from inclination and their orientation on the sky. Models show that tails can have a wide range of both small- and large-scale thicknesses along our line of sight \citep{Smith94}; thus surface densities and other properties discrepant by a factor of $\sim$3 can all correspond to one ``in-tail" physical density. Highly inclined mergers can also dramatically increase the apparent thickness of a tail. We see this especially with NGC~1487 and NGC~2992, where, for many pixels, their ``true" \nhi\ values may be lower by an order of magnitude.

The translation of other line-of-sight properties to true physical properties might be even more complicated. If the local velocity dispersions within tails are not always isotropic (and they are certainly not likely to be), the proper interpretation of \vdisp\ will vary pixel-to-pixel. Streaming motions of multiple \HI\ components along our line-of-sight and beam smearing along the plane of the sky will further affect our measurements of \vdisp, not to mention \shear. That quantity bears the additional warning that artificial pixel-to-pixel variations that arise in trying to define \vuse\ in a region of low \HI\ signal (the tail boundaries) will produce high \shear\ values. This is seen in a few tails in Figure Sets~\ref{fig:tailmap} and \ref{fig:contours}. Furthermore, \ke\ is defined as an isotropic energy density, which is a less accurate description of the actual dynamic state of the \HI\ gas as beam smearing and ordered gas motions becomes more prominent.   

\mbox{Table 3} furthermore reveals that several especially distant tails suffer from completeness effects, and our source lists for them are incomplete by 50\% or slightly more. In these cases, all pixels have a slightly higher probability that they contain an SCC, with different adjustments for the different completeness curves of in-tail and out-of-tail areas. The completeness curves of M11 show that out-of-tail areas are complete to fainter magnitudes than in-tail regions (by $\sim$0.5 mag); thus for a source luminosity bin close to the completeness limits, we are likely to detect more sources out of tail than in-tail in distant systems. By this logic our computed contamination fractions are overestimated, and the $p_{{\rm real}}$ probabilities for SCC-carying pixels derived from $f_{{\rm SC}}$ fractions listed in \mbox{Table 3} are too low. Since this effect does not artificially increase the probability that pixels with SCCs contain observable clusters, we consider our existing $p_{{\rm real}}$ work to be suitable lower limits.

Lastly and most importantly, a cursory glance at \mbox{Table 1} reveals that, by necessity, data from these tails are quite heterogenous. They arise from a number of different observation program designs, array configurations and weightings in the reduction process, and therefore exhibit a range of output channel widths and pixel/beam sizes. Not all tails are sensitive to the same velocity dispersions and physical scales, so additional care must be taken in assessing gas properties and possible thresholds across the tail sample more homogeneously.


\subsection{Comparing \HI, Global SCC Populations, and Interaction Characteristics}
\label{sec:smooth}

In this section we address the issue of heterogeneous physical resolution in our \HI\ data and facilitate a better comparison between the tails in our sample. To make our \HI\ cubes uniform in pixel scale, we repeated the work of Section 3--4 for the tails, but with their cubes re-gridded to the largest physical scale of the tail sample, i.e.\ the 2.1~kpc pixel scale of NGC~3921S. We exclude the NGC~6872E/W tails in this analysis, as their physical beam sizes ($\approx$16 kpc) are much larger than the rest of the sample (several kpc) and probe \HI\ gas on correspondingly different spatial scales. 

For this task, we used the IDL function {\em frebin}, which conserves flux for every spatial slice of the cubes. In all calculations requiring the beam size, the beam dimensions of each original cube was used, with the exception of the calculation of \shear\ (Equation~(\ref{eqn:shear})). Here, we used an annulus whose inner and outer radii are 4 and 12 kpc in the determination of $N_{{\rm pix}}$. This simulates using an 8 kpc beam, which is physically the size of the largest beams in the sample (excluding NGC~6872E/W) and allowed a more direct comparison across the tail sample. 

We then sorted all the analyzed tail pixels into two categories based on pertinent characteristics of the interactions---pixels from the list of tails with widespread cluster populations as determined by M11 vs.\ those from the tails that did not, pixels from tails that resulted from early-stage and late-stage interactions, and pixels from tails produced in interactions from different mass ratios of their progenitor galaxies. This approach allows a more significant assessment of \HI\ and merger properties with respect to cluster populations than additional individual studies would allow, especially for the closest half of the tail sample where the large physical pixel scales prohibit meaningful individual study. In the subsections below we therefore discuss these distinctions and the distributions of \HI\ properties in greater detail, accompanying Figures~\ref{fig:hist_combined}--\ref{fig:cont_mr}.

In Figures~\ref{fig:hist_combined}, \ref{fig:hist_age}, and \ref{fig:hist_mr}, we outline the normalized distributions of  \nhi, \shear, \vdisp, and \ke\ for pixels from these types of tails. Subcategories of SCC richness, age, and mass ratio are represented as blue, filled histograms, and orange, outlined histograms. Cumulative distributions of these pixels are plotted over the histograms as blue and orange curves, respectively. In all cases, bin sizes were set by the number of pixels in each tail, and the standard deviation of their various distributions. We also indicate the fraction of pixels that were unfitted by the measuring techniques as $f_{{\rm unfit}}$; as before these are typically pixels that fall within the $optical$ tail, but have ill-defined \HI\ content. 

We also perform K--S tests between the (unbinned) blue and orange distributions in each panel of these plots. In {\it all} cases we calculate extremely low probabilities ($p_{{\rm KS}}$ $<$ 10$^{-4}$) that these pairs of distributions are statistically indistinguishable. Thus, the differences in \HI\ properties between tail types we note below are likely real and significant.  

Figures~\ref{fig:cont_combined}, \ref{fig:cont_age}, and \ref{fig:cont_mr} are meanwhile analogous to the contour plots of Online Figure Set~\ref{fig:contours}, excluding the presentation of $f_{{\rm SC}}$, which changes tail-to-tail. Individual pixel $p_{{\rm real}}$ probabilities were computed as before for each pixel containing at least one SCC from each tail and are reflected here. The fraction of unfitted SCC-carrying pixels $f_{{\rm unfit}}^{{\rm SCC}}$ is also shown. Contour bins are also the same as in Figure~\ref{fig:contours}, as are all definitions of \HI\ thresholds. In each of these figures, we juxtapose the \nhi--\vdisp\ and \ke--\shear\ parameter space of the two types of tails considered for visual comparison.

In the following subsections, it is important to note that projection effects are still an issue. However, a reasonably random set of tail orientations would artificially change values of properties by different amounts and blur distinctions between distributions of an \HI\ variable between two types of tails. We therefore contend that, while statistically similar distributions may be inconclusive about physical differences between types of debris, statistically different distributions are still significant even against irrevocable geometric effects. Any differences between tail types are likely telling of real physical dissimilarities in the \HI\ content between them.

\subsubsection{\HI\ and Tail-wide SCCs}
\label{sec:HIandM11}

Of the M11/\citet{K03} tail sample with \HI\ coverage, M11 identified significant tail-wide cluster populations (to 2.5$\sigma$ confidence) in NGC~2993, NGC~2782E, NGC~2535, NGC~6872E/W, NGC1614N/S, and NGC~3256W. While NGC~2992 did not have a strictly global SCC population, many of its cluster candidates are found towards the tip of the optical tail, implying (along with their photometry and present statistics) that many could be real clusters. Therefore, we include that system along with the rest of those listed as tails with very likely widespread cluster populations. We therefore compared the \HI\ properties of the re-gridded pixels of these tails with the remaining, globally SCC-poor, tail sample as in Section \ref{sec:histograms}.

\begin{figure*}[p]
\centering
\includegraphics[height=0.75\textheight]{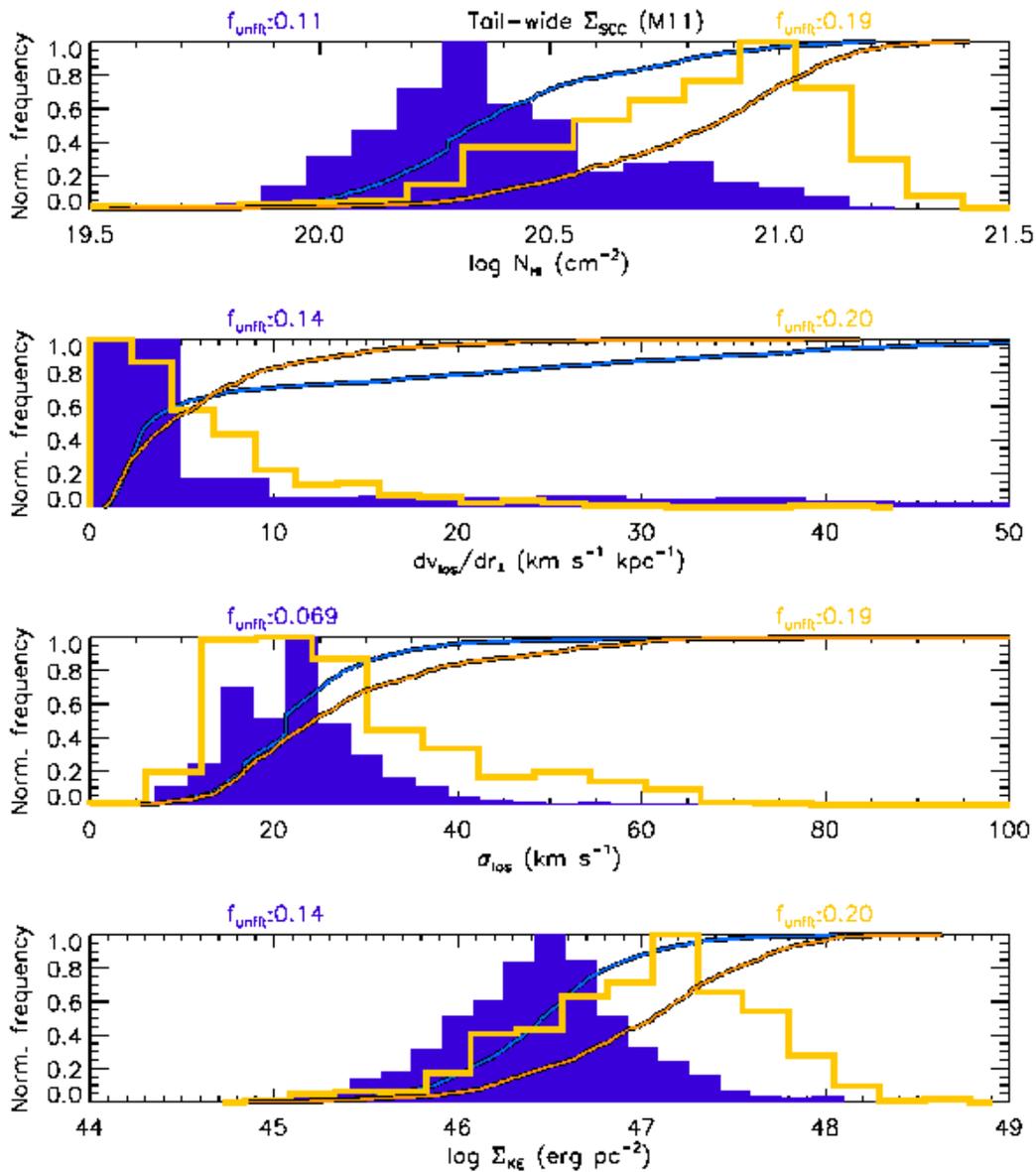}
\caption{Normalized histograms of \nhi, \shear, \vdisp, and \ke\ for identically-sized pixels from tails that do not have global SCCs populations (blue) and those that do (orange).The fraction of unfitted pixels $f_{{\rm unfit}}$ is also recorded (see \S\ref{sec:HIandM11} for details). A full resolution version of this figure is available in {\it The Astrophysical Journal.}\label{fig:hist_combined}}
\addtocounter{figcount}{1}
\end{figure*}

We present the results of this investigation in \mbox{Figure \ref{fig:hist_combined}}, with solid blue and empty orange normalized histograms representing pixels from SCC-poor and SCC-rich tails, respectively. Most strikingly, the distributions of \nhi\ and \ke\ are very clearly different between putatively cluster-rich and cluster-poor tails---the former peaks at about 0.5 dex higher in \nhi\ than the latter, with large percentage of its measurable \HI\ pixels above the \citet{aparna07} threshold. Tails with large numbers of SCCs also on average have higher \HI\ kinetic energy densities on $\sim$kpc scales (again, by $\sim$0.5 dex). Despite individual cases of localized, probable clusters in tails with low column densities ($\approx$20\% of the pixels of SCC-rich tails are unfitted for individual \HI\ properties, compared to $\approx$10\% for SCC-poor tails), \nhi\ and \ke\ are still apparently important variables in the holistic appraisal of tidal tail cluster formation activity. These variables could be tied to physical \HI\ densities and pressures, which likely play prominent roles in the formation of star clusters and other bound structures (e.g., \citealp{elmegreen93}; \citealp{elmegreen97}; \citealp{elmegreen08}).  

%
%

\begin{figure*}[p]
\centering
\includegraphics[width=1.0\textwidth]{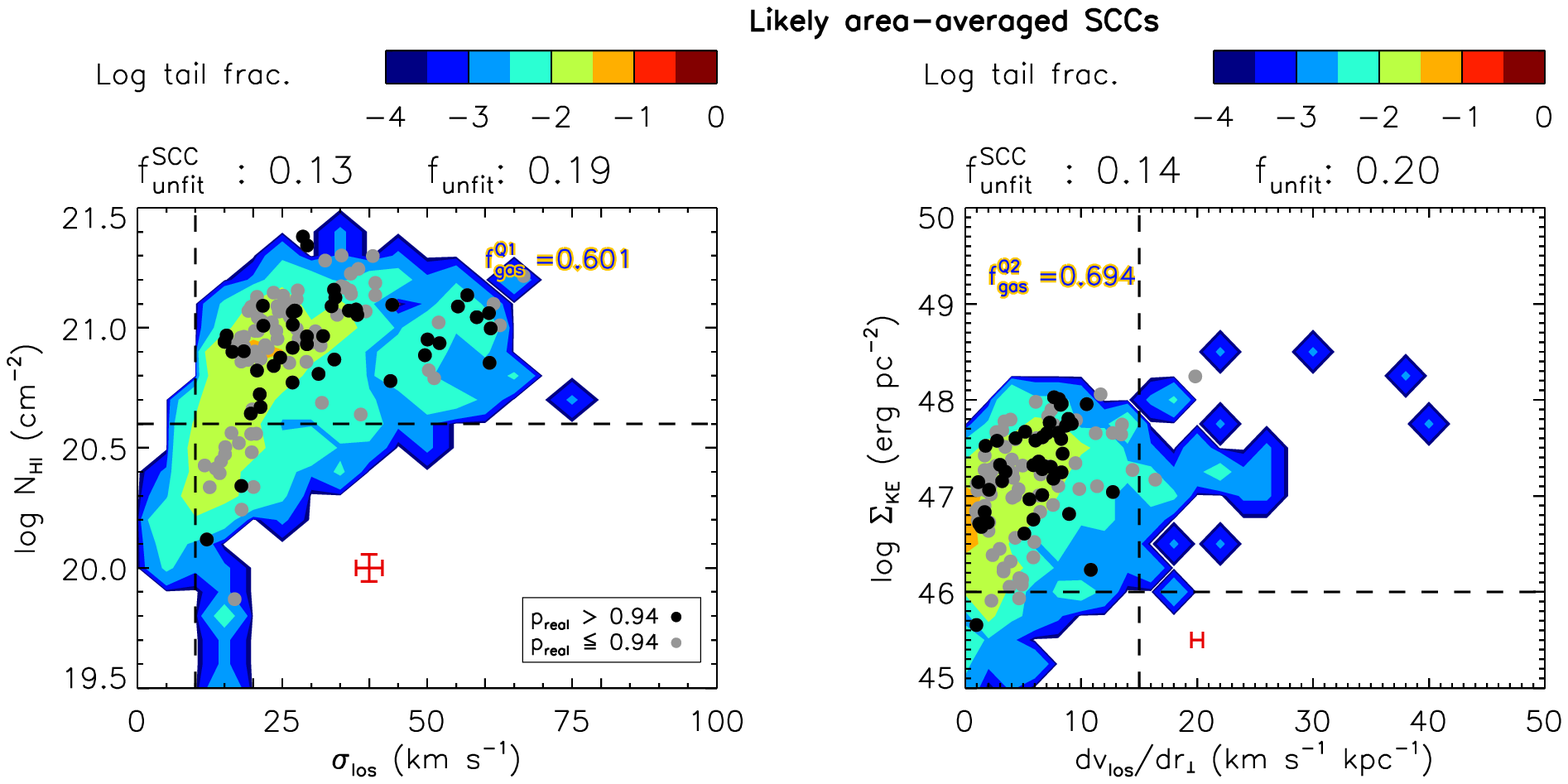}
\includegraphics[width=1.0\textwidth]{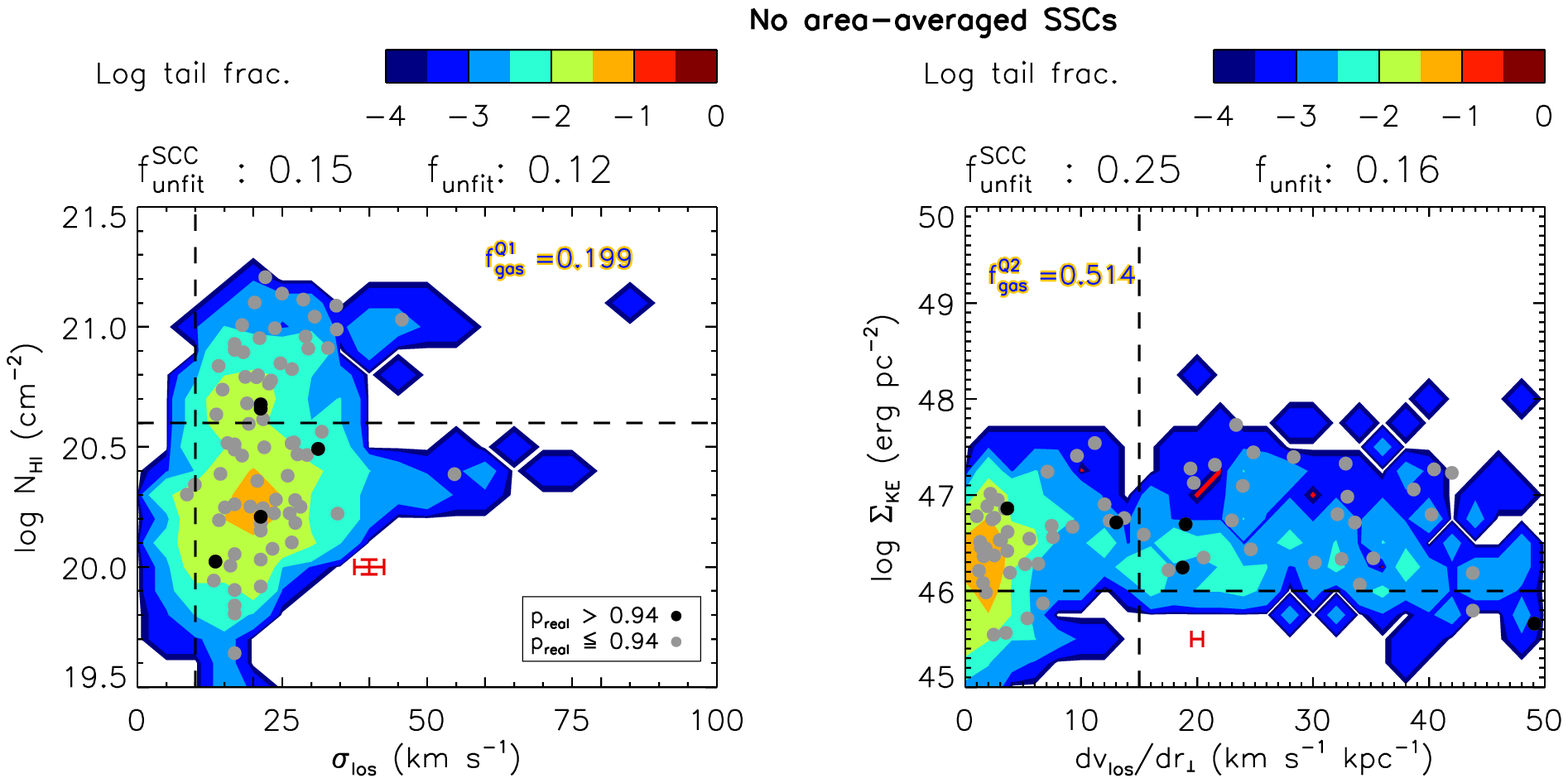}
\caption{Combined, contoured distribution of measured \HI\ quantities for tails with widespread SCC populations (M11; top row), and those without (bottom row) with levels corresponding to the fraction of the tail subsample within the indicated parameter space. Pixels with SCCs most likely to have star clusters are plotted in black ($p_{{\rm real}} > $ 0.94), and those less likely in gray ($p_{{\rm real}} <$ 0.94). The fraction of unfitted pixels, and the sample fraction within the specified quadrants Q1 (left) and Q2 (right) are provided. These are defined by the intersection of lines of potential threshold column density (Maybhate et al.\ 2007) with the thermal-turbulent velocity dispersion in the first plots; and their combined energy density with the value of shear in the solar neighborhood in the right plots. See \S\ref{sec:HIandM11} for details.\label{fig:cont_combined}}
\addtocounter{figcount}{1}
\end{figure*}

Similarly, SCC-rich tails have an extended distribution of \vdisp, with many pixels showing values in the 30--75 \kms\ range. These tails seem to have more highly disturbed gas. While relatively more \HI\ in SCC-poor tails appears to have line-of-sight velocity gradients $\lesssim$10 \kmskpc, the velocity gradient measurement distribution trails off to generally higher values than that of the SCC-rich tails. This is also evident in contrasting the contour plots of \mbox{Figure \ref{fig:cont_combined}}; the top row displays the contoured distributions for SCC-rich tails, and the bottom row the equivalent plots for SCC-poor tails. While about 60\% of SCC-rich tail \HI\ is measurably turbulent and of high column density, the percentage for the SCC-poor tails is only about 20\%.

%
%

Understandably, SCC-rich tails have more overplotted points than SCC-poor tails, and especially more pixels containing high probability clusters. Again, many of the highest probability clusters lie in regions of high column and kinetic energy density. However, as noted in the Appendix, many tails have probable clusters in unfitted or low-\nhi\ pixels. Given the statistics of many of the sources, it is likely that many of the gray points also indicate positions of real clusters in this \HI\ parameter space. Evidently, these plots provide a compelling argument for the importance of kpc-scale \HI\ in tidal tail star cluster formation as a whole, but there are individual exceptions to the rule. 

\subsubsection{\HI\ and Tail Dynamical Age}
\label{sec:HIandage}

There is much evidence that tail age or interaction stage has a pronounced effect on star formation phenomenology. In their sample of compact groups, \citet{martinez12} find evidence for a general decrease of molecular gas deficiency with the evolutionary phase (age) of the interacting group. They interpret this as an enhanced conversion of atomic to molecular gas early in the interactions, followed by gradual stripping and dispersal of \HI\ in their later phases. In their models of compressive tides in pairwise galaxy mergers, \citet{renaud09} additionally show that, while regions with tidal compression can appear in interactions at multiple stages, they often occur at greater frequency in younger stages. Other models of star-forming tidal debris often indicate that strong bursts of star formation occur early in interactions, soon after periapse (\citealp{chien10}; \citealp{dim08}, \citealp{m&h94}). Perhaps as a consequence, M11 noted that the most successfully star cluster-forming tails of their sample were all relatively young ($\lesssim$250 Myr), although they did not find a statistically significant difference in the age distributions of SCC-rich and SCC-poor tails.

\begin{figure*}[p]
\includegraphics[height=0.75\textheight]{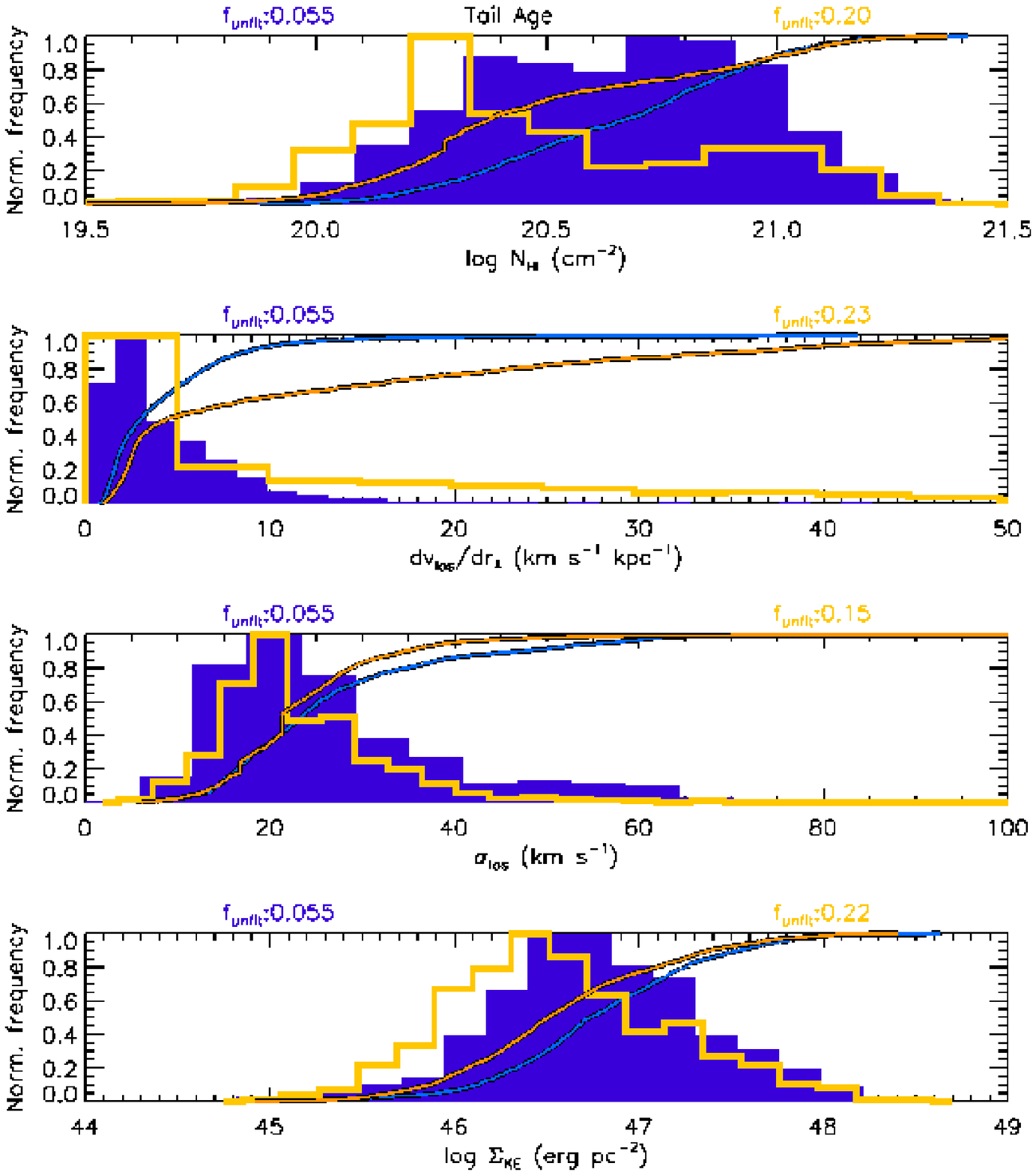}
\centering
\caption{Normalized histograms of \nhi, \shear, \vdisp, and \ke\ for identically-sized pixels from young ($<$250 Myr; blue) and old ($\geq$250 Myr; orange) tail regions in our sample. The fraction of unfitted pixels $f_{{\rm unfit}}$ is also recorded (see \S\ref{sec:HIandage} for details). A full resolution version of this figure is available in {\it The Astrophysical Journal.}\label{fig:hist_age}}
\end{figure*}

In \mbox{Figure \ref{fig:hist_age}} we exhibit normalized histograms of \HI\ properties of identically sized pixels from young ($<$250 Myr; blue) and old ($\geq$250 Myr; orange) tails. Ages are from M11 and sources therein (see Table~1); the age cut was selected to approximate the median of the M11 age distribution ($\approx$300 Myr). For all examined properties, these two categories of tails are typically quite different; younger tails enjoy higher column densities, have less gas with high velocity gradients ($\gtrsim$15 \kmskpc) over relatively large scales ($\sim$10 kpc), but have a wider range of local ($\sim$kpc-scale) velocity dispersions. There is more gas with \vdisp\ measurements $\gtrsim$20 \kms. The fractions of unfitted pixels for these \HI\ characteristics tend also to be lower in younger tails ($f_{{\rm unfit}}$ $\approx$ 0.06, compared to $\approx$0.15--0.23 for older tails). Consequently, the \HI\ of younger tails have higher mechanical energy densities on average. In all, the \HI\ gas in younger tails may be under higher pressures and is more kinematically disturbed than that of older tails.

%
%

\addtocounter{figcount}{1}

\begin{figure*}[p]
\includegraphics[width=1.0\textwidth]{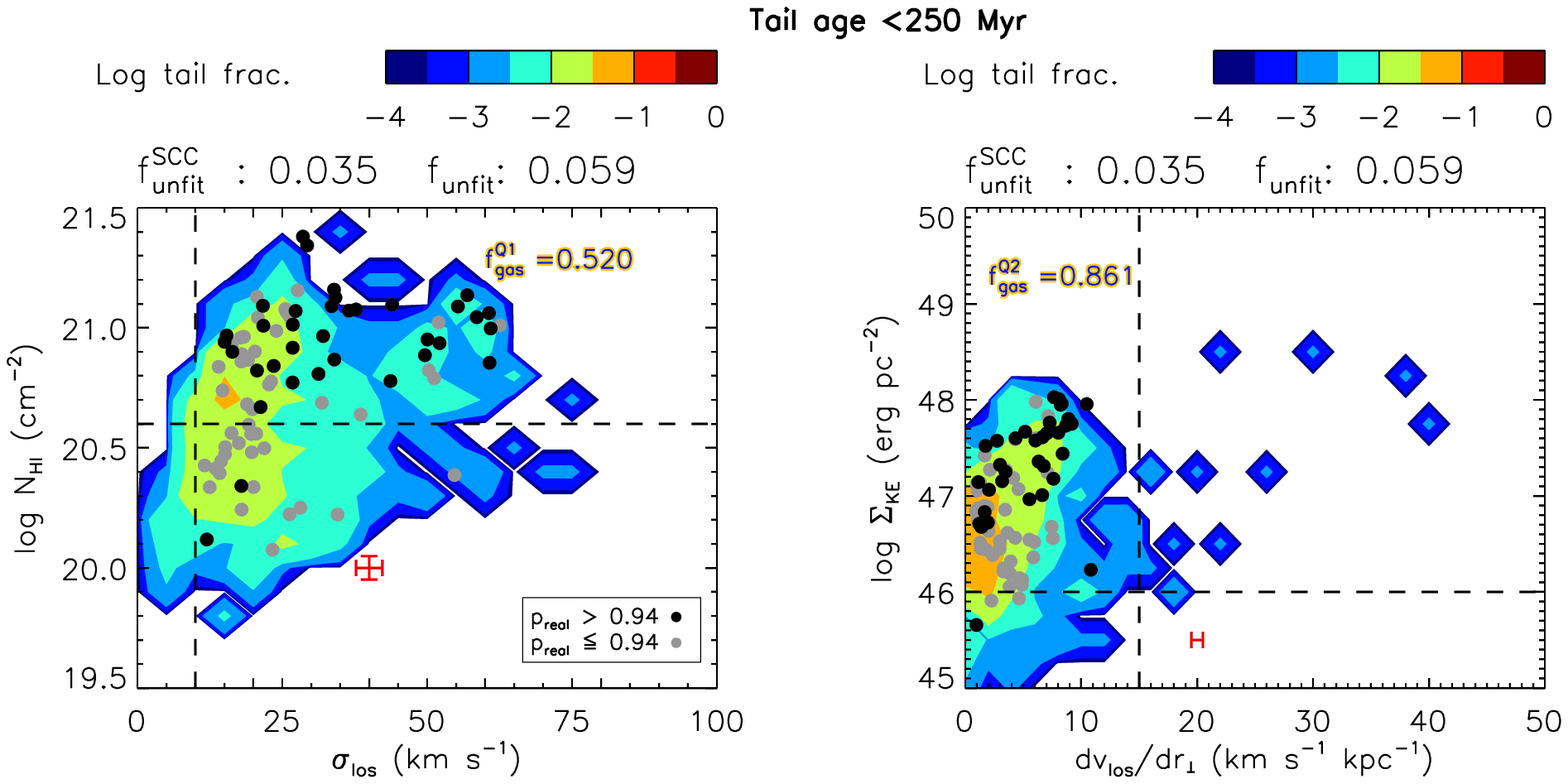}
\includegraphics[width=1.0\textwidth]{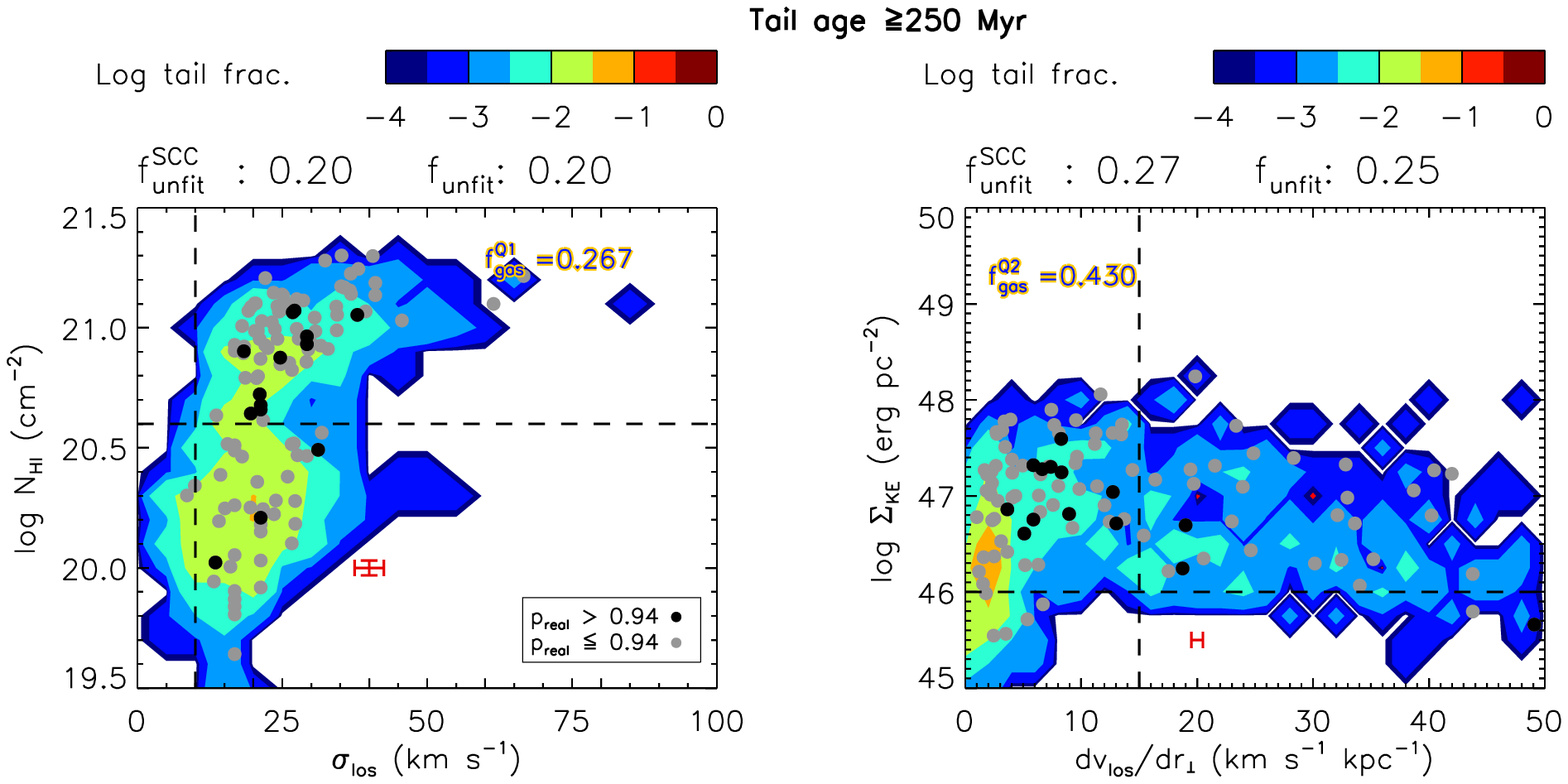}
\centering
\caption{Combined, contoured distribution of measured \HI\ quantities for young ($<$250 Myr) tails (top row), and old ($\geq$250 Myr) tails (bottom row), with levels corresponding to the fraction of the tail subsample within the indicated parameter space. Pixels with SCCs most likely to have star clusters are plotted in black ($p_{{\rm real}} > $ 0.94), and those less likely in gray ($p_{{\rm real}} <$ 0.94). The fraction of unfitted pixels, and the sample fraction within the specified quadrants Q1 (left) and Q2 (right) are provided. These are defined by the intersection of lines of potential threshold column density (Maybhate et al.\ 2007) with the thermal-turbulent velocity dispersion in the left plots; and their combined energy density with the value of shear in the solar neighborhood in the right plots. See \S\ref{sec:HIandage} for details.\label{fig:cont_age}}
\end{figure*}

\mbox{Figure \ref{fig:cont_age}} contrasts the contoured \HI\ parameter space of young and old tails in the top and bottom rows, respectively, along with their SCC-carying pixels. There are more probable SCCs in younger tails, despite the larger number of older tails in our sample (by a factor of $\approx$2). Their host pixels follow the higher percentage of higher column density and turbulent gas ($\approx$52\% compared to 27\%), as well as the higher percentage of high mechanical energy densities and low large-scale velocity gradients ($\approx$86\% compared to 43\%). As before, there are pixels with possible SCCs in low-\nhi\ areas in tails of all ages; the fraction of these with unfitted \HI\ characteristics is higher in older tails than in younger tails.

%
%

\addtocounter{figcount}{1}


Debris like NGC~1614N/S are examples of older tidal regions with ongoing or prolonged star cluster formation, broadening the distributions of \HI\ properties of older tails and contributing high-probability cluster-carrying pixels in the bottom plots of Figure~\ref{fig:cont_age}. The unplotted tail NGC~6872W shows most of the black points in the ``subcritical" regions of the left plot of its contour diagram in the Online Figure Set~\ref{fig:contours}, perhaps as a consequence of its large beam size (diluting smaller-scale \nhi\ concentrations) or rapid \HI\ depletion. Both of these cases are described in the Appendix. But in general these results are consistent with findings in the literature reporting the significance of young ages in establishing the necessary conditions---ambient densities and pressures---for star forming behavior.

Shearing kinematics in tidal debris also help explain the dispersion of SCC positions in especially older tails. Given the witnessed velocity gradients of galactic disks and debris regions examined here ($\approx$0--20 \kmskpc\ on average), it is likely that star clusters migrate a significant distance from their ``original" \HI\ pixels by the time they are observed. An order-of-magnitude estimate reveals that clusters can move several pixels (several kpc) away from their original pixel by the cutoff age of 250 Myr for these plots. Combined with age-dependent \HI\ depletion and the tendency of clusters to fade as their stellar populations evolve (e.g., M11 and references therein), a tighter correlation between SCC detections and \HI\ column densities and kinematics should logically be expected for young tails than older tails.  

\subsubsection{\HI\ and Progenitor Mass Ratio}
\label{sec:HIandMR}

\begin{figure*}[p]
\includegraphics[height=0.75\textheight]{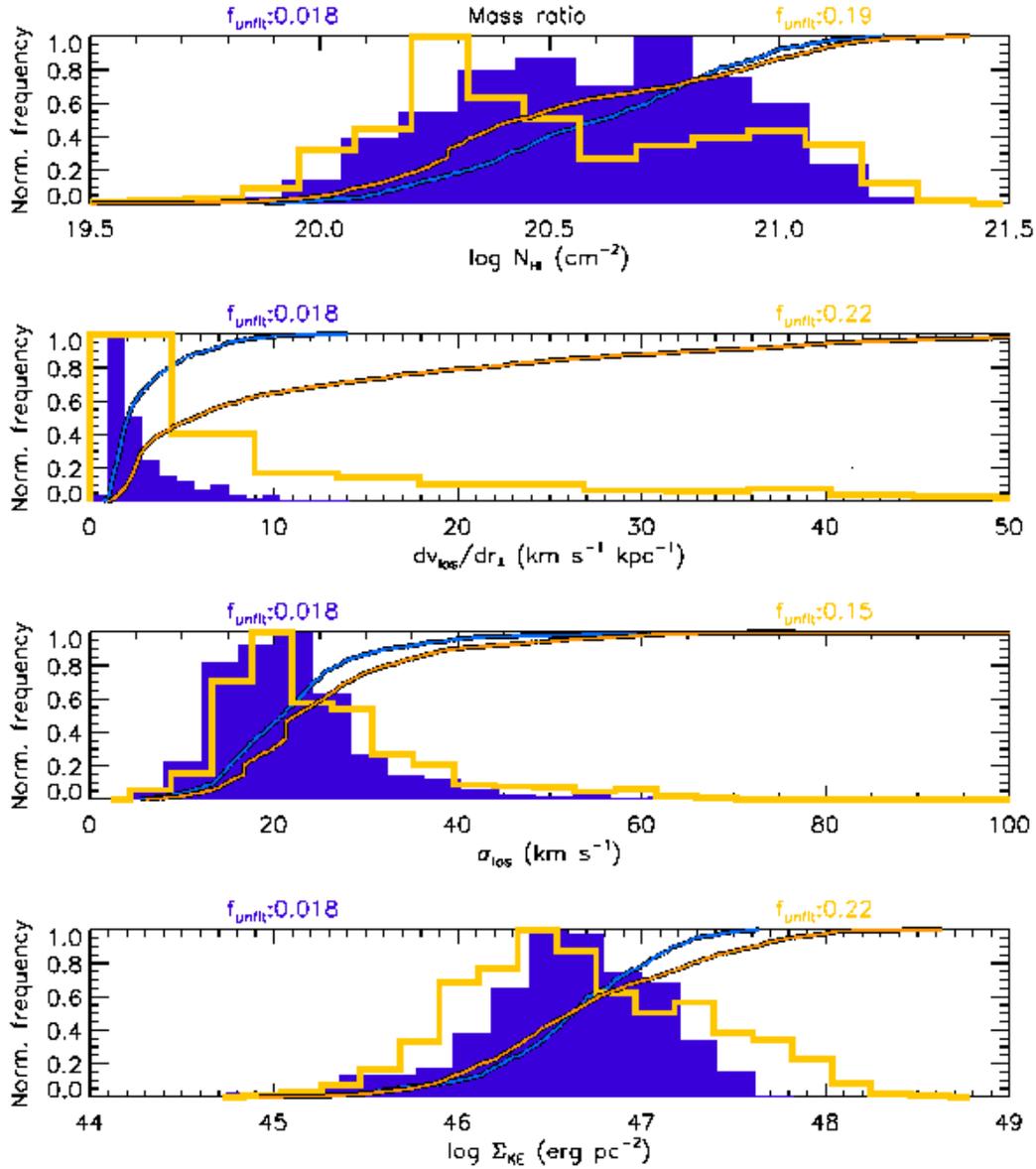}
\centering
\caption{Normalized histograms of \nhi, \shear, \vdisp, and \ke\ for identically-sized pixels from tails from minor ($M_2$/$M_1$ $<$ 1; blue) and major ($M_2$/$M_1$ $\geq$ 1; orange) encounters in our sample. The fraction of unfitted pixels $f_{{\rm unfit}}$ is also recorded (see \S\ref{sec:HIandMR} for details). A full resolution version of this figure is available in {\it The Astrophysical Journal.}\label{fig:hist_mr}}
\end{figure*}

Major mergers outnumber minor mergers in our sample by a factor of $\approx$2. The normalized histograms of \mbox{Figure \ref{fig:hist_mr}} show that tails from minor mergers seem to have a larger fraction of relatively high column density \HI\ than do tails from major mergers. Most of their gas, however, has line-of-sight velocity dispersions $<$40 \kms, while major mergers have a slightly more extended distribution of \vdisp\ out to $\approx$50 \kms. The overall appearance of the \vdisp\ histograms for these two types of tails is somewhat similar, but appears statistically independent. Given the higher fraction of \HI\ with column densities $\gtrsim$10$^{20.6}$ \cmtwo\ in our subsample of tails from minor encounters, this tail type also enjoys a bump in kinetic energy density histogram towards log \ke\ $\gtrsim$ 46.5--47 erg pc$^{-2}$.

%
%

\addtocounter{figcount}{1}


The distributions of \shear\ are the most strikingly different between tails from major and minor encounters. Velocity gradients appear largely confined to values below $\approx$10 \kmskpc\ on $\approx$10 kpc scales for the latter category, while our set of tails from major mergers has a drastically broader range of \shear\ values. Furthermore, tails from major mergers have a consistently higher fraction of unfitted pixels ($\approx$20\%) than tails from minor mergers ($\approx$2\%).   

\begin{figure*}[p]
\includegraphics[width=1.0\textwidth]{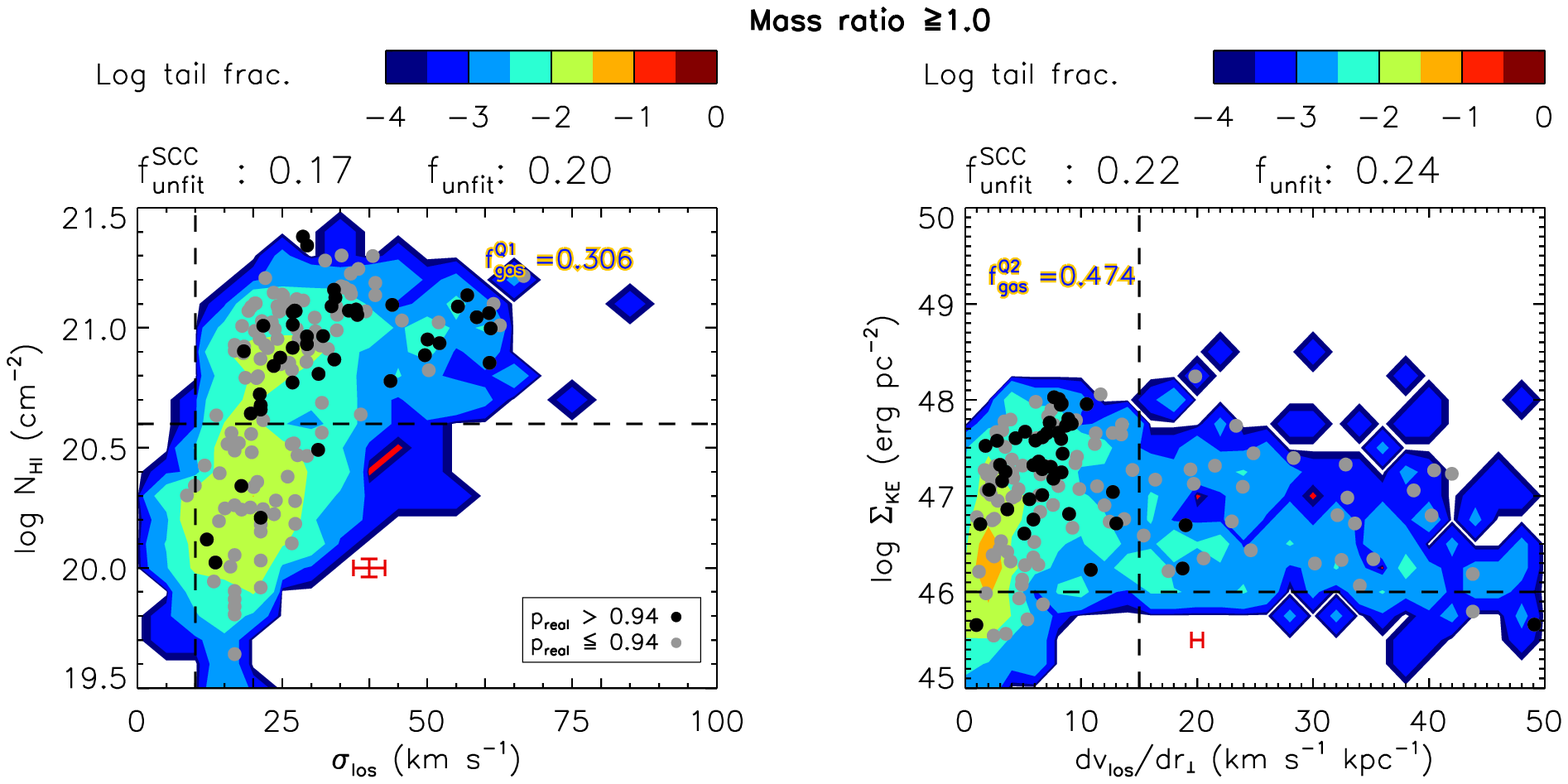}
\includegraphics[width=1.0\textwidth]{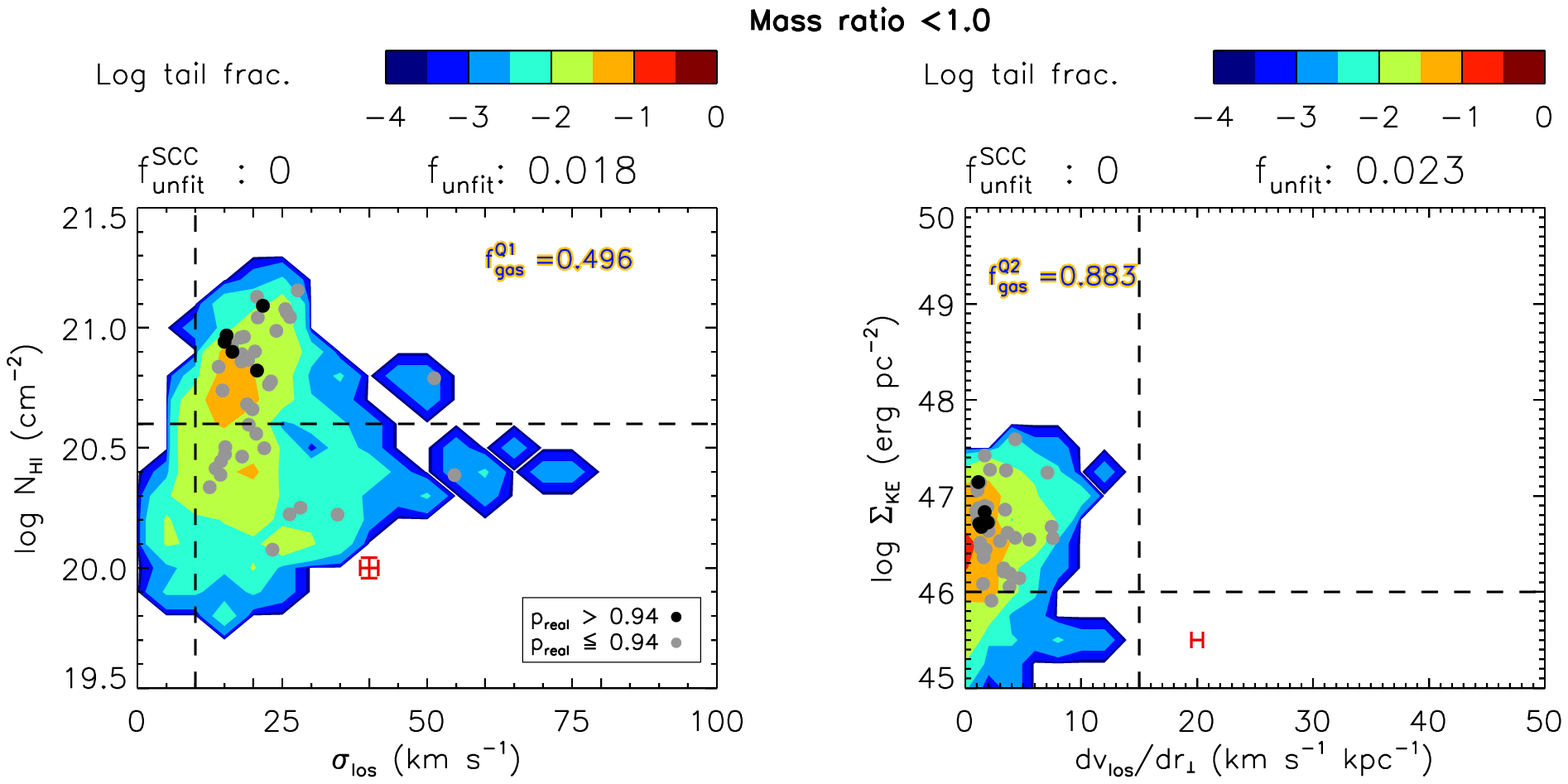}
\centering
\caption{Combined, contoured distribution of measured \HI\ quantities for tails from major ($M_2$/$M_1$ $\geq$ 1; top row) and minor ($M_2$/$M_1$ $<$ 1; bottom row) encounters, with levels corresponding to the fraction of the tail subsample within the indicated parameter space. Pixels with SCCs most likely to have star clusters are plotted in black ($p_{{\rm real}} > $ 0.94), and those less likely in gray ($p_{{\rm real}} <$ 0.94). The fraction of unfitted pixels, and the sample fraction within the specified quadrants Q1 (left) and Q2 (right) are provided. These are defined by the intersection of lines of potential threshold column density (Maybhate et al.\ 2007) with the thermal-turbulent velocity dispersion in the left plots; and their combined energy density with the value of shear in the solar neighborhood in the right plots. See \S\ref{sec:HIandMR} for details.\label{fig:cont_mr}}
\end{figure*}

Many of these trends are of course reflected in the associated contoured distributions of these two types of tails in \mbox{Figure \ref{fig:cont_mr}}, with the addition of SCC-carrying pixels as before. There are clearly more cluster candidates of both relatively high and low probability in the tails of major interactions than in minor ones, but again tails from major interactions outnumber those from minor interactions 14 to 6 in this figure. Despite much of the \HI\ gas inhabiting the ostensibly subcritical column density region of the parameter space explored in the top row (major encounters) of \mbox{Figure \ref{fig:cont_mr}}, many of the pixels with high probabilities of having at least one cluster still lie in the higher column density region.

%
%

\addtocounter{figcount}{1}


Between the evident trend of minor merger tails having higher \nhi\ overall and/or a smaller range of \shear\ for this medium, it is possible that major mergers are prone to more intense interactions. A higher \HI\ dispersal efficiency may be possible in major interactions; the larger fractions of unfitted pixels in tails from these encounters might testify to this. Stronger interactions may also engender greater disruptions to the velocity fields on $\sim$10 kpc scales. For example, \citet{keel03} find that the major merger NGC~6621/2 exhibits strong H$\alpha$ velocity perturbations and has an extensive population of star clusters. This contrasts with their observations of NGC~5752/4, a minor merger with evidently less kinematic disruption. While the interaction ages of these two collisions are different ($\sim$100 Myr for the former and $\sim$250 Myr for the latter), \citeauthor{keel03} contend that star cluster formation depends on a mass ratio-modulated strength of the interaction more than the interaction stage. However, they also examine cluster populations across the progenitor galaxy disks rather than just the tidal debris, making these studies difficult to compare. 

But most critically, it must be noted that the single and combined distributions of \HI\ properties of tail pixels from minor and major mergers bear a qualitative resemblance to those from young and old tails, respectively. This can be seen by comparing \mbox{Figure \ref{fig:cont_age}} with \mbox{Figure \ref{fig:cont_mr}}. Table~1 shows that this results from the limited sample size; while the M11 sample is reasonably mixed in terms of age and mass ratio, the subsample attributed to \citet{K03} supplies a large number of older tails from major encounters. Thus, while our observations and analyses can be viewed through the lens of interaction age and/or mass ratio, our finite sample size makes it difficult to ascertain the effects of one or the other. Removing the \citet{K03} sample for this work would only further decrease the sample size and make any results more sensitive to particular tails and their unstudied additional effects that certainly contribute to their SCC populations. We therefore opt not to do that here. Rather, we maintain that a {\it combination} of tail age and progenitor mass ratio (and likely a number of other variables not studied here) is at least partly responsible for the \HI\ and SCC phenomenology we present.

\subsection{Turbulence: A Cause or Consequence of Star Formation?}
\label{sec:sfr}

Supernovae (SNe), particularly Type II, are generally regarded as the dominant source of feedback in the ISM in galactic contexts (e.g., \citealp{vdb}; \citealp{tasker08}; \citealp{tasker06}; \citealp{spitzer90}). It then becomes a question for this project of whether the supernovae produced in prior star formation events are the primary source of the mechanical energies observed in the neutral \HI\ in these tidal tails. That is, do the kinetic energy densities (and presumably pressures) we measure result from star and cluster formation, or are they the cause of star and cluster formation?

Using UV/IR-derived star formation rates (SFRs) of a sample of relatively low inclination disk galaxies in the THINGS survey, \citet{tamburro} calculated the pixel-by-pixel kinetic energy densities that would be produced from feedback from core-collapse SNe. In comparing these to measurements of the actual \HI\ energy densities, they find that the efficiency of energy injection from SNe would have to be unrealistically high in order for their measured \ke\ values to be supplied by star formation. Not having a uniform survey of UV, IR, or H$\alpha$ diagnostics of star formation rates for our sample, we instead aim to determine the potential star formation rates of our tidal tails from their \ke\ and suggest whether the results are reasonable for this kind of environment.

As in \citet{tamburro}, we begin by equating the kinetic energy density to that theoretically provided by supernovae; i.e.\ \ke\ = $\epsilon_{{\rm SN}} \dot{E}_{{\rm SN}} \tau_{{\rm SN}}$. $\dot{E}_{{\rm SN}}$ is the rate at which energy is released by SNe, which is a function of the SFR; $\epsilon_{{\rm SN}}$ is the efficiency with which that energy is converted to the observed turbulence, and $\tau_{{\rm SN}}$ is the turbulence dissipation rate. Combining their Equations~(5)--(7), characterizing the dissipation timescale and fraction of the stellar populations that produce SNe, we derive the following expression for the pixel-by-pixel SFR ($\Upsilon$):
\begin{align}
\frac{\Upsilon}{\msun~{\rm yr}^{-1}}~=~2.03 \times 10^{-3}~\frac{\alpha}{\epsilon_{{\rm SN}}(1+f_{{\rm 1a}})}\nonumber\\
\times \left(\frac{f_{{\rm SN}}}{\langle M/\msun \rangle}\right)~\left(\frac{\lambda}{100~{\rm pc}}\right)^{-1}\nonumber \\ 
\times \left(\frac{M_{\mbox{\scriptsize{\HI}}}}{\msun}\right)~\left(\frac{\vdisp}{10~{\rm km~s}^{-1}}\right)^{3}~.   
\label{eqn:sfr}
\end{align}
\noindent As before, $\alpha$ corresponds to the degree of isotropy (here assumed to be 3/2), and the factor 1/(1+$f_{{\rm 1a}}$) is a term included in our derivation to account for the fractional contribution of type-1a SNe. Following \citet{mannucci05}, we consider $f_{{\rm 1a}}$ values in the range 0--0.33. The term $f_{{\rm SN}}$ indicates the fraction of (recently formed) stars that end as core-collapse SNe, while $\langle M/\msun\rangle$ denotes the average mass of the stellar population within the examined region. Dividing these two terms in the equation, we consider values in the range 0.9 $\times$ 10$^{-2}$ to 1.3 $\times$ 10$^{-2}$, as shown by \citet{tamburro} for an upper IMF mass limit of 20--50 \msun, with a mass function of the form $\phi(m) = m^{-\alpha_0}$; $\alpha_0$ = 1.3 for 0.1--0.5 \msun\ and 2.3 for 0.5--120 \msun\ (e.g., \citealp{calzetti07} and references therein). We use values in the range 0.1--0.5 for the efficiency $\epsilon_{{\rm SN}}$; the lower limit is suggested by numerical simulations \citep{thornton98}, and the upper limit provides a conservative lower limit to $\Upsilon$. The turbulence driving scale is given by $\lambda$, which simulations predict to be 100$\pm$30 pc (\citealp{joung06}; \citealp{deavillez07}). The mass of \HI\ is $M_{\mbox{\scriptsize{\HI}}}$, and \vdisp\ as defined here. Note the additional factor of \vdisp; this results from the implicit expression of $\tau_{{\rm SN}}$ in terms of $\lambda$ and \vdisp. The constant 2.03 $\times$ 10$^{-3}$ incorporates all unit conversions and the 10$^{51}$ erg that are assumed to be released from each type II SN event \citep{heiles87}.

We then mapped out the pixel-by-pixel SFR of all the tails using our maps of \mhi\ and \vdisp, for the combinations of aforementioned constants that yield the smallest and greatest result. We performed this for both original maps and those produced at the constant pixel scale. In practice, changing the efficiency of injection $\epsilon_{{\rm SN}}$ has the dominant affect in this equation; we added the results of all pixels belonging to individual tails and record the results in Table~5. We posit that if the tail-wide ongoing star formation rate (of similar timescale as the turbulence dissipation timescale) could fall between the ranges quoted in this table, it is possible that star formation and the SNe explosions that result from it generally caused the turbulent kinetic energy densities we observe rather than was an effect of it. Ideally, uniformly deep, wide-field imaging in H$\alpha$ of our tail sample would provide the star formation rate on the corresponding timescale ($\sim$10 Myr), allowing a direct and straightforward comparison. Unfortunately, no such survey exists.

%
%
%

\begin{table}[tb]

{\scriptsize
\begin{tabular}{l|rr|rr}
\multicolumn{5}{c}{\textbf{{Table 5.}}}\\ 
\multicolumn{5}{c}{{ Hypothetical Star Formation Rates}} \\
\hline\hline
\multicolumn{1}{l}{} & \multicolumn{2}{l}{Variable Pix. Scale} & \multicolumn{2}{l}{Const. Pix. Scale} \\
\hline
 
Tail & min. $\Upsilon$ & max. $\Upsilon$ & min. $\Upsilon$ & max. $\Upsilon$ \\
 & (\msun\  & (\msun\  & (\msun\ & (\msun\ \\
  & yr$^{-1}$)  & yr$^{-1}$)  & yr$^{-1}$)  & yr$^{-1}$) \\
 
\hline
 
NGC 1487E & 0.056 & 1.0 & 0.15 & 2.7 \\
NGC 1487W & 0.035 & 0.62 & 0.069 & 1.2 \\
NGC 4747 & 0.0079 & 0.14 & 0.0053 & 0.095 \\
NGC 520 & 0.080 & 1.4 & 0.080 & 1.4 \\
NGC 2992 & 9.1 & 160 & 11. & 190 \\
NGC 2993 & 3.0 & 53. & 2.8 & 50. \\
NGC 2782E & 2.6 & 47. & 3.1 & 56. \\
NGC 2782W & 0.53 & 9.4 & 0.54 & 9.7 \\
NGC 2444 & 1.1 & 20. & 1.3 & 23. \\
NGC 2535 & 1.5 & 28. & 1.9 & 33. \\
NGC 6872E & 18. & 320 & 18. & 320 \\
NGC 6872W & 15. & 260 & 14. & 260 \\
NGC 1614N & 0.44 & 7.9 & 0.48 & 8.6 \\
NGC 1614S & 0.24 & 4.3 & 0.48 & 8.6 \\
NGC 4038A & 0.022 & 0.39 & 0.0082 & 0.15 \\
NGC 4038B & 0.0011 & 0.020 & 0.0019 & 0.033 \\
NGC 4038C & 2.2e-4 & 0.0039 & 2.4e-4 & 0.0042 \\
NGC 3256E & 2.3 & 41. & 3.2 & 57. \\
NGC 3256W & 6.0 & 110 & 7.4 & 130 \\
NGC 7252E & 0.26 & 4.6 & 0.26 & 4.6 \\
NGC 7252W & 2.4 & 43. & 2.9 & 52. \\
NGC 3921S & 1.5 & 26. & 1.5 & 26. \\
\hline\hline
\end{tabular} 
}
\end{table}


There are a few potentially elucidating case studies, however. \citet{K12} find a H$\alpha$-derived SFR for the entire western tail of NGC~2782 of 9 $\times$ 10$^{-6}$ \msun\ yr$^{-1}$ kpc$^{-2}$. Adjusting for the in-tail area used here, we calculate a rate $\Upsilon$ $\approx$2 $\times$ 10$^{-3}$ \msun\ yr$^{-1}$. This is several orders of magnitude lower than the values reported in Table~5; thus the \HI\ turbulence is not likely caused by star formation in this case. Moreover, the H$\alpha$ data of \citet{mihos93} indicate that most current star formation in NGC~6872 is confined to its tidal tails, and is occurring at a rate $\sim$3 \msun\ yr$^{-1}$. The individual rates for the eastern and western tails must be smaller; but in either case the rate is far below the minimum $\Upsilon$ needed ($\sim$15 \msun\ yr$^{-1}$ for either tail and pixel size) for the observed turbulent energy to be supplied by feedback from star formation overall. 

While the turbulence we observe is purely on $\sim$kpc scales, it is likely crucial to the kinematic nature of the ISM on smaller, star-forming ($\sim$pc) scales. Turbulence is injected on hundreds of pc to $\sim$kpc-scales (e.g., \citealp{maclow99}; \citealp{elmegreen93}; \citealp{block10}), and cascades down to smaller scales until it is dissipated into thermal energy by viscosity (\citealp{elmegreen03}; \citealp{kolmogorov41}). There are a number of observed radiative pathways available to eliminate this energy effectively and faster than the turbulence decay timescale by several orders of magnitude; e.g., collisionally-excited lines of \CIV\ \citep{wolfire03}, rovibrational transitions of H$_2$ (eg.\ \citealp{cluver09}), and radio/X-ray continuum emission \citep{kaufman11}.   

In effect, we are left with a top-down evolutionary process, where gas kinematics at observably large scales---likely influenced by a combination of tail age and progenitor mass ratio as seen in Section \ref{sec:smooth}---influences star formation on the otherwise unresolved, smaller scales. Turbulent gas generates a log-normal density probability density function (PDF) in the local ISM \citep{wada02}, whose width---dispersion in density---increases mainly with turbulent Mach number \citep{krumholz07}. Thus, turbulence on large scales affects the density distribution of gas on small scales. This is exacerbated in mergers, where additional, expeditious kinematic perturbations create deviations in the log-normal distribution, favoring more high-density gas \citep{bournaud11}. 

In all, our results present a compelling case for the importance of interaction-established, kpc-scale \HI\ kinematics and column densities in determining the capacity of pc-scale star cluster formation in tidal debris. Other, currently unsurveyed interaction properties---e.g., dark matter halo potentials, interaction speeds and impact parameters, H$_2$ content, and dust-to-gas ratios---may help set the stage for luminous star cluster formation in tidal tails and may help account for the variety and distribution of SCC populations witnessed here. However, the character of the neutral hydrogen medium, described by its densities, pressures, and kinematics, also appears to be an invaluable factor in this process. We have shown that high-confidence SCC detection often occurs in tails with large fractions of high column density, turbulent gas (Section \ref{sec:HIandM11}). Both the putative star cluster populations and the underlying \HI\ medium clearly varies depending on the age (Section \ref{sec:HIandage}) and/or the progenitor galaxy mass ratio (Section \ref{sec:HIandMR} ) of the interaction, indicating that these dynamical properties are important in tidal tail star cluster formation as well.

%

\section{Conclusions}
\label{sec:conclusions}

In this paper, we have performed a pixel-by-pixel characterization and analysis of the \HI\ content of 22 tidal debris fields from pairwise galaxy interactions, comparing these properties between pixels that contain star cluster candidates (SCCs) and those that do not. In our tail sample we measured \HI\ column density \nhi, mass density \mhi, line-of-sight velocities \vuse\ and velocity dispersions \vdisp, as well as mechanical energy densities \ke\ to the best pixel-scale spatial resolutions ($\sim$kpc) our archival data afford. We also constructed a tracer of the locally-centered line-of-sight velocity gradient \shear\ for these regions, examining the velocity fields of the tails over larger, $\sim$10 kpc scales.

While tail gas kinematics and geometry, small-scale star and cluster formation events, and feedback likely produce dispersion in our results, it appears that many ``successful" kpc-scale triggered clustered star formation events often take place in $\sim$kpc-scale regions of high \HI\ gas column and kinetic energy densities. Kinetic energy density should generally trace \HI\ pressures (integrated along the line of sight), so this would imply that clustered star formation as detected by M11 requires high pressures (high \ke\ from moderately turbulent gas), but also high \nhi\ on these scales. It is possible that these conditions could arise from intermittent, compressive tides (e.g., \citealp{renaud09}) or gravitational instabilities (e.g., \citealp{bournaud10}).

We repeated our analysis for data regridded to a constant pixel scale, and compared the distributions of \HI\ properties between tails of different global cluster populations, dynamical ages, and progenitor mass ratios. Despite variation in individual tail phenomenology and SCC populations, several clear trends and contrasts are evident in the \HI\ content of these types of tails:

\begin{enumerate}

\item Tails with global cluster populations (as identified in M11) tend to have \HI\ gas of higher column and kinetic energy densities than tails with no apparent tail-wide cluster populations, with much of it ($\gtrsim$50\%) above the putative log \nhi\ = 20.6 \cmtwo, log \ke\ = 46 erg pc$^{-2}$ thresholds mentioned here. The distribution of velocity dispersions also has a greater contribution of high values (\vdisp\ $\gtrsim$ 30 \kms).

\item Relatively young tails (dynamical age $<$250 Myr) have more pixels with high probability clusters than older tails ($\geq$250 Myr), and contain gas with larger column and kinetic energy densities on average. There is also more gas overall at high velocity dispersions (\vdisp\ $\gtrsim$ 30 \kms) in these types of tails. Older tails are more likely to be affected by age-dependent fading in their cluster populations as well. 

\item Tails from minor encounters appear to have more relatively high column density \HI\ than tails from major interactions. When measured over a common, $\sim$10 kpc-scale velocity field, major encounters appear typically slightly more capable of driving the gas to higher ($>$10 \kmskpc) velocity gradients. The strength of the encounter, gauged in part by the interacting mass ratio (among other variables) may help control the efficiency of \HI\ dispersal and disruption of the velocity fields in its resulting tidal debris. Major mergers may drive gas to high kpc-scale velocity dispersions, but may also encourage stronger local depletion and larger-scale shearing motions in certain cases.

\item Because of the strong degeneracy in our sample between old tails and tails from major interactions, it cannot be well determined whether the observations noted above are explained by primarily age or mass ratio, or a combination of the two. Other unstudied properties like interaction speed, dark matter halo structure, and ISM dust-to-gas ratios are likely to also play roles in star cluster formation in tidal tails. 

\item In calculating the required star formation rate (SFR) for the turbulent energies to be provided by feedback from star formation, we deduce that it is unlikely that the kinetic energy densities of the \HI\ medium are a consequence of star formation, but rather a pre-existing environmental condition. However, we can only verify this for a few systems that have data of SFR indicators (i.e.\ H$\alpha$) that trace activity around the same timeframe as the $\sim$10 Myr turbulence injection and dissipation.

\end{enumerate}

While these results are promising, much work remains in resolving certain issues. Additional optical photometry (e.g., broadband $UB$ coverage to add to existing $VI$ images) is required to accurately age-date cluster candidates for a more precise comparison to \HI\ properties than the probabilistic work incorporated here. A larger tail sample would also allow a more robust comparison between \HI\ measurements between certain tail types, and additional subcategories within. A homogeneous H$\alpha$ survey of the tidal debris sample would facilitate a quantitative comparison between hypothetical star formation rates calculated here, and their measured equivalents over the appropriate $\approx$10 Myr timescale. Lastly, consistent modeling of all interacting systems would provide a direct link from the projected quantities we observe here to real physical properties like \HI\ number density and local pressure.

\acknowledgments
We thank the anonymous referee for helpful comments and suggestions that greatly improved the clarity and results of this paper. This project was supported by a grant from the Space Telescope Science Institute (grant no.\ HST-GO-11134.05-A). Funding was also provided by the National Science Foundation under award AST--0908984.The Institute for Gravitation and the Cosmos is supported by the Eberly College of Science and the Office of the Senior Vice President for Research at the Pennsylvania State University. 

{\it Facilities:} \facility{VLA}, \facility{ATCA}, \facility{WSRT}




\appendix
\twocolumn

\section{Notes on Individual Tails}
\label{sec:tailnotes}

Below we present an overview of the observed \HI\ properties of each tail as evinced in \mbox{Figure \ref{fig:alltails}} and online Figure Sets~\ref{fig:tailmap} and \ref{fig:contours}. For more information about these systems and their SCCs, consult M11 and references therein; several dynamical and observed properties from those works are occasionally mentioned here.

\subsection{NGC~1487E} 

This eastern tail of a $\approx$500 Myr old merger remnant between similarly massed galaxies \citep{lee05}, NGC~1487E contains a single SCC that is not likely real ($f_{{\rm SC}}$ = 0). This system is the closest of the tail sample, so optical images from M11 reveal that star formation may have occurred in detectable, fainter structures. Some of these may be $\sim$10$^4$ \msun\ star clusters formed \textit{in situ}. This tail clearly favors the formation of dispersed, lower-mass structures over more massive star clusters.  

Figures 4.1 and 5.1 display a high column density \HI\ tail on $\sim$kpc scales, with line-of-sight velocity dispersions typically $\sim$20 km s$^{-1}$. The velocity field of this tail varies smoothly and produces line-of-sight velocity gradients within this tail all $\gtrsim$5 km s$^{-1}$ kpc$^{-1}$. Figure~5.1 indicates that NGC~1487E has a high fraction of turbulent and dense gas, but M11 suspected that inclination effects artificially increase the apparent \HI\ densities measured for this system. This can easily elevate \nhi\ by a factor of 3 (0.5 dex) alone, so the actual fraction of gas above the \citet{aparna07} threshold (and at high mechanical energies) may be small. Consequently, the idea of a critical \HI\ column density for massive cluster formation may still be supported in this case. 

\subsection{NGC~1487W} 

This tail has no SCCs according to the present optical criteria. Like the eastern tail, optical detection of point sources in M11 demonstrate that this tail may preferentially exhibit a star forming morphology devoid of massive clusters. Instead, this tail hosts a large number of fainter blue ($V-I$ $\sim$ 0.0) sources across the WFPC2 FOV (M11). The distributions of \nhi, \vdisp, and \ke\ are very similar to those of this tail's eastern counterpart, though again it is likely that projection effects may artificially enhance the apparent column densities and affect other measured quantities. Overall, \shear\ in this tail appears constrained to values generally $\lesssim$20 km s$^{-1}$ kpc$^{-1}$, unlike in NGC~1487E, which shows several regions with larger velocity gradients. Otherwise, $\sim$kpc-scale mechanical energies across the \HI\ debris are similar between these two tails.
     
\subsection{NGC~4747}

The debris seen in NGC~4747 results from a $\approx$320 Myr old merger between its progenitor galaxy and the more massive NGC~4725 (\citealp{haynes79}; \citealp{wevers}). As in NGC~1487E/W, sources were observed at fainter magnitudes in M11, but few sources (6 in tail) were detected at M$_{{\rm V}} <$ -8.5. These also have a good probability of being real clusters based on star counts; $f_{{\rm SC}}$ = 0.80 overall, while one pixel has a $p_{{\rm real}}$ value $>$0.94 evident in \mbox{Figure 5.3}. 

The WSRT data used in this analysis is not as deep as the VLA/ATCA cubes for other tails, but accounting for the beam dimensions, it is still sensitive with our channel masking to log $\nhi$ $\gtrsim$ 19.6 (cm$^{-2}$). A small amount of low column density gas is detected towards the center of the WFPC2 FOV,and has relatively low velocity dispersions ($\lesssim$30 km$^{-1}$) on its $\approx$6 kpc beam scale. We measure high velocity gradients across this \HI\ gas (\shear\ $>$ 30 \kmskpc), but the few surrounding pixels for each measurement makes this quantity prone to uncertainty. A large fraction of pixels ($\approx$50\%) are marked as unfitted for at least one of the \HI\ properties plotted, appearing in optical images but showing no clear \HI\ emission here. This tail appears optically prominent but relatively \HI-poor.

The fact that this intermediate-stage merger has hardly any \HI\ to the best of its cube's offered resolution is peculiar. It may be possible that it has been stripped out and/or dispersed by its recent encounter with NGC~4725. What gas remains is characterized by low kinetic energy densities ($\sim$10$^{46}$ erg pc$^{-2}$) and low column densities. However, the compliment of blue ($V-I$ $<$ 0.5) point sources seen in M11 may indicate that less massive star clusters ($\sim$10$^{4}$ \msun) have formed in the past few hundreds of Myr. Thus, it appears that the formation of these structures would require a lower column density threshold than the fiducial log \nhi\ $\approx$ 20.6 (cm$^{-1}$) proposed for more massive clusters \citep{aparna07}, and/or the star formation that has occurred may have helped efficiently consume/disperse an existing \HI\ reservoir over the past few hundred Myr.

 \subsection{NGC~520} 

The tip of NGC~520 tail, the \HI\ and optical condensation seen here and in M11 could be a TDG from a $\approx$300 Myr old interaction or a dwarf galaxy entrained in the \HI\ tidal stream \citep{stanford91}. Within, fainter sources than the SCC magnitude cutoff were originally detected. Two sufficiently bright objects of the M11 source lists an be considered SCCs and are scattered in the \HI\ debris, but they have low probabilities of being real clusters ($f_{{\rm SC}}$ = 0.06).

The very center of the potential TDG contains high column density gas (\nhi\ $\gtrsim$ 10$^{20.6}$ cm$^{-1}$) on $\approx$3 kpc scales, and has a low line-of-sight velocity gradient (\shear\ $<$ 5 \kmskpc). However, the \ke\ distribution peaks barely above 10$^{46}$ erg pc$^{-2}$ from the moderately turbulent nature of the gas, and overall the fraction of the tail \HI\ that is both of high column density and turbulent on $\approx$3 kpc scales is $\approx$17\%. Once more, this environment may be somehow favorable for a smoother, diffuse star formation morphology that does not permit the formation of massive star clusters.

 \subsection{NGC~2992} 

NGC~2992 is $\approx$100 Myr into its interaction with NGC~2993 \citep{NGC2992_pub}. It is heavily inclined along our line of sight, enhancing the tail column density by a factor of $\approx$3 (0.5 dex). The putative dwarf galaxy candidate at the tail tip was previously confirmed to be kinematically distinct and therefore is likely a real TDG (\citealp{brinks04}; \citealp{bournaud04}). Most of the 10 SCCs detected in this system can be found there, at least in projection. Though M11 found that across the debris, NGC~2992 is deficient in clustered star formation (to $M_{{\rm V}} <$ -8.5), individually we find that about 72\% of the detected sources are probably massive clusters, especially given their nonuniform distribution.

The \nhi\ measurements for in-tail pixels are all high in projection, with almost 80\% of the gas having apparent ($\approx$4 kpc-scale) column densities log \nhi\ $\gtrsim$ 20.6 (\cmtwo). This is also evident in the \mhi\ map, where regions with mass densities $\gtrsim$10 \msun\ pc$^{-2}$ are apparent, despite the saturation of \HI\ known to occur around that (deprojected) density. We detect a reasonably uniform \vdisp\ distribution, with values ranging from $\approx$20--80 \kms, while our map of \shear\ highlights a region with a locally shallow ($<$5 \kmskpc) velocity gradient within the TDG and its SCCs.

In terms of the SCCs, K--S tests reveal that the distributions of tail pixels with and without at least one SCC are similar for all properties except \nhi, owing to the cluster candidates situated in the high column density TDG (though note the small sample size for the former category). Again, though it is not clear from Figure~5.5, these SCCs appear superimposed over a region of especially shallow large-scale velocity gradient in \mbox{Figure 4.5}. And even if the \nhi\ values were decreased by 0.5 dex to roughly account for tail geometry and orientation, Figure 5.5 would show that these SCCs would still be embedded in \HI\ gas with higher column densities than the \citet{aparna07} threshold.

 \subsection{NGC~2993} 

The interacting, similarly massed counterpart to NGC~2992, NGC~2993 presents a relatively face-on orientation. High \nhi\ areas on $\approx$4 kpc scales also register high moment-measured velocity dispersions (\vdisp\  $\gtrsim$ 40 \kms) and kinetic energy densities (log \ke\ $\gtrsim$ 47 erg pc$^{-2}$), and there appears to be a strong dependence of \vdisp\ on \nhi\ in \mbox{Figure 5.6}. About 70\% of these sources should be clusters, but only half lie in areas of high \nhi. Most are still found in regions of high kinetic energy density (log \ke\ $\gtrsim$ 46 erg pc$^{-2}$). However, K--S tests show that the distributions of SCC-empty and SCC-laden pixels are not independent for any measured quantity.

 \subsection{NGC~2782E}

A $\approx$200 Myr-old minor merger, NGC~2782 exhibits an eastern tail that may be the distorted remnant of the smaller galaxy \citep{Smith94}. Interestingly, the optical tail extends beyond the \HI\ tail. Approximately 57\% of the $\approx$4 kpc-scale tail pixels exhibit \HI\ of supercritical column densities and turbulent velocity dispersions. Many pixels are unfitted ($f_{{\rm unfit}} \approx$ 0.24), so the fraction of high-column density, turbulent gas relative to the total \HI\ reservoir is higher.

The velocity field of the \HI\ debris varies smoothly across the eastern side of the debris (\shear\ $\lesssim$ 5 \kmskpc), and changes more sharply on the western side (\shear\ $\gtrsim$ 10 \kmskpc). Velocity dispersions can also range from $\sim$10 \kms\ on the eastern side to $\gtrsim$60 \kms\ in the west. The \ke\ map follows this trend accordingly. We compute a high $f_{{\rm SC}}$ value (0.96) for the 56 SCCs in this tail; thus most should be clusters. These nearly all populate the turbulent--high \nhi\ region of \mbox{Figure 5.7}. They also lie in pixels with log \ke\ $>$ 46.5 erg pc$^{-2}$, and span a range of \shear\ measurements (between 0--30 \kmskpc). 

The pixels with SCCs also comprise different distributions of \nhi, \vdisp, \ke, \shear, and \ke, with K--S tests producing results $<$5$\times$ 10$^{-4}$ in all cases. Cluster formation may occur in regions of preferentially higher column density, higher line-of-sight velocity dispersions, high kinetic energy densities, and a broad range of $\sim$10 kpc-scale velocity gradients in this tail. These gradients may be too broadly constructed (measuring differences over areas larger than the radio beam) to accurately trace the kinematic structure of local SCC-conducive environments, however, particularly where the velocity field varies rapidly towards the central galaxy.

\subsection{NGC~2782W} 

If NGC~2782E is the consumed minor galaxy of the NGC~2782 interacting pair, this western tail may be the \HI-rich tidal debris of the primary galaxy. While \citet{smith99} detect CO emission in the eastern tail, none is found in NGC~2782W by \citet{braine01}. Perhaps correspondingly, far fewer SCCs (6) are detected in this tail. However, as mentioned in Section \ref{sec:probs}, the work of \citet{TF12} and our star counts imply that $\approx$52\% of these sources should be star clusters. In particular, \citet{TF12} find these sources are $\approx$10$^4$ \msun, $\approx$ a few Myr old clusters. This is consistent with H$\alpha$ observations of \citet{K12} and their conclusion that ongoing star formation does occur in this tail, but at low efficiency compared to other extragalactic environments. For no properties do the distributions of SCC-carrying and SCC-devoid pixels appear statistically distinct.

The central debris shows large regions with very shallow velocity gradients (\shear\ $\lesssim$ 5 \kmskpc), and generally a less extensive range of line-of-sight velocity dispersions (\vdisp\ $\approx$ 10--30 \kms) on $\approx$4 kpc scales. Compared to NGC~2782E, This tail has qualitatively similar gas fractions in the turbulent--high \nhi\ quadrant of \mbox{Figure 5.8}.  The western tail has a lower fraction of unfitted pixels, however, ($f_{{\rm unfit}} \approx$ 0.1), and a higher percentage of gas with relatively low values of \shear. The distributions of \nhi, \vdisp, and \ke\ also peak at broadly lower values for NGC~2782W than for the eastern debris.

\subsection{NGC~2444}

A young ($\approx$100 Myr old) tail, the debris of NGC~2444 formed quickly from a possible high-speed encounter (\citealp{NGC2444_pub}; c.f.\ \citealp{koopmann}). Gas profiles indicate the \HI\ is kinematically disturbed (\vdisp $\approx$ 20--80 \kms\ for most of the pixels) on $\approx$8 kpc scales.There are no pixels with \nhi\ above the fiducial threshold, and the \shear\ map of \mbox{Figure 4.9} shows the velocity gradients across the tail are mostly below $\approx$10 \kmskpc.

5 of the SCCs detected lie within the tidal debris; $\approx$3 of them may be star clusters based on star counts ($f_{{\rm SC}}$ = 0.63). Moreover, the pixels upon which they appear superimposed do not seem to have different distributions of \nhi, \vdisp, \shear, or \ke\ than do all tail pixels (to the extent to small sample of SCC-containing pixels allows). Most ($\approx$95\%) of the $\sim$8 kpc-scale \HI\ has kinetic energy densities above 10$^{46}$ erg pc$^{-2}$ and has a relatively shallow velocity gradient, but does not have a widespread cluster population. If there are indeed a few clusters in the SCC source list for this system, either they did not need kpc-scale high column density \HI\ gas to form, our resolution is too poor to see more local concentrations of \HI, or previous star formation dispersed the gas efficiently in the past 100 Myr. Given the optical faintness of this tail (it does not register on WFPC2 images), it may be unlikely star formation feedback has played much of a role in this case. 

\subsection{NGC~2535} 

The approximately face-on NGC~2535 debris results from a young ($\approx$100 Myr) interaction with NGC~2536 \citep{hancock07}. Measurements of $\approx$6 kpc-scale velocity dispersions for these profiles ranges from 10--50 \kms, typically, with the regions of highest \vdisp\ and lowest \shear\ ($\lesssim$ 4 \kmskpc) closely mirroring the most optically prominent regions seen in M11. Most of the 36 in-tail SCCs also fall in these regions, $\approx$93\% of which should be star clusters. From K--S tests, the pixels that contain them appear to have generally higher values of \nhi, and lower values of \shear, but do not arise from different distributions of \vdisp\ and \ke\ compared to all tail pixels. Several pixels have multiple SCCs ``within," so their $p_{{\rm real}}$ values are higher ($>$0.94); these occupy the regions of turbulent, high \nhi, low \shear, and high \ke\ in \mbox{Figure 5.10}.

\subsection{NGC~6872E}

NGC~6872 is $\approx$150 Myr into an interaction with the smaller companion galaxy IC 4970 (\citealp{horellou}; \citealp{NGC6872_pub}). NGC~6872E is the easternmost FOV of the enormous eastern tail, featuring a moderate percentage of high column density \HI\ (about 50\% of the gas has log \nhi\ $>$ 20.6 cm$^{-2}$ on $\approx$16 kpc scales). The line-of-sight velocity dispersion ranges from $\approx$20 \kms\ from the tail edges to $\approx$100 \kms\ in the interior.

The central region of the debris is also host to an extensive population of SCCs (over a hundred are detected here), which is also coincident with areas of shallow large-scale velocity gradients (\shear\ $\lesssim$ 5 \kmskpc). For all considered properties except \shear, the pixels hosting SCCs comprise different distributions than those of tail pixels that do not; their distributions tend to peak at higher values of \nhi, \vdisp, and therefore also \ke. According to \mbox{Table 3}, about 94\% of these sources should be clusters, and pixels with multiple SCCs have even higher probabilities of containing at least one. As seen in \mbox{Figure 5.11}, of these pixels of highest cluster-containing probability, almost all have kinetic energy densities $>$10$^{47}$ erg pc$^{-2}$, and column densities $>$10$^{20.6}$ \cmtwo. The positions of the SCCs are also aligned with regions of H$\alpha$ emission (\citealp{bastian05}; \citealp{mihos93}).

There are still cluster candidates in pixels with ``subcritical" \nhi, however---including roughly 6--10\% of them that lie in regions unfitted by several \HI\ properties---and many of these may be real. It is then again a question of whether this \nhi\ threshold is important, these clusters had formed in previously high column density \HI\ in the past 150 Myr that has since dispersed or been consumed, or instrumental effects prevent an assessment of \HI\ properties at crucially lower (pc--kpc) scales. A non-negligible fraction of the tail pixels are unfitted ($f_{{\rm unfit}} \approx 0.15$), so these effects are certainly possible. For NGC~6872, the physical scale may be especially important, as the cube for this system has a particularly large beam, both in angular and physical extent. 

\subsection{NGC~6872W} 

This western debris highlights the base of the eastern tail of the NGC~6872 system, displaying a central clump of moderate \HI\ emission around the \citet{aparna07} limit, immersed in a tail of lower column density gas ($\approx$80\% of the in-tail pixels) that follows the optical structure seen in M11. Flanking the \HI\ tail is bright optical debris that both contains SCCs and is devoid of detectable neutral hydrogen. This helps account for the large fractions of unfitted SCC-hosting and SCC-empty pixels ($\approx$0.4 and 0.3, respectively).  The densest regions exhibit velocity dispersions up to $\approx$100 \kms. Here the velocity structure is not of a smoothly rotating, extended disk, but has a number of kinematically distinct components, as elaborated in the modeling of \citet{NGC6872_pub}. Thus, the \shear\ map, showing measurements of line-of-sight velocity gradients across pixels separated by more than one beam, displays areas with seemingly steep gradients. Internally, however, the local gradients may be shallower.

In either case, the 101 SCCs detected appear as widely distributed as the kinematically varied \HI\ debris. K--S tests demonstrate that the pixels that contain them are generally not different in terms of \nhi, \vdisp, or \ke, and while their distribution of \shear\ values appears statistically distinct, the aforementioned spatial resolution of these NGC~6872 data make this difficult to interpret. 18 of these sources exist in exclusively optically defined debris, and the remaining sources mostly in low-\nhi\ areas are also likely to be actual clusters ($f_{{\rm SC}}$ = 0.93). Even the pixels with the highest probabilities of having at least one cluster generally have \nhi\ values below the fiducial cutoff, while their high \vdisp\ measurements ensure that their mechanical energy densities remain all above log \ke\ = 46 erg pc$^{-2}$.

\subsection{NGC~1614N} 

The northern region of the NGC~1614 tidal debris is \HI-poor ($f_{{\rm unfit}} \approx$ 0.89), with concrete detections only at the very northern and southern edges of the central host galaxy. It is conversely prominent in the optical, with galactic material arranged in warped rings around the merger remnant of an interaction thought to have begun $\approx$750 Myr ago (\citealp{neff90} and references therein). The few pixels with detectable \HI\ emission show high \vdisp\ ($\approx$75 \kms). These isolated clumps of neutral hydrogen have high column densities on $\approx$5 kpc scales, and appear to have larger-scale velocity gradients \shear\ $\sim$10 \kmskpc, though the paucity of \HI\ pixels makes this comparative measurement admittedly uncertain.

19 SCCs are found interspersed in the optical debris, with only 3 lodged in the \HI\ debris (several occupy the same pixels). Thus, K--S tests cannot be performed on the different pixel distributions. But the majority of the SCCs should be clusters ($f_{{\rm SC}}$= 0.95), so clusters are detected in this tail regardless of \HI\ content. NGC~1614N results from a very late-stage interaction, so these sources may also be several hundred Myr old and had initially formed while surrounded by an \HI\ medium that has since dispersed, been ionized, or consumed. While the photometry of M11 cannot age-date the SCCs, their red $V-I$ colors ($V-I$ $\sim$ 1) may imply these clusters are older (or reddened by extinction). 

NGC~1614 is luminous at infrared wavelengths and contains a powerful nuclear starburst \citep{alonso01}. It is possible that the \HI\ in the northern tail has been eliminated by central winds, photoionized by the central starburst, or photoionized by past star formation. \citet{hibbard2000} contends that a mean R-band surface brightness of $\approx$23 mag arcsec$^{-2}$---depending on the \HI\ column density and dimensions of the tidal debris---is required for the stellar population to effectively rid itself of the warm neutral medium on these scales. NGC~1614N is one of the few debris fields in the M11 sample to exceed this surface brightness (its F606W-band surface brightness is $\approx$22 mag arcsec$^{-2}$, which in terms of photometric bandpasses is close to $R$), so it is possible mass \textit{in situ} photoionization of the tidal \HI\ has occurred in the past. 

\subsection{NGC~1614S} 

The \HI\ of the southern tail of NGC~1614 is confined to a ``stub" extending from the southernmost tip of the merger remnant; most tail pixels are optically defined ($f_{{\rm unfit}} \approx$ 0.89). Its \HI\ characteristics are similar to those of the northern debris; the measurable \HI\ here is of comparably high column density (log \nhi\ $\gtrsim$ 20.6 \cmtwo) and with line-of-sight velocity dispersions ranging from the instrumental limit to $\approx$75 \kms. The ranges of velocity gradients are also similar, with \shear\ $\approx$ 5--15 \kmskpc. 

The main difference is the number and distribution of SCCs. 30 sources are located in the combined optical and \HI\ tail, with most of them falling in the \HI\ stub. Once more, a high fraction of them are probable star clusters ($f_{{\rm SC}}$ = 0.96, with $p_{{\rm SC}}$ exceeding this number for many pixels that contain more than one SCC). The sources here are very blue ($V-I$ $\lesssim$ 0; M11), indicating star formation might be relatively recent and/or ongoing. These sources trace out the high-\nhi, high-\ke, and turbulent regions of the \HI\ parameter space of \mbox{Figure 5.14}, and the pixels that have them do not appear different with respect to those properties than \HI\ pixels overall. 

It is however curious that the positions of these SCCs is offset by a few arcseconds from the regions of highest column density as seen in the \nhi\ map of Figure~4.14. The \HI\ may be in the process of being locally consumed or swept out by current star formation. Approximately 42\% of the equally high probability SCCs are found in optically bright yet \HI-unfitted regions, so these sources may be further examples of past feedback and \HI\ dispersal or consumption sometime over the course of this tail's $\approx$750 Myr lifetime. Unfortunately, $VI$ observations of M11 do not permit accurate age-dating of cluster candidates to provide evidence for this age-dependent \HI\ removal. If generous K--S probability cutoff is employed, the distributions of \shear\ between pixels and without SCCs appear marginally independent. But again the low spatial resolution for this especially distant object may confuse this quantification of the large-scale velocity field. 

\subsection{NGC~4038A/B/C}

This well-studied merger has two large, symmetric tails of ages $\approx$450 Myr (K03 and references therein). WFPC2 images of the ABC positions along these tails reveal them to be optically faint, diffuse, and all but devoid of SCCs (K03; M11). As a whole they are poor in \HI, with very little gas at or above the \citet{aparna07} threshold on $\approx$kpc scales. Across the WFPC2 pointings, it appears $\approx$30--80\% of the tail pixels are classified as unfitted by at least one \HI\ characteristic. This gas is relatively kinematically unperturbed, with \vdisp\ $\lesssim$ 25 \kms, and much of the gas showing dispersions less than the turbulent-thermal cutoff. Consequently, less than $\approx$5--20\% of the gas has kinetic energy densities $>$10$^{46}$ erg pc$^{-2}$. 

3 SCCs are found in NGC~4038A, and 1 is found in the NGC~4038B and NGC~4038C FOVs each. They all have low values for $f_{{\rm SC}}$ and their host pixels have correspondingly low values of $p_{{\rm real}}$ (0--0.24); thus they are are likely point source contaminants. They populate too few pixels to conduct K--S tests between distributions of \HI\ characteristics. Unlike the active, luminous interior and bridge regions of the merger, the tails of this system are extraordinarily quiescent in terms of ongoing star formation.

\subsection{NGC~3256E/W} 

Dynamically, the tails of NGC~3256 are about 400 Myr old (K03), resulting from an ongoing major merger of two similarly massed galaxies. The channel width of the NGC~3256 data cube is wide ($\approx$33.5 \kms), so each pixel has few channels of signal. Combined with the beam size and sensitivity of the data, uncertainties in \HI\ properties are high for pixels with \nhi\ $\lesssim$ 10$^{20}$ \cmtwo. The vertical ``tail" of constant \vdisp\ in Figures 5.18 and 5.19 represents the instrumental limit in determining velocity dispersions. Much of the \HI\ gas in these tails is of high column density and line-of-sight velocity dispersions on scales of $\approx$5 kpc; we find $\approx$72\% and 88\% of tail pixels has those properties in the eastern and western tails, respectively. The distribution of \shear\ values in both tails is very broad, ranging from 0--50 \kmskpc, while the distribution of that property is skewed towards lower values for NGC~3256W.

The  western tail has fewer pixels with at least one unfitted \HI\ characteristic (by a factor of $\approx$4--10). 32 SCCs are located in the eastern tail, and 51 are in the west. Given the calculated values for $f_{{\rm SC}}$ (0.50--0.60), a little over a dozen sources in the east and about 30 sources in the west may be genuine star clusters. This strong level of contamination results from this system's low galactic latitude ($b \approx 11^{\circ}$), and led K03 to conclude that the eastern tail did not have a statistically significant widespread cluster population, and the western tail did. This is still true; the eastern tail may have only a handful of real clusters, but they are spread throughout the $\approx$1720 kpc$^2$ tail area and do not constitute a significant population.

Most of the SCCs in either tail populate the high \nhi\ and high \vdisp\ parameter space of the neutral medium. According to K--S tests, pixels containing SSCs in the eastern tail may have typically higher line-of-sight velocity dispersions, but do not comprise unique distributions compared to all pixels in terms of other measured properties. Conversely, the pixels hosting cluster candidates in NGC~3256W might have generally shallower velocity gradients with a generous interpretation of that K--S result. The central ``spine" of the western debris is a region of low \shear\ ($\lesssim$5 \kmskpc), an attribute absent in the eastern tail. Perhaps a combination of this and the higher fraction of high column density gas contributes to the factor of $\sim$3 increase of possible clusters in the western tail over the east.

\subsection{NGC~7252E/W}

NGC~7252 is a late-stage ($\approx$730 Myr old; K03) merger between two comparably massed galaxies. The channel width of this system's data cube is the widest of the debris sample ($\approx$42.5 \kms). The \HI\ in both debris fields is relegated to the tips of the straight and narrow eastern and western tails, and is characterized by low column and mass densities on $\approx$1.5 kpc scales and steep velocity gradients (\shear\ $\gtrsim$ 15 \kmskpc) on larger scales. The latter may be attributed in part to the velocities ascribed to the low-signal edge pixels, which can be uncertain. The moderately turbulent kinematics of the gas helps to offset the low mass densities and maintains the kinetic energy densities of much of the \HI\ above 10$^{46}$ erg pc$^{-2}$.

Comparable numbers of SCCs are in either tail (11 and 17 for east and west, respectively), with fairly comparable values of $f_{{\rm SC}}$ (0.80 and 0.89). In 7252E these sources are widely distributed, many of which are situated in the \HI-poor optical debris ($f_{{\rm unfit}}^{{\rm SCC}} \approx 0.40$). Two neighboring sources are located in a clump of diffuse optical emission, which is slightly offset from the central high-\nhi\ region seen in \mbox{Figure 4.20}. Pixels that contain the SCCs in this tail do not seem to arise from dissimilar distributions of \nhi, \vdisp, \ke, or \shear\ than all \HI\ tail pixels.

The same is true for the western tail, except for the velocity dispersion, for which SCC-laden pixels seem to have a more narrow range of values (\vdisp\ $\approx$ 10--30 \kms). 11 sources are dispersed throughout the primarily optically defined tidal tail, with $\approx$40\% in especially \HI\-poor regions (in comparison, $f_{{\rm unfit}} \approx$ 0.30). A few \HI\ pixels in the only high mass density region at the end of the tail contain 6 of them. This region of the debris has the lowest \shear, and is the only part of \mbox{Figure 5.21} that shows \nhi\ $>$ 10$^{20.6}$ \cmtwo. These sources belong to the prominent optical/\HI-defined TDG in the western tail \citep{NGC7252_pub}. The fact that this one region of obvious clustered star formation is the only region of high column density \HI\ and contains a large number of SCCs (the superimposed black points in \mbox{Figure 5.21}) indicates that, at least in this case, the \citet{aparna07} \nhi\ threshold may be important in forming star clusters. These few \HI\ pixels do not stand out in the distribution of \ke; a combination of high mechanical energy densities and column densities may be observationally important in finding SCCs in situations like this.

\subsection{NGC~3921S}  

NGC~3921 is the most distant interaction of the sample (84.5 Mpc), and accordingly its \HI\ cube has the worst spatial resolution of the debris surveyed here ($\approx$2 kpc for each pixel; roughly 4 times that for the size of the beam). The optical completeness limit for this tail as analyzed by K03 is also the worst of the sample, with a $V$-band 50\% completeness limit at $M_{{\rm V}}$ = -8.8. Therefore, it is probable that some SCCs are missing from the source list.

Its \HI\ debris has velocity dispersions on this scale of $\approx$5--30 \kms, and has a smoothly varying velocity structure with shallow larger-scale velocity gradients (\shear\ $\lesssim$ 5 \kmskpc). Approximately 10\% of the tail pixels has a high column density, and is confined to the long tail-shaped structure that extends from the north to the south of the \nhi\ map. Of the 14 SCCs in the \HI\ or optical debris, only two are embedded in this relatively higher-\nhi\ material, and most are dispersed throughout the debris. About 87\% of these sources should be clusters, and one pixel that contains two sources---part of a short chain of sources that extends from the north of the optical image---correspondingly has a high chance of containing at least one massive cluster.

These SCC-supporting pixels do not appear to have statistically unique distributions of measured \HI\ quantities compared to all \HI\ pixels, and all but one of them lie in regions of low column density (log \nhi\ $<$ 20.6 \cmtwo). Their moderate velocity dispersions mean that their kinetic energies are still greater than the fiducial 10$^{46}$ erg pc$^{-2}$. As is the case with other tails, the low spatial resolution, feedback from star formation (certainly possible over the $\approx$460 Myr lifetime of the tail; K03), and/or a revised \nhi\ threshold might at least partially explain the detection of these handful of reasonably probable clusters.The particularly poor resolution makes it especially possible that sub-kpc clumps of denser \HI\ gas are being effectively diluted by the size of the beam for this system, especially when they are found in closer, better resolved tails.



\begin{thebibliography}{}
\setlength{\parskip}{0.5pt}




\bibitem[Ajhar, Blakeslee, \& Tonry(1994)]{Ajhar} Ajhar, E.~A., Blakeslee, J.~P., \& Tonry, J.~L.\ 1994, \aj, 108, 2087
\bibitem[Alonso-Herrero et al.(2001)]{alonso01} Alonso-Herrero, A., Engelbracht, C.~W., Rieke, M.~J., Rieke, G.~H., \& Quillen, A.~C.\ 2001, \apj, 546, 952
\bibitem[Anders et al.(2009)]{anders} Anders, P., Lamers, H.~J.~G.~L.~M., \& Baumgardt, H.\ 2009, \aap, 502, 817  
\bibitem[Appleton et al.(1987)]{NGC2444_pub} Appleton, P. N., Ghigo, F. D, van Gorkom, J. H.,Schombert, J. M., \& Struck-Marcell, C. 1987, Nature, 330, 140








\bibitem [Bastian et al.(2005)] {bastian05} Bastian, N., Hempel, M., Kissler-Patig, M.,Homeier, N.L., \& Trancho, G.~2005, A\&A, 435, 65
\bibitem[Bastian \& Gieles(2008)]{bastian08} Bastian, N., \& Gieles, M.\ 2008, Mass Loss from Stars and the Evolution of Stellar Clusters, 388, 353 
\bibitem[Bigiel et al.(2010)]{bigiel10} Bigiel, F., Leroy, A., Walter, F., Blitz, L., Brinks, E., de Blok, W.~J.~G., \& Madore, B.\ 2010, \aj, 140, 1194 
\bibitem[Block et al.(2010)]{block10} Block, D.~L., Puerari, I., Elmegreen, B.~G., \& Bournaud, F.\ 2010, \apjl, 718, L1
\bibitem[Blitz \& Rosolowsky(2004)]{blitz04} Blitz, L., \& Rosolowsky, E.\ 2004, \apjl, 612, L29
\bibitem[Blitz \& Rosolowsky(2006)]{blitz06} Blitz, L., \& Rosolowsky, E.\ 2006, \apj, 650, 933 
\bibitem[Bournaud(2010)]{bournaud10} Bournaud, F.\ 2010, Galaxy Wars: Stellar Populations and Star Formation in Interacting Galaxies, 423, 177 
\bibitem[Bournaud et al.(2004)]{bournaud04} Bournaud, F., Duc, P.-A., Amram, P., Combes, F., \& Gach, J.-L.\ 2004, \aap, 425, 813 
\bibitem[Bournaud(2011)]{bournaud11} Bournaud, F.\ 2011, EAS Publications Series, 51, 107
\bibitem[Braine et al.(2001)]{braine01} Braine, J., Duc, P.-A., Lisenfeld, U., Charmandaris, V., Vallejo, O., Leon, S., \& Brinks, E.\ 2001, \aap, 378, 51
\bibitem[Briggs et al.(1999)]{robust} Briggs, D.~S., Schwab, F.~R., \& Sramek, R.~A.\ 1999, Synthesis Imaging in Radio Astronomy II, 180, 127 
\bibitem[Brinks et al.(2004)]{brinks04} Brinks, E., Duc, P.-A., \& Walter, F.\ 2004, Recycling Intergalactic and Interstellar Matter, 217, 532









\bibitem[Calzetti et al.(2007)]{calzetti07} Calzetti, D., Kennicutt, R.~C., Engelbracht, C.~W., et al.\ 2007, \apj, 666, 870
\bibitem[Casasola et al.(2004)]{casasola04} Casasola, V., Bettoni, D., \& Galletta, G.\ 2004, \aap, 422, 941
\bibitem[Chien(2010)]{chien10} Chien, L.\ 2010, Astronomical Society of the Pacific Conference Series, 423, 197 
\bibitem[Cluver et al.(2010)]{cluver09} Cluver, M.~E., Appleton, P.~N., Boulanger, F., et al.\ 2010, \apj, 710, 248









 
\bibitem[de Avillez \& Breitschwerdt(2007)]{deavillez07} de Avillez, M.~A., \& Breitschwerdt, D.\ 2007, \apjl, 665, L35
\bibitem[de Grijs(2010)]{RdG09} de Grijs, R.\ 2010, Royal Society of London Philosophical Transactions Series A, 368, 693
\bibitem[de Grijs et al.(2002)]{RdG} de Grijs, R., Lee, J.~T., Mora Herrera, M.~C., Fritze-v.~Alvensleben, U., \& Anders, P.\ 2002, New Astronomy, 8, 155
\bibitem[Di Matteo et al.(2008)]{dim08} Di Matteo, P., Bournaud, F., Martig, M., Combes, F., Melchior, A.-L., \& Semelin, B.\ 2008, \aap, 492, 31
\bibitem[Duc et al.(2000)] {NGC2992_pub} Duc, P.-A., Brinks, E., Springel, V., Pichardo, B., Weilbacher, P., \& Mirabel, I.F. 2000, AJ, 120, 1238
\bibitem[Duc et al.(2004)]{duc04} Duc, P.-A., Bournaud, F., \& Masset, F.\ 2004, \aap, 427, 803 





\bibitem[Efremov \& Ivanov(1987)]{efremov87} Efremov, I.~N., \& Ivanov, G.~R.\ 1987, \apss, 129, 39
\bibitem[Elmegreen(1993)]{elmegreen93} Elmegreen, B.~G.\ 1993, \apj, 411, 170
\bibitem[Elmegreen(2008)]{elmegreen08} Elmegreen, B.~G.\ 2008, \apj, 672, 1006
\bibitem[Elmegreen \& Efremov(1997)]{elmegreen97} Elmegreen, B. G., \& Efremov, Y. N. 1997, ApJ, 480, 235
\bibitem[Elmegreen et al.(2003)]{elmegreen03} Elmegreen, B.~G., Elmegreen, D.~M., \& Leitner, S.~N.\ 2003, \apj, 590, 271
\bibitem[English \& Freeman(2001)]{NGC1487_pub} English, J., \& Freeman, K.~C.\ 2001, Gas and Galaxy Evolution, 240, 858
\bibitem[English et al.(2003)]{NGC3256_pub} English, J., Norris, R.~P., Freeman, K.~C., \& Booth, R.~S.\ 2003, \aj, 125, 1134













\bibitem[Getts(2001)]{getts01} Getts, T.~J.\ 2001. PhD Thesis, University of Virginia.
\bibitem[Gieles \& Bastian(2008)]{gieles08} Gieles, M., \& Bastian, N.\ 2008, \aap, 482, 165
\bibitem[Gil de Paz et al.(2005)]{gdp05} Gil de Paz, A., Madore, B.~F., Boissier, S., et al.\ 2005, \apjl, 627, L29 
\bibitem[Girardi et al.(2008)]{girardi} Girardi, L., et al.\ 2008, \pasp, 120, 583 
\bibitem[Guillard et al.(2012)]{guillard12} Guillard, P., Boulanger, F., Pineau des For{\^e}ts, G., et al.\ 2012, \apj, 749, 158 








\bibitem[Hancock et al.(2007)]{hancock07} Hancock, M., Smith, B.~J., Struck, C., Giroux, M.~L., Appleton, P.~N., Charmandaris, V., \& Reach, W.~T.\ 2007, \aj, 133, 676
\bibitem[Haynes(1979)]{haynes79} Haynes, M.~P.\ 1979, \aj, 84, 1830 
\bibitem[Heiles(1987)]{heiles87} Heiles, C.\ 1987, \apj, 315, 555
\bibitem[Hibbard et al.(1994)]{NGC7252_pub} Hibbard, J.~E., Guhathakurta, P., van Gorkom, J.~H., \& Schweizer, F.\ 1994, \aj, 107, 67
\bibitem[Hibbard et al.(2001)]{NGC4038_pub} Hibbard, J.~E., van der Hulst, J.~M., Barnes, J.~E., \& Rich, R.~M.\ 2001, \aj, 122, 2969 
\bibitem[Hibbard \& van Gorkom(1996)] {NGC520_pub} Hibbard, J.E. \& van Gorkom, J.H., 1996, AJ, 111,655
\bibitem[Hibbard et al.(2000)]{hibbard2000} Hibbard, J.~E., Vacca, W.~D., \& Yun, M.~S.\ 2000, \aj, 119, 1130 
\bibitem[Hibbard \& Yun(1996)]{NGC1614_pub} Hibbard, J.~E., \& Yun, M.~S.\ 1996, Cold Gas at High Redshift, 206, 47 
\bibitem[Holtzman et al.(1995)]{holtzman95} Holtzman, J.~A., Burrows, C.~J., Casertano, S., Hester, J.~J., Trauger, J.~T., Watson, A.~M., \& Worthey, G.\ 1995, \pasp, 107, 1065
\bibitem[Horellou \& Koribalski(2003)] {horellou} Horellou, C., \& Koribalski, B.\ 2003, Ap\&SS, 284, 499
\bibitem[Horellou \& Koribalski(2007)]{NGC6872_pub} Horellou, C. \& Koribalski, B.\ 2007, \aap, 464, 155









\bibitem[Joung \& Mac Low(2006)]{joung06} Joung, M.~K.~R., \& Mac Low, M.-M.\ 2006, \apj, 653, 1266







\bibitem[Kaufman et al.(1997)]{NGC2535_pub} Kaufman, M., Brinks, E., Elmegreen, D.~M., Thomasson, M., Elmegreen, B.~G., Struck, C., \& Klaric, M.\ 1997, \aj, 114, 2323
\bibitem[Kaufman et al.(2012)]{kaufman11} Kaufman, M., Grupe, D., Elmegreen, B.~G., et al.\ 2012, \aj, 144, 156 
\bibitem[Keel \& Borne(2003)]{keel03} Keel, W.~C., \& Borne, K.~D.\ 2003, \aj, 126, 1257
\bibitem[Kennicutt(1989)]{kennicutt89} Kennicutt, R.~C., Jr.\ 1989, \apj, 344, 685
\bibitem[Kennicutt(1998)]{KSlaw} Kennicutt, R.~C., Jr.\ 1998, \apj, 498, 541
\bibitem[Knierman(2007)]{K07} Knierman, K.\ 2007, Island Universes - Structure and Evolution of Disk Galaxies, 307
\bibitem[Knierman et al.(2003)]{K03} Knierman, K.A., Hunsberger, S.D., Gallagher, S.C., Charlton, J.C., Whitmore, B., Kundu, A., Hibbard, J., \& Zaritsky, D. 2003, AJ, 126, 1227
\bibitem[Knierman et al.(2012)]{K12} Knierman, K., Knezek, P.~M., Scowen, P., Jansen, R.~A., \& Wehner, E.\ 2012, \apjl, 749, L1
\bibitem[Kolmogorov(1941)]{kolmogorov41} Kolmogorov, A.\ 1941, Akademiia Nauk SSSR Doklady, 30, 301
\bibitem[Konstantopoulos(2010)]{iraklis09} Konstantopoulos, I.~S.\ 2010, Galaxy Wars: Stellar Populations and Star Formation in Interacting Galaxies, 423, 135
\bibitem[Koopmann et al.(2008)]{koopmann} Koopmann, R.~A., et al.\ 2008, \apjl, 682, L85 
\bibitem[Krumholz et al.(2009)]{krumholz09} Krumholz, M.~R., McKee, C.~F., \& Tumlinson, J.\ 2009, \apj, 693, 216
\bibitem[Krumholz \& Thompson(2007)]{krumholz07} Krumholz, M.~R., \& Thompson, T.~A.\ 2007, \apj, 669, 289
\bibitem[Kundu \& Whitmore(2001)]{Kundu} Kundu, A., \& Whitmore, B. C. 2001, AJ, 121, 2950 








\bibitem[Larsen \& Richtler(2000)]{larsen00} Larsen, S.~S., \& Richtler, T.\ 2000, \aap, 354, 836
\bibitem[Lee \& Lee(2005)]{lee05} Lee, H.~J., \& Lee, M.~G.\ 2005, Journal of Korean Astronomical Society, 38, 345 
\bibitem[Leroy et al.(2008)]{leroy08} Leroy, A.~K., Walter, F., Brinks, E., et al.\ 2008, \aj, 136, 2782
\bibitem[Liu et al.(2011)]{liu11} Liu, G., Koda, J., Calzetti, D., Fukuhara, M., \& Momose, R.\ 2011, \apj, 735, 63









\bibitem[Mac Low(1999)]{maclow99} Mac Low, M.-M.\ 1999, \apj, 524, 169
\bibitem[Ma{\'{\i}}z Apell{\'a}niz(2009)]{maiz09} Ma{\'{\i}}z Apell{\'a}niz, J.\ 2009, \apj, 699, 1938
\bibitem[Mannucci et al.(2005)]{mannucci05} Mannucci, F., Della Valle, M., Panagia, N., et al.\ 2005, \aap, 433, 807 
\bibitem[Martinez-Badenes et al.(2012)]{martinez12} Martinez-Badenes, V., Lisenfeld, U., Espada, D., et al.\ 2012, \aap, 540, A96 
\bibitem[Maybhate et al.(2007)]{aparna07} Maybhate, A., Masiero, J., Hibbard, J.~E., Charlton, J.~C., Palma, C., Knierman, K.~A., \& English, J.\ 2007, \mnras, 381, 59 
\bibitem[McKee \& Ostriker(1977)]{mckee77} McKee, C.~F., \& Ostriker, J.~P.\ 1977, \apj, 218, 148
\bibitem[Mihos et al.(1993)]{mihos93} Mihos, J.~C., Bothun, G.~D., \& Richstone, D.~O.\ 1993, \apj, 418, 82 
\bibitem[Mihos \& Hernquist(1994)]{m&h94} Mihos, J.~C., \& Hernquist, L.\ 1994, \apjl, 431, L9
\bibitem[Mullan et al.(2011)]{M11} Mullan, B., Konstantopoulos, I.~S., Kepley, A.~A., et al.\ 2011, \apj, 731, 93










\bibitem[Neff et al.(2005)]{neff05} Neff, S.~G., et al.\ 2005, \apjl, 619, L91
\bibitem[Neff et al.(1990)]{neff90} Neff, S.~G., Hutchings, J.~B., Standord, S.~A., \& Unger, S.~W.\ 1990, \aj, 99, 1088 
%








\bibitem[Peterson(1993)]{Peterson}  Peterson, C. J. 1993, in \textit{ASP Conf. Ser. 50, Structure and Dynamics of Globular Clusters}, ed. S. G. Djorgovsky \& G. Meylan (San Francisco: ASP), 337
\bibitem[Portegies Zwart et al.(2010)]{pzwart10} Portegies Zwart, S.~F., McMillan, S.~L.~W., \& Gieles, M.\ 2010, \araa, 48, 431












\bibitem[Renaud et al.(2009)]{renaud09} Renaud, F., Boily, C.~M., Naab, T., \& Theis, C.\ 2009, \apj, 706, 67 
\bibitem[Robin et al.(2003)]{starcounts} Robin, A.~C., Reyl{\'e}, C., Derri{\`e}re, S., \& Picaud, S.\ 2003, \aap, 409, 523








 

\bibitem[Sault et al.(1995)]{miriad} Sault, R.~J., Teuben, P.~J., \& Wright, M.~C.~H.\ 1995, Astronomical Data Analysis Software and Systems IV, 77, 433
\bibitem[Saviane, Hibbard, \& Rich(2004)]{saviane} Saviane, I., Hibbard, J.~E., \& Rich, R.~M.~R.\ 2004,AJ, 127, 660
\bibitem[Schweizer(1978)]{schweizer78} Schweizer, F.\ 1978, in IAU Symposium, Vol.\ 77, Structure and Properties of Nearby Galaxies, ed.\ E.\ M.\ Berkhuijsen \& R.\ Wielebinski, 279
\bibitem[Schweizer et al.(1996)]{schweizer96} Schweizer, F., Miller, B.~W., Whitmore, B.~C., \& Fall, S.~M.\ 1996, \aj, 112, 1839
\bibitem[Schaye(2004)]{schaye} Schaye, J.\ 2004, \apj, 609, 667
\bibitem[Schlegel, Finkbeiner, \& Davis(1998)]{schlegel} Schlegel, D.~J., Finkbeiner, D.~P., \& Davis, M.\ 1998, \apj, 500, 525 
\bibitem[Sengupta et al.(2012)]{sengupta11} Sengupta, C., Saikia, D.~J., \& Dwarakanath, K.~S.\ 2012, \mnras, 420, 2 
\bibitem[Simkin et al.(1987)]{simkin87} Simkin, S.~M., van Gorkom, J., Hibbard, J., \& Su, H.-J.\ 1987, Science, 235, 1367
\bibitem[Skillman(1987)]{skillman87} Skillman, E.~D.\ 1987, in Lonsdale Persson C.~J., ed., Star Formation in Galaxies. NASA Conference Publication, 2466, 263
\bibitem[Skillman \& Bothun(1986)]{skillman86} Skillman, E.~D., \& Bothun, G.~D.\ 1986, \aap, 165, 45
\bibitem[Sellwood \& Balbus(1999)]{sellwood99} Sellwood, J.~A., \& Balbus, S.~A.\ 1999, \apj, 511, 660
\bibitem[Smith(1991)]{Smith91} Smith, B.~J.\ 1991, \apj, 378, 39 
\bibitem[Smith(1994)]{Smith94} Smith, B.~J.\ 1994, \aj, 107, 1695
\bibitem[Smith(1997)]{Smith97} Smith, B.~J.\ 1997, \aj, 114, 2177 
\bibitem[Smith et al.(1999)]{smith99} Smith, B.~J., Struck, C., Kenney, J.~D.~P., \& Jogee, S.\ 1999, \aj, 117, 1237 
\bibitem[Spitzer(1990)]{spitzer90} Spitzer, L., Jr.\ 1990, \araa, 28, 71
\bibitem[Stanford \& Balcells(1991)]{stanford91} Stanford, S.~A., \& Balcells, M.\ 1991, \apj, 370, 118










\bibitem[Tamburro et al.(2009)]{tamburro} Tamburro, D., Rix, H.-W., Leroy, A.~K., et al.\ 2009, \aj, 137, 4424
\bibitem[Tasker \& Bryan(2006)]{tasker06} Tasker, E.~J., \& Bryan, G.~L.\ 2006, \apj, 641, 878
\bibitem[Tasker \& Bryan(2008)]{tasker08} Tasker, E.~J., \& Bryan, G.~L.\ 2008, \apj, 673, 810
\bibitem[Thilker et al.(2005)]{thilker05} Thilker, D.~A., Bianchi, L., Boissier, S., et al.\ 2005, \apjl, 619, L79
\bibitem[Thornton et al.(1998)]{thornton98} Thornton, K., Gaudlitz, M., Janka, H.-T., \& Steinmetz, M.\ 1998, \apj, 500, 95 
\bibitem[Toomre(1977)]{toomre77} Toomre, A. 1977, in ``The Evolution of Galaxies and Stellar Populations'', eds. B. M. Tinsley and R. B. Larson (Yale University Press, New Haven), p.~401
\bibitem[Torres-Flores et al.(2012)]{TF12} Torres-Flores, S., de Oliveira, C.~M., de Mello, D.~F., Scarano, S., \& Urrutia-Viscarra, F.\ 2012, \mnras, 421, 3612
\bibitem[Tran et al.(2003)]{tran} Tran, H.~D., et al.\ 2003, ApJ, 585, 750
\bibitem[Trancho et al.(2012)]{trancho12} Trancho, G., Konstantopoulos, I.~S., Bastian, N., et al.\ 2012, \apj, 748, 102 









\bibitem[van den Bosch(2000)]{vdb} van den Bosch, F.~C.\ 2000, \apj, 530, 177
\bibitem[van der Hulst(1979)]{vdh79} van der Hulst, J.~M.\ 1979, \aap, 71, 131
\bibitem[van der Hulst et al.(2001)]{NGC4747_pub} van der Hulst, J.~M., van Albada, T.~S., \& Sancisi, R.\ 2001, Gas and Galaxy Evolution, 240, 451
\bibitem[Verschuur \& Kellermann(1988)]{HIeqn} Verschuur, G.~L., \& Kellermann, K.~I.\ 1988, Galactic and extra-galactic radio astronomy. Berlin: Springer, 1988, 2nd ed., edited by Verschuur, Gerrit L.; Kellermann, Kenneth I.









\bibitem[Wada et al.(2002)]{wada02} Wada, K., Meurer, G., \& Norman, C.~A.\ 2002, \apj, 577, 197
\bibitem[Walter et al.(2008)]{things} Walter, F., Brinks, E., de Blok, W.~J.~G., et al.\ 2008, \aj, 136, 2563
\bibitem[Walter et al.(2006)]{walter06} Walter, F., Martin, C.~L., \& Ott, J.\ 2006, \aj, 132, 2289
\bibitem[Weidner et al.(2010)]{weidner10} Weidner, C., Bonnell, I.~A., \& Zinnecker, H.\ 2010, \apj, 724, 1503 
\bibitem[Wevers et al.(1984)]{wevers} Wevers, B.~M.~H.~R., Appleton, P.~N., Davies, R.~D.,\& Hart, L.\ 1984, A\&A, 140, 125
\bibitem[Whitmore, Chandar, \& Fall(2007)]{whitmore07} Whitmore, B.~C., Chandar, R., \& Fall, S.~M.\ 2007, \aj, 133, 1067 
\bibitem[Whitmore et al.(1995)]{whitmore95} Whitmore, B.~C., Sparks, W.~B., Lucas, R.~A., Macchetto, F.~D., \& Biretta, J.~A.\ 1995, \apjl, 454, L73 
\bibitem[Whitmore et al.(1999)] {whitmore99} Whitmore, B., Zhang, Q., Leitherer, C., Fall, S.M., Schweizer, F., \& Miller, B. 1999, AJ, 118, 1551
\bibitem[Wolfire et al.(1995)]{wolfire95} Wolfire, M.~G., Hollenbach, D., McKee, C.~F., Tielens, A.~G.~G.~M., \& Bakes, E.~L.~O.\ 1995, \apj, 443, 152
\bibitem[Wolfire et al.(2003)]{wolfire03} Wolfire, M.~G., McKee, C.~F., Hollenbach, D., \& Tielens, A.~G.~G.~M.\ 2003, \apj, 587, 278
\bibitem[Wong \& Blitz(2002)]{wong02} Wong, T., \& Blitz, L.\ 2002, \apj, 569, 157









\bibitem[Yun et al.(1994)]{yun94} Yun, M.~S., Ho, P.~T.~P., \& Lo, K.~Y.\ 1994, \nat, 372, 530








\end{thebibliography}
\end{document}